January 2015

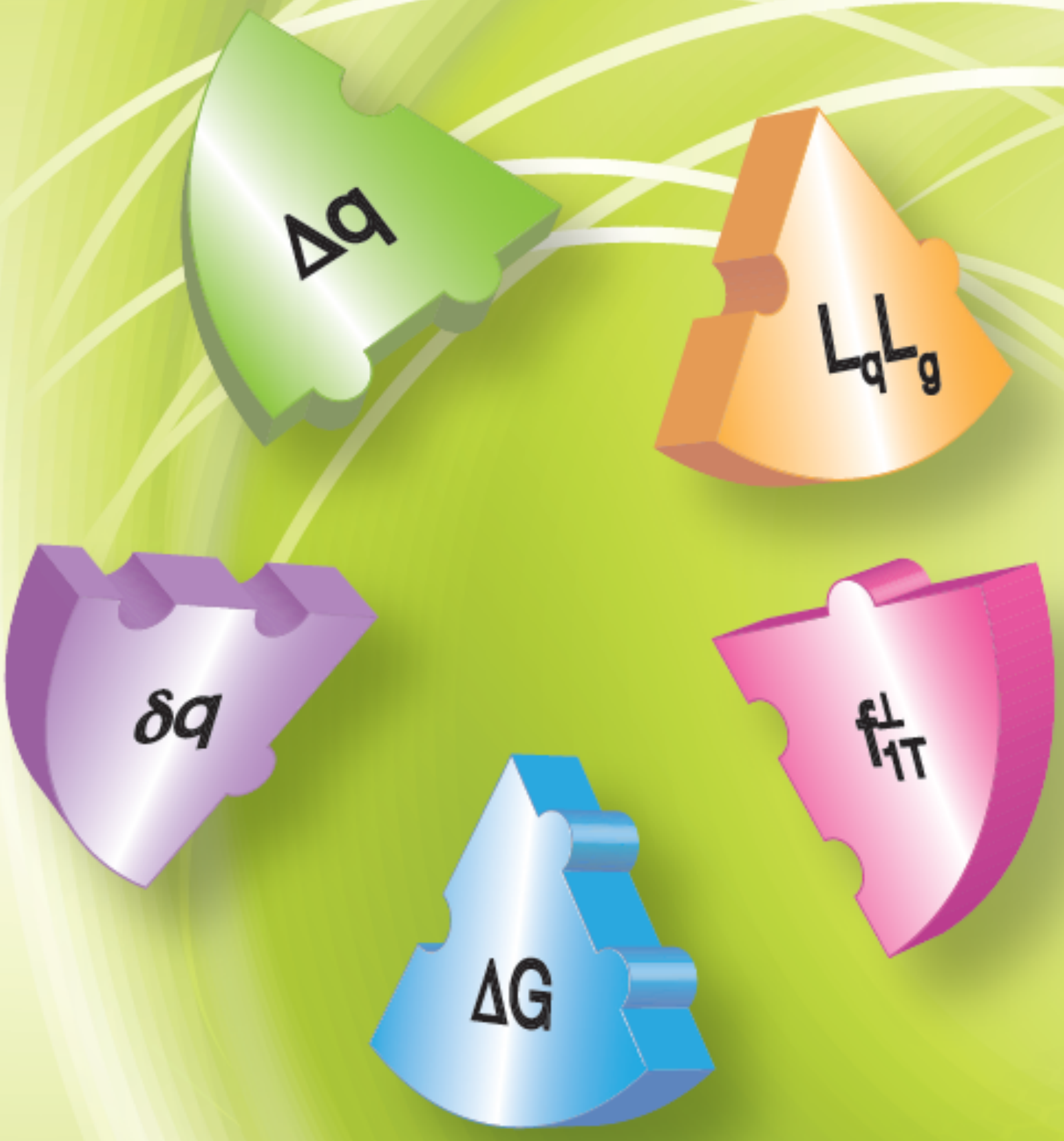

# The RHIC SPIN Program

## Achievements and Future Opportunities

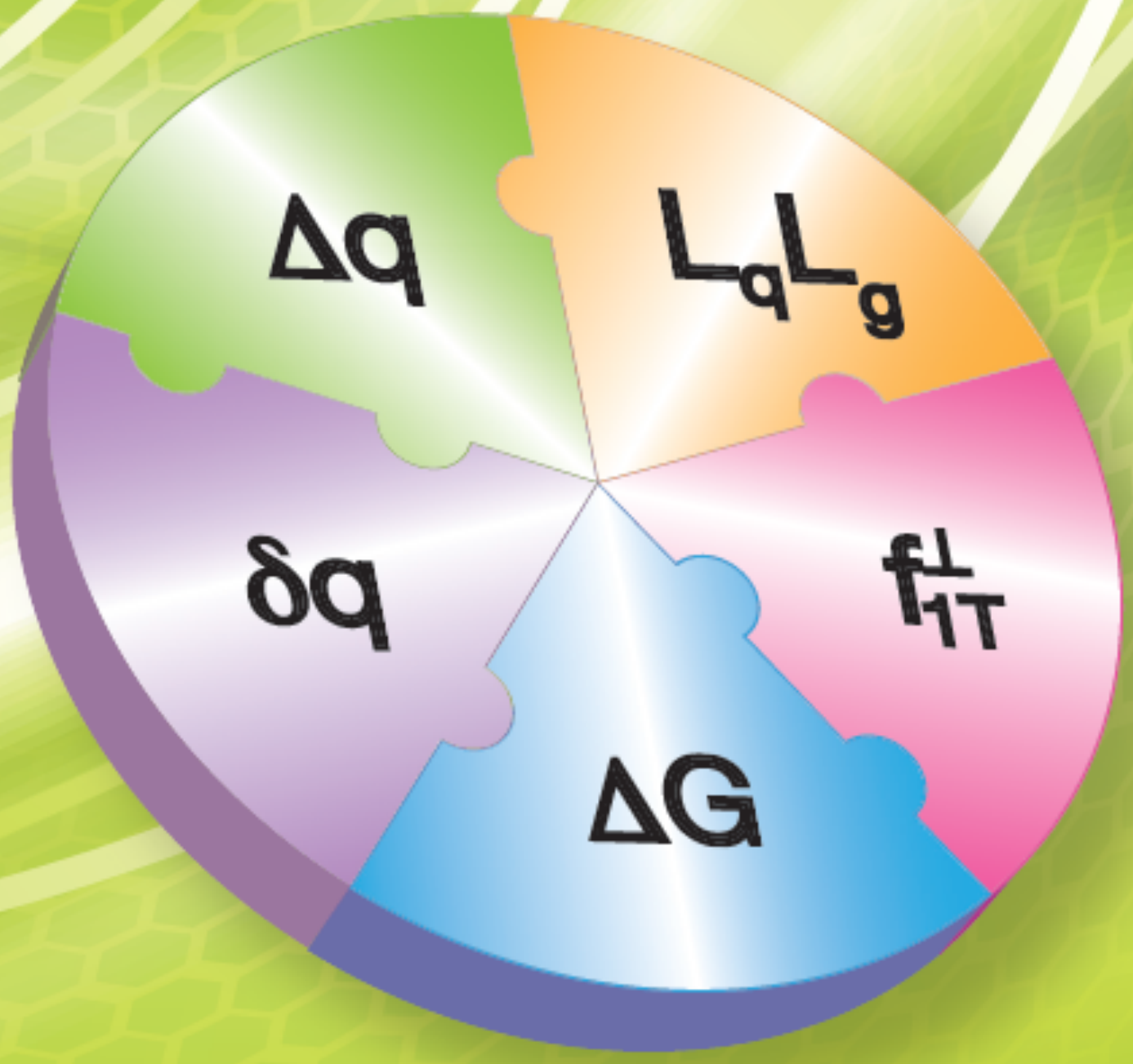

Cover design by J. Abramowitz, BNL

# The RHIC SPIN Program
## Achievements and Future Opportunities

## Authors, for the RHIC SPIN Collaboration[1]

Elke-Caroline Aschenauer[2] (BNL), Alexander Bazilevsky (BNL), Markus Diehl (DESY), James Drachenberg (Valparaiso U.), Kjeld Oleg Eyser (BNL), Renee Fatemi (Kentucky U.), Carl Gagliardi (Texas A&M), Zhongbo Kang (Los Alamos), Yuri V. Kovchegov (Ohio State U.), John Lajoie (Iowa State U.), Jeong-Hun Lee (BNL), Emanuele-R. Nocera (Milano U. & INFN), Daniel Pitonyak (BNL), Alexei Prokudin (JLab), Rodolfo Sassot (Buenos Aires U.), Ralf Seidl (Riken), Ernst Sichtermann (LBNL), Matt Sievert (BNL), Bernd Surrow (Temple U.), Marco Stratmann (Univ. of Tuebingen), Werner Vogelsang (Univ. of Tuebingen), Anselm Vossen (Indiana U.), Scott W. Wissink (Indiana U.), and Feng Yuan (LBNL)

[1] The **RHIC Spin Collaboration** consists of the spin working groups of the RHIC collaborations**,** many theorists and members of the BNL accelerator department.

[2] Chair

# 1 EXECUTIVE SUMMARY

Spin is one of the most fundamental and subtle concepts in physics, deeply rooted in the symmetries and structure of space-time. Spin determines whether a particle follows Fermi or Bose statistics, which has profound implications for the structure of matter and the stability of many-body systems, and lays the foundations for the fields of chemistry and biology. Despite their quantum-mechanical and relativistic origins, spin effects are evident even at large scales and play a critical role in everyday applications such as nuclear magnetic resonance imaging or spintronics-based memory chips. Except for the recently discovered Higgs boson, all elementary particles we know of today carry spin, among them the particles that are subject to the strong interactions: the spin-1/2 quarks and the spin-1 gluons. Spin, therefore, also plays a central role in our theory of the strong interactions, Quantum Chromodynamics (QCD), and the study of spin phenomena in QCD will help to further our understanding of QCD itself. The primary goal of the spin physics program at RHIC is to use spin as a unique probe to unravel the internal structure and the QCD dynamics of nucleons with unprecedented precision.

Protons and neutrons, which make up all nuclei and hence most of the visible mass in the universe, themselves carry spin-1/2. As has been known for over eight decades now, they also possess internal structure. This insight came directly due to spin, through the measurement of the unexpected "anomalous" magnetic moment of the proton. In fact, there is an important lesson to be learned from this discovery: measuring the magnetic moment of the proton was not viewed as an important step at the time, because the answer was already assumed to be "known" to fairly high precision. However, this turned out to be false – and as a result we learned that the proton has substructure. This was just the first of numerous surprises related to spin in strong interaction physics, culminating in the proton "spin crisis" uncovered by the EMC experiment in the late 1980s. The EMC discovery that quark and antiquark spins provide only a small fraction of the proton spin, once again, proved previous expectations to be incorrect and showed that proton substructure was much richer than we had imagined.

Our modern view of the proton is that of a complex system of quarks and transient quark-antiquark pairs, bound together by gluons (see Figure 1-1). The study of the inner structure of such systems that are composed of quarks and gluons is at the heart of investigating QCD in the regime where quarks and gluons interact so strongly that they are confined within hadrons. Spin plays a dual role in this context, foremost to study proton structure in its own right and also as a tool for uncovering properties of the strong interactions. In a broad sense, RHIC investigates how spin phenomena in QCD arise at the quark and gluon level. A particularly important question, and a key focus ever since the EMC measurements, is how quarks and gluons conspire to provide the proton's spin-1/2 through their spin and orbital angular momentum contributions.

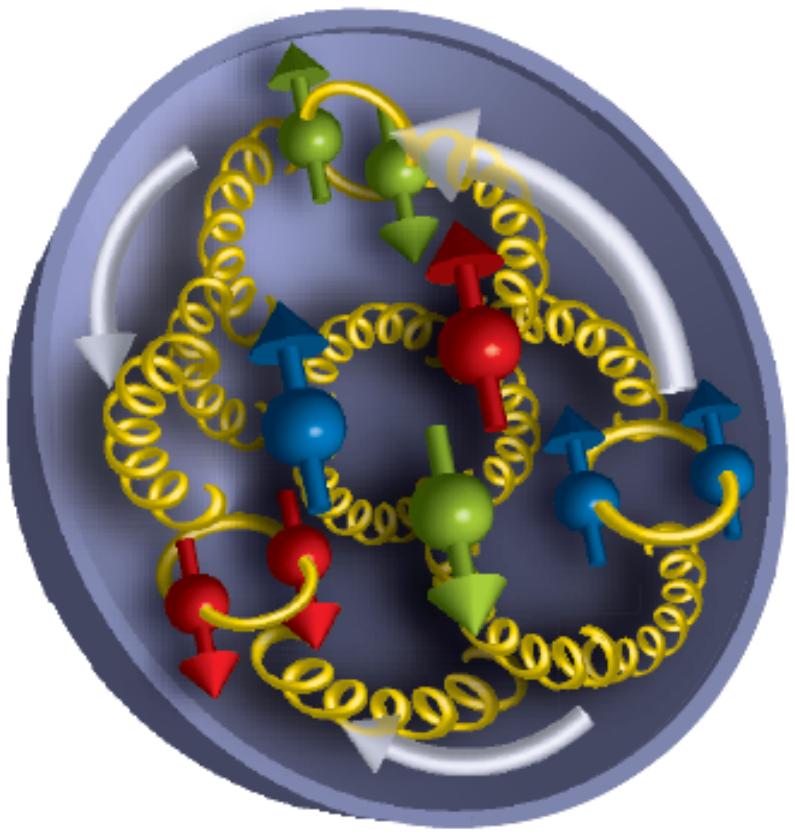

Figure 1-1: Schematic view of the proton built from quarks, quark-antiquark pairs, and gluons.

RHIC addresses these topics in various complementary ways, making use of the tremendous versatility of the machine, which makes readily available both longitudinal and transverse polarization of the protons relative to their momenta. The focus of RHIC spin is on the following key questions:

### How do gluons contribute to the proton spin?

The polarization of gluons in the proton has long been expected to be a source of significant contributions to the proton spin. Indeed, the latest longitudinally polarized data from RHIC have, for the first time, provided evidence that gluons do show a preferential alignment of their spins with the proton's spin. This is a milestone for the field, offering new clues on the proton spin decomposition and on the nature of the strong force fields inside a proton. The current generation of RHIC measurements is providing information about the polarization of those gluons that carry around 1% or more of the proton's momentum. Detector upgrades during the next few years will extend this sensitivity to gluons with even smaller momenta.

### What is the "landscape" of the polarized sea in the nucleon?

In order to understand the dynamics of the quark-antiquark fluctuations in the proton, one needs to learn about the up, down, and strange quark and antiquark densities, individually. This is expected to provide insight into the question of why it is that the total quark plus antiquark contribution to the proton spin was found to be so small. It is also important for models of nucleon structure, which generally make clear qualitative predictions about, for example, the flavor asymmetry in the light quarks in the proton sea. Such predictions are often related to fundamental concepts such as the Pauli principle. At RHIC one uses a powerful technique based on the violation of parity in weak interactions. The $W^{\pm}$ bosons naturally select left quark handedness and right antiquark handedness and hence are ideal probes of nucleon helicity structure. Data from RHIC have now reached the precision needed for obtaining meaningful constraints on the distributions, and a significant further increase in precision is anticipated for the coming years. Comparison with data from semi-inclusive lepton scattering offers tests of basic concepts of high-energy perturbative QCD, such as the universality of parton densities.

### What do transverse-spin phenomena teach us about proton structure?

The past decade has seen tremendous activity and progress, both theoretically and experimentally, in this area. Among the quantities of interest are parton distribution functions that may be accessed in spin asymmetries for hard-scattering reactions involving transversely polarized protons. One of these distributions, known as the "Sivers function", is particularly interesting because it encapsulates the correlations between a parton's transverse momentum inside the proton, and the proton spin vector. As such it contains information on orbital motion of partons in the proton. It was found that the Sivers functions are not universal in hard-scattering reactions. This by itself is nothing spectacular; however, closer theoretical studies have shown that the non-universality has a clear physical origin that may broadly be described as a rescattering of the struck parton in the color field of the remnant of the polarized proton. Depending on the process, the associated color Lorentz forces will act in different ways on the parton (see Figure 1-2).

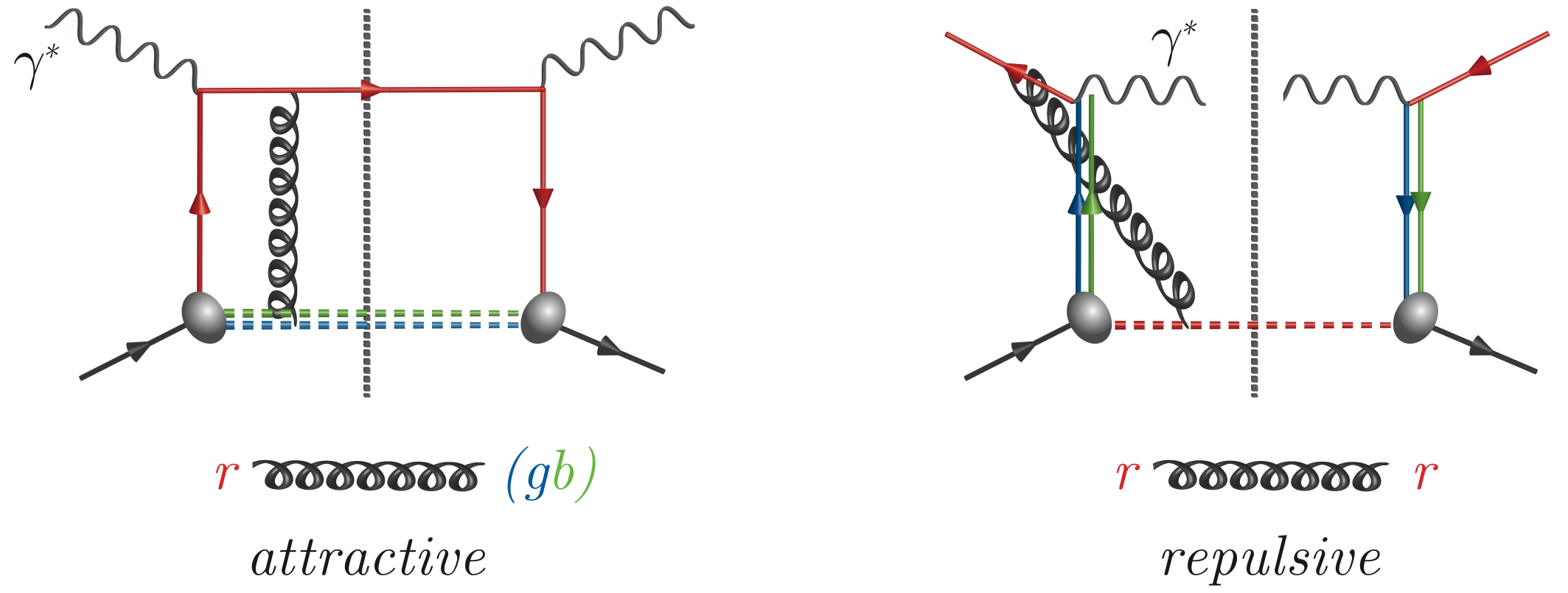

Figure 1-2: Final-state and initial-state color interactions in deep-inelastic lepton nucleon scattering (left hand side) and the Drell Yan process (right hand side).

In deep-inelastic lepton proton scattering, the final-state interaction between the struck parton and the nucleon remnant is attractive. In contrast, for the Drell-Yan process it becomes an initial-state interaction and is repulsive. As a result, the Sivers functions contribute with opposite signs to the single-spin asymmetries for these two processes. This is a fundamental prediction about the nature of QCD color interactions, directly rooted in the quantum nature of the interactions. The Sivers effect exhibits direct physical implications of the gauge potential in a similar way as the Aharonov-Bohm effect does in Electrodynamics. It tests all the concepts for analyzing hard-scattering reactions that we know of. Studies have begun at RHIC aiming at verifying this prediction, which would be a milestone for the field of hadronic physics.

Time and again, spin has been a key element in the exploration of fundamental physics. Spin-dependent observables have often revealed deficits in the assumed theoretical framework and have led to novel developments and concepts. Spin is exploited in many parity-violating experiments searching for physics beyond the Standard Model or studying the nature of nucleon-nucleon forces. The RHIC spin program plays a special role in this grand scheme: it uses spin to study how a complex many-body system such as the proton arises from the dynamics of QCD. Many exciting results from RHIC spin have emerged to date, most of them from RHIC running after the 2007 Long Range Plan. In this document we present highlights from the RHIC program to date and lay out the roadmap for the significant advances that are possible with future RHIC running.

# 2 INTRODUCTION

Quantum Chromodynamics (QCD), the theory of strong interactions, is a cornerstone of the Standard Model of modern particle physics. It explains all strongly interacting matter in terms of point-like quarks interacting by the exchange of gauge bosons, known as gluons. Over the past several decades, a rich phenomenology has come to light, with several overarching questions remaining unanswered that have been, and continue to be, addressed by the RHIC program with polarized proton beams:

- *What is the nature of the spin of the proton?*
- *How do quarks and gluons hadronize into final-state particles?*
- *How can we describe the multidimensional landscape of nucleons and nuclei?*
- *What is the nature of the initial state in nuclear collisions?*

Much of our present knowledge of nucleon structure comes from deep-inelastic lepton-nucleon scattering (DIS) experiments, with a great wealth of data on the unpolarized structure of the proton being available from HERA [1]. From HERA we have learned that quarks carry roughly 50% of the momentum of the proton, with the other half being carried by gluons, which dominate at momentum fractions[3] $x$ below 0.1.

Despite all the knowledge that has been acquired through DIS measurements, studying nucleon structure in a wide variety of reactions is essential in order to piece together a complete picture and to verify the underlying theoretical framework. In this respect, hadron-hadron interactions offer several advantages. Direct access to gluons is possible through several gluon-dominated hard scattering processes such as inclusive high-$p_T$ jet or hadron production, making the determination of the contribution of gluons to the spin of the proton a key component of the RHIC spin program. $W$-boson production and the Drell-Yan process are both golden probes to cleanly access antiquark distributions in hadron-hadron collisions. Comparing observations from DIS and hadronic interactions also allows us to test the key assumption of ***universality*** in describing hadron structure and hadronization across different hard scattering processes within the framework of perturbative QCD (pQCD). At high enough energies, quarks and gluons can be treated as nearly free particles moving collinearly with their parent hadron. Hence, in pQCD hadronic interactions are assumed to factorize into a) parton distribution functions (PDFs) for each of the initial-state hadrons, b) calculable partonic hard-scattering cross sections, and c) fragmentation functions (FFs) describing the hadronization of a scattered parton into an observed hadron.

This framework has been tremendously successful over the past several decades in describing quantitatively a wealth of hadronic cross sections characterized by a high momentum transfer and ranging from fixed-target to collider energies. Despite significant progress both experimentally and theoretically, there remain fundamental aspects of the nucleon partonic structure that are still rather poorly determined. One example is the nature of the nucleon spin; another is the quest to go beyond our current, one-dimensional picture of parton densities by correlating, for instance, the information on the individual parton contribution to the spin of the nucleon with its transverse momentum and spatial position. Figure 2-1 shows the non-trivial connections between different quantities describing the distribution of partons inside the proton. The functions given here are for unpolarized partons in an unpolarized proton; analogous relations hold for polarized quantities.

The full spatial ($b_T$) and transverse momentum ($k_T$) dependent structure of partons in the nucleon as a function of the longitudinal momentum fraction of the partons is encoded in the Wigner function $W(x,k_T,b_T)$, which currently is not accessible in experiments.

[3] All kinematic variables are explained in Table 7-1

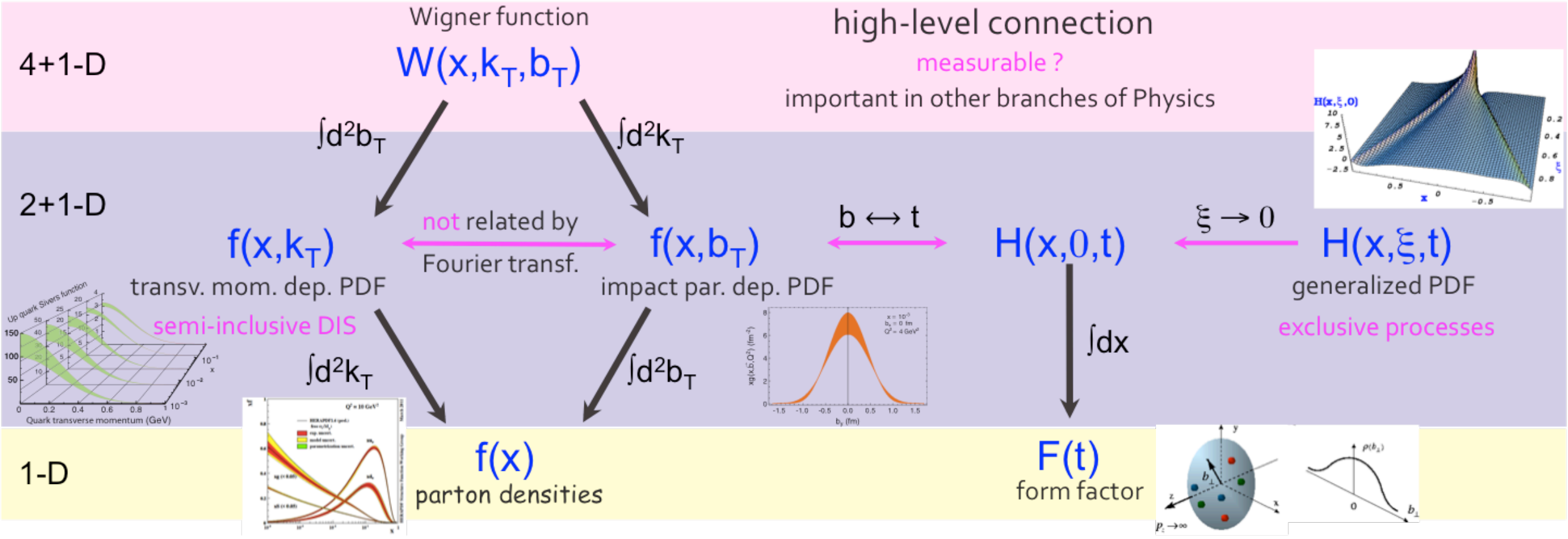

Figure 2-1: Connections between different quantities describing the distribution of partons inside the proton. The functions given here are for unpolarized partons in an unpolarized proton; analogous relations hold for polarized quantities.

Integrating $W(x,k_T,b_T)$ over either the transverse momentum $b_T$ or position $k_T$ will result in the transverse momentum dependent parton distribution functions (TMDs) $f(x,k_T)$ and in the impact parameter dependent parton distribution functions $f(x,b_T)$, respectively.

These distribution functions can be accessed through experimental measurements. The impact parameter dependent parton distribution functions can be related to generalized distribution functions (GPDs) $H(x,\xi,t)$ by Fourier transforming the impact parameter dependence $b_T$ into the Mandelstam variable $t$ and by extrapolating a GPD from finite to zero skewness $\xi$. The well-known collinear parton distribution functions can be obtained by integrating over the transverse momentum dependence for TMDs and the spatial dependence for $f(x,b_T)$.

The collider energies available at RHIC place high-$p_T$ reactions comfortably within a regime described by the factorization theorem of pQCD. It is worth noting that the relevant hard scale in DIS is usually taken to be the photon virtuality $Q$, while in hadron-hadron interactions it is often the transverse momentum $p_T$ of the produced jet or particle. While both $Q^2$ and $x$ are accessible in DIS, in hadron-hadron measurements the $p_T$ of the produced particle is correlated with $x$ in a complicated way and any given $p_T$ bin typically samples a broad range of $x$ values (see Figure 2-2). At the same moment, selecting different $p_T$ values allows one to enhance particular underlying partonic subprocesses contributions (see Figure 2-3).

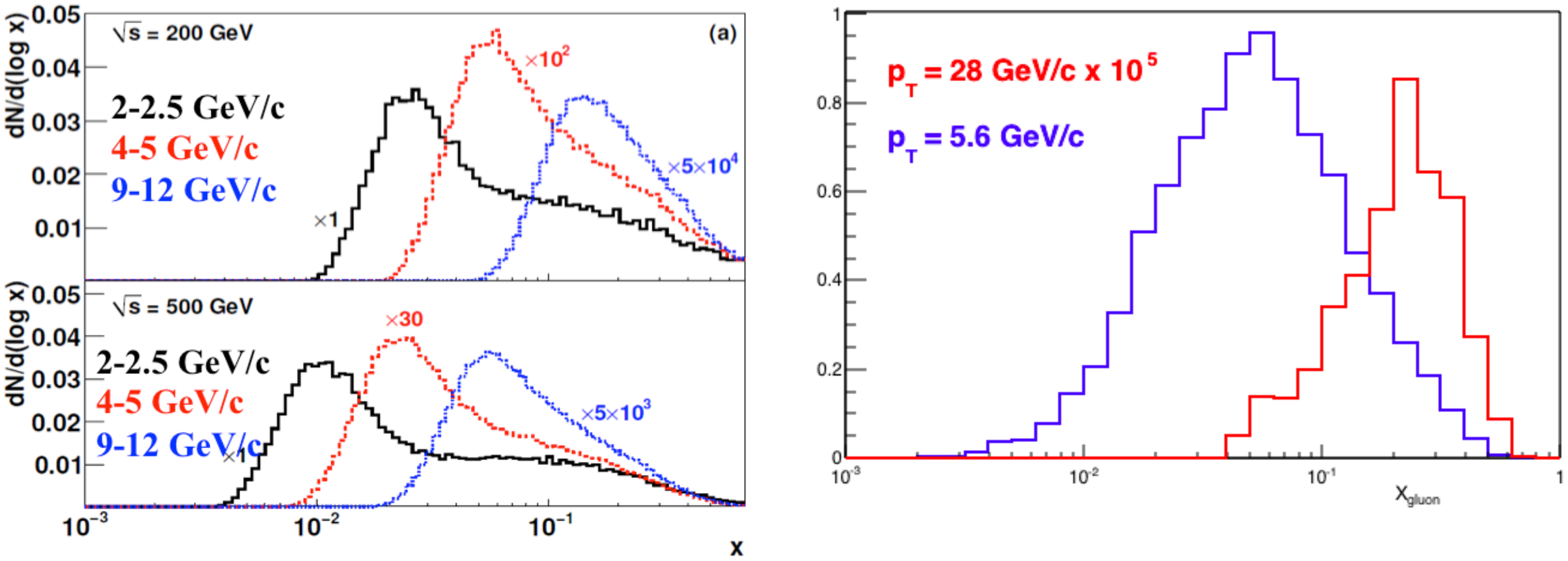

Figure 2-2 Left panel: distributions of gluon momentum fractions $x$ sampled in three $p_T$ bins obtained from a NLO pQCD simulation of $\pi^0$ production at $\sqrt{s}$=200 GeV and 500 GeV. The right panel shows the relative contributions of gluons with a given momentum fraction x to high $p_T$ inclusive jet cross production in $p+p$ collisions at mid rapidity for $\sqrt{s}$=200 GeV [2].

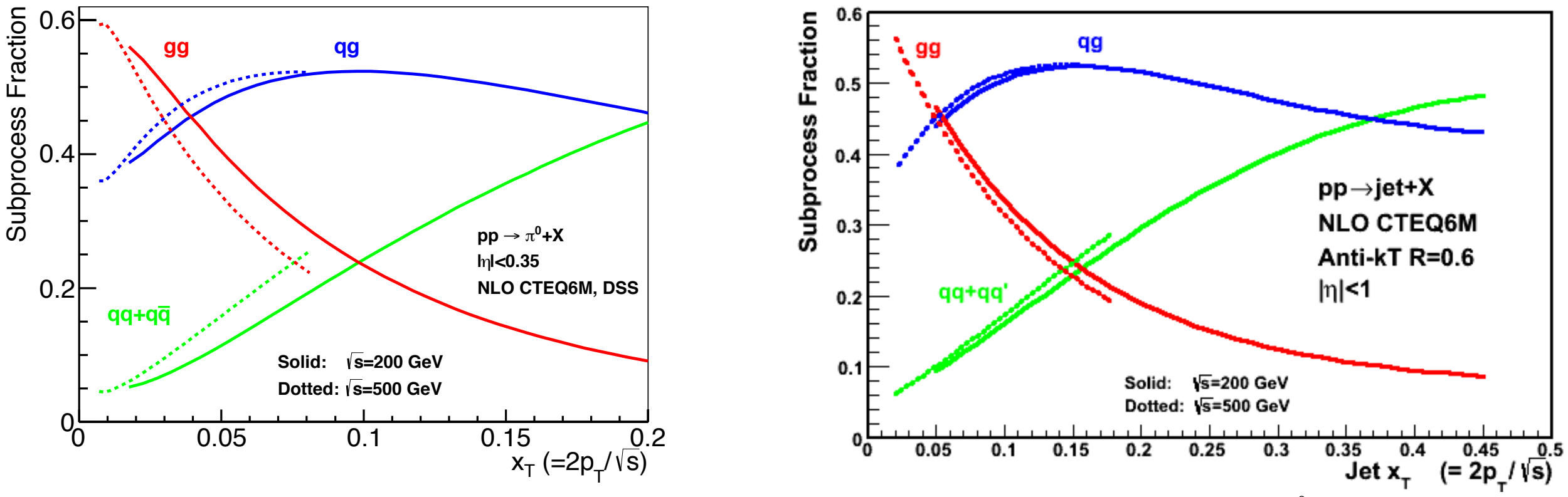

Figure 2-3: Relative contributions of different partonic subprocesses contributing to inclusive $\pi^0$ (left panel) and jet production (right panel) as a function of $x_T$. Only minor differences can be seen when going from √s=200 GeV to 500 GeV. At low $x_T$ gluon-gluon scattering dominates, followed by quark-gluon scattering at higher $x_T$. At very high $x_T$, quark-quark scattering eventually becomes the dominant production channel.

Questions about the nature of the nucleon spin and the transverse momentum and spatial structure of partons in the nucleon have also manifested themselves in the Nuclear Physics performance milestones from DOE. Table 2-1 lists the current Nuclear Physics performance milestones related to the RHIC *p+p* physics program. In the following sections we will describe how these questions have been and will be addressed by the RHIC spin physics program in the next years.

| Year | # | Milestone |
|---|---|---|
| 2013 | HP8 | Measure flavor-identified $q$ and $\bar{q}$ contributions to the spin of the proton via the longitudinal-spin asymmetries of W production |
| 2013 | HP12 (update of HP1 met in 2008) | Utilize polarized proton collisions at center of mass energies of 200 and 500 GeV, in combination with global QCD analyses, to determine if gluons have appreciable polarization over any range of momentum fraction between 1 and 30% of the momentum of a polarized proton. |
| 2015 | HP13 (new) | Test unique QCD predictions for relations between single-transverse spin phenomena in p-p scattering and those observed in deep –inelastic lepton scattering |

Table 2-1: Current nuclear physics performance milestones related to the RHIC *p+p* physics program.

# 3 THE HELICITY STRUCTURE OF THE PROTON

Helicity-dependent parton densities encode to what extent quarks and gluons with a given momentum fraction $x$ tend to have their spins aligned with the spin direction of a longitudinally polarized nucleon. The most precise knowledge about these non-perturbative quantities, along with estimates of their uncertainties, is gathered from comprehensive global QCD analyses [3,4] of all available data taken in spin-dependent proton-proton collisions and DIS, with and without additional identified hadrons in the final state.

Apart from being essential for a comprehensive understanding of the partonic structure of hadronic matter, helicity PDFs draw much of their relevance from their relation to one of the most fundamental and basic but yet not satisfactorily answered questions in hadronic physics, namely how the spin of a nucleon is composed of the spins and orbital angular momenta of quarks and gluons. The integrals of helicity PDFs over all momentum fractions $x$ (first moments) at a resolution scale $Q$, provide information about the contribution of a given parton flavor $f$ to the spin of the nucleon.

The RHIC spin physics program was founded to advance our understanding of the spin and flavor structure of the proton in terms of its constituent quarks and gluons by exploiting the unique capability of RHIC to provide access to high-energy polarized $p+p$ collisions. Using longitudinally polarized beams, one can probe the helicity preferences of gluons and (flavor-separated) quarks and antiquarks in a broad range of momentum fraction $x$, to gain detailed quantitative insight into their respective contributions to the spin of the proton.

RHIC has completed very successful polarized $p+p$ runs both at $\sqrt{s}$ = 200 GeV and 500(510) GeV. Table 3-1 summarizes the luminosities recorded by PHENIX and STAR and the average beam polarization (as measured by the H-jet polarimeter) for runs since 2006.

| **Year** | **$\sqrt{s}$ (GeV)** | **Recorded Luminosity for longitudinally polarized *p+p* STAR** | **Recorded Luminosity for longitudinally polarized *p+p* PHENIX** | **<P> in %** |
|---|---|---|---|---|
| 2006 | 62.4 | -- $pb^{-1}$ | 0.08 $pb^{-1}$ | 48 |
| | 200 | 6.8 $pb^{-1}$ | 7.5 $pb^{-1}$ | 57 |
| 2009 | 200 | 25 $pb^{-1}$ | 16 $pb^{-1}$ | 55 |
| | 500 | 10 $pb^{-1}$ | 14 $pb^{-1}$ | 39 |
| 2011 | 500 | 12 $pb^{-1}$ | 18 $pb^{-1}$ | 48 |
| 2012 | 510 | 82 $pb^{-1}$ | 32 $pb^{-1}$ | 50/53 |
| 2013 | 510 | 300 $pb^{-1}$ | 155 $pb^{-1}$ | 50/53 |
| 2015 | 200 | 50 $pb^{-1}$ | -- | 60 |

Table 3-1: Recorded luminosities for collisions of longitudinally polarized proton beams at the indicated center-of-mass energies for past RHIC runs since 2006. The PHENIX numbers are for $|vtx| < 30$cm. The bottom row reflects the STAR and PHENIX beam use request for 2015.

# 3.1 THE POLARIZED GLUON DISTRIBUTION

The measurement of the gluon polarization in a longitudinally polarized proton has been a major emphasis and strength of the spin physics program at RHIC since its inception [5]. In 2009, with the improved luminosity and polarization as well as upgraded triggering and data acquisition systems for STAR, the uncertainties on the published inclusive $\pi^0$ and jet longitudinal double spin asymmetries $A_{LL}$ at $\sqrt{s}$ = 200 GeV were improved considerably. Figure 3-1 and Figure 3-2 compare the published inclusive $\pi^0$ and jet $A_{LL}$ (blue points) vs. $x_T$ at mid-rapidity for $|\eta| < 0.35$ [6] and $|\eta| < 1$ [7], respectively, to predictions from several NLO global analyses [4,8,9] (blue curves). The error bars represent the point-to-point uncertainties. The gray/golden bands illustrate the size of the correlated systematic uncertainties. The impact of the new jet data on the polarized gluon distribution and its truncated $x$-integral was studied using the reweighting method adopted by the NNPDF collaboration [9]. The integral of $\Delta g(x,Q^2=10$ GeV$^2)$ in the range $0.05 < x < 0.2$ is $0.05 \pm 0.15$ for the original NNPDF fit and $0.17 \pm 0.06$ when the fit is reweighted using the 2009 STAR jet data for $\sqrt{s}$=200 GeV. It is shown that the new results are completely consistent with PHENIX $\pi^0$ $A_{LL}$ data. The DSSV group has performed a new global analysis [4] including both the 2009 PHENIX and STAR results for $A_{LL}$. They find that the integral of $\Delta g(x,Q^2=10$ GeV$^2)$ in the region $x > 0.05$ is $\mathbf{0.20^{+0.06}_{-0.07}}$ at 90% C.L., consistent with the NNPDFpol1.1 fit [9].

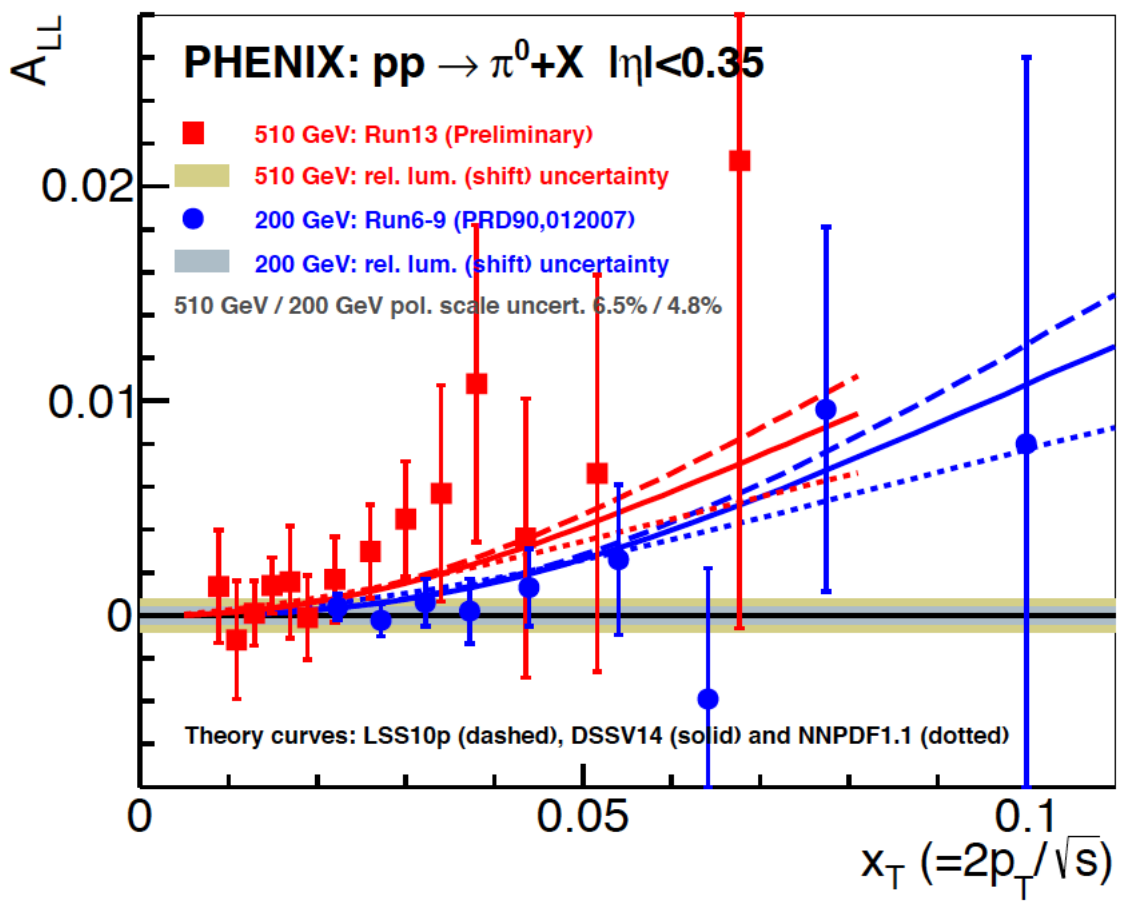

Figure 3-1: $A_{LL}$ vs. $x_T$ for $\pi^0$-mesons production at mid rapidity with the point-to-point uncertainties in 200 GeV (blue circles) and 510 GeV (red squares) $p$+$p$ collisions, compared to predictions from three recent NLO global analyses [4,8,9] (blue curves for 200 GeV and red curves for 510 GeV). The gray/gold bands give the correlated systematic uncertainties.

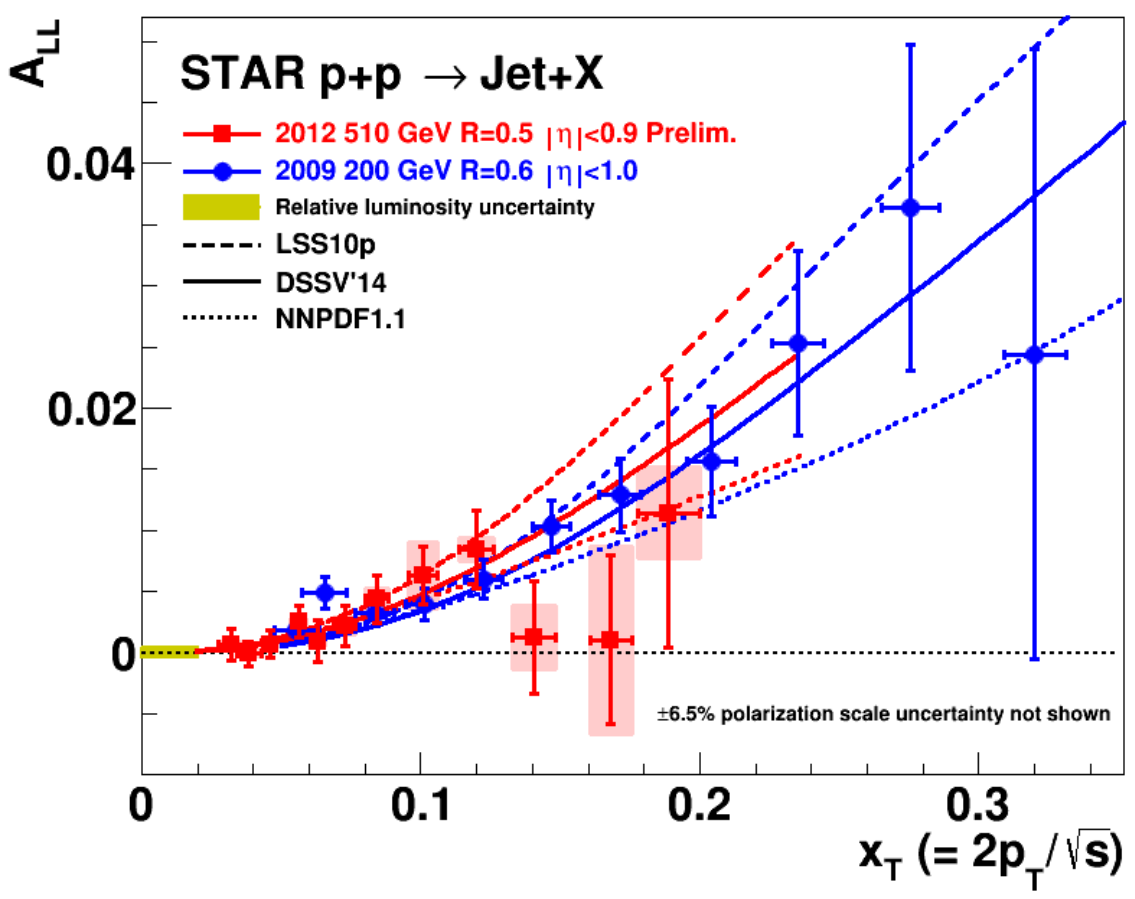

Figure 3-2: $A_{LL}$ vs. $x_T$ for inclusive jet production at mid-rapidity in 200 GeV (blue circles) [7] and 510 GeV (red squares) $p$+$p$ collisions, compared to predictions from three recent NLO global analyses [4,8,9] (blue curves for 200 GeV and red curves for 510 GeV).

Figure 3-3 represents the 90% C.L. areas in the plane spanned by the truncated integrals of $\Delta g$ computed for $0.05 < x < 1$ and $0.001 < x < 0.05$ at a scale $Q^2$=10 GeV$^2$. Results for DSSV [3], DSSV* and the new DSSV [4] analysis, with the symbols denoting the respective values of each central fit, are shown. It is obvious that the 2009 $A_{LL}$ data from PHENIX and STAR taken at $\sqrt{s}$ = 200 GeV have reduced the uncertainties for $x > 0.05$ significantly. For the first time there is clear evidence for a **nonzero** gluon contribution to the spin of the proton from the x region covered by current RHIC data. But the plot also shows clearly that the uncertainties in the unmeasured small-x region, $x < 0.05$, still remain very large.

Preliminary mid-rapidity inclusive $\pi^0$ and jet $A_{LL}$ data from RHIC collected in Run-2012 and 2013 at both $\sqrt{s}$ = 200 GeV and 510 GeV, as well as future measurements at forward rapidities ($\eta$ > 3) will further reduce the uncertainties on the truncated moment of $\Delta g$. In particular, the latter measurements will significantly extend the kinematic reach in $x$ towards values of a few times $10^{-3}$ where $\Delta g$ is unconstrained so far. Figure 3-1 and Figure 3-2 also include the preliminary PHENIX and STAR results for inclusive $\pi^0$ and jet $A_{LL}$ measurements at $\sqrt{s}$ = 510 GeV from Run-2013 and Run-2012 (red points), respectively, as a function of $x_T$ compared to three recent QCD global analyses (red curves). Taking $x_T$ as a rough proxy for the smallest momentum fraction $x$ at which $\Delta g$ can be sampled by a given data point, illustrates the importance of the 510 GeV RHIC measurements in extending the kinematic reach. Both new data sets also exhibit high statistical precision and small systematic uncertainties, which is crucial for making a significant impact in upcoming global QCD analyses of helicity PDFs.

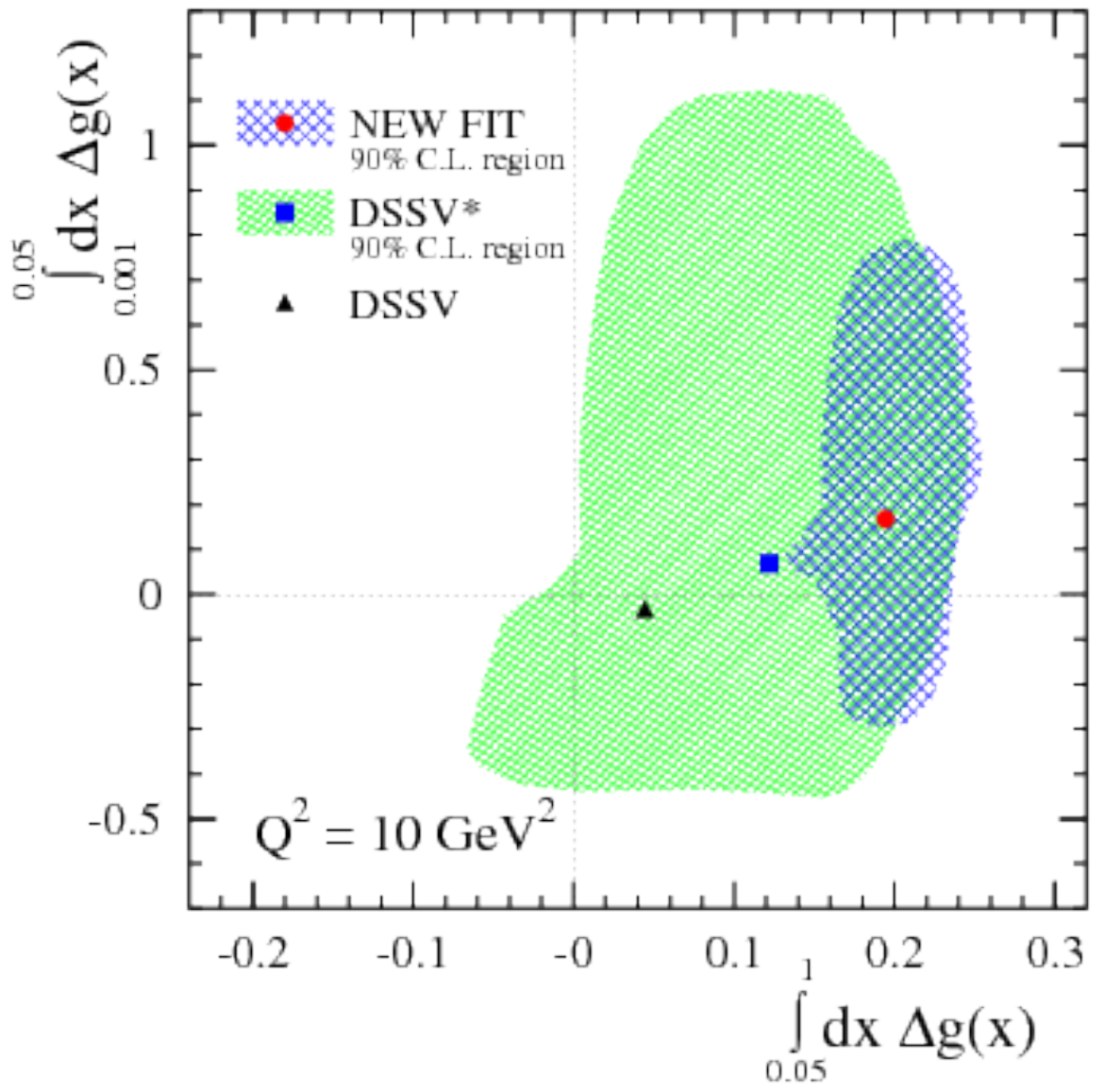

Figure 3-3: 90% C.L. areas in the plane spanned by the truncated moments of $\Delta g$ computed for $0.05 < x < 1$ and $0.001 < x < 0.05$ at $Q^2$=10 GeV$^2$. Results for DSSV, DSSV* (including preliminary and final RHIC data up to Run-2006, respectively) and the new DSSV analysis based on RHIC data up to Run-2009 are shown, with the symbols corresponding to the respective central values of each central fit.

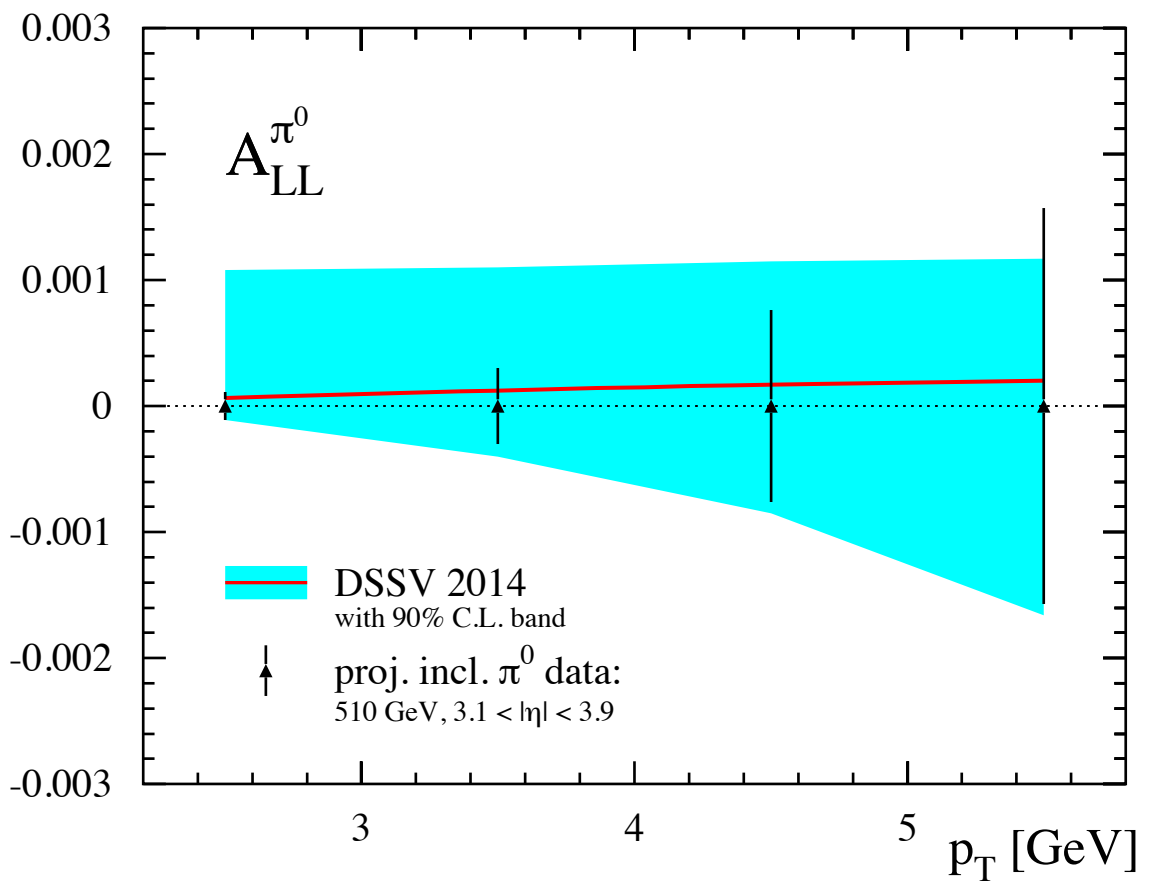

Figure 3-4: The anticipated statistical precision for inclusive $\pi^0$ $A_{LL}$ measurements at $\sqrt{s}$=510 GeV in the pseudorapidity range $3.1 < \eta < 3.9$ with the PHENIX MPC in 2013 compared to the theoretical expectation based on the latest DSSV-2014 PDFs including 90% C.L. uncertainties.

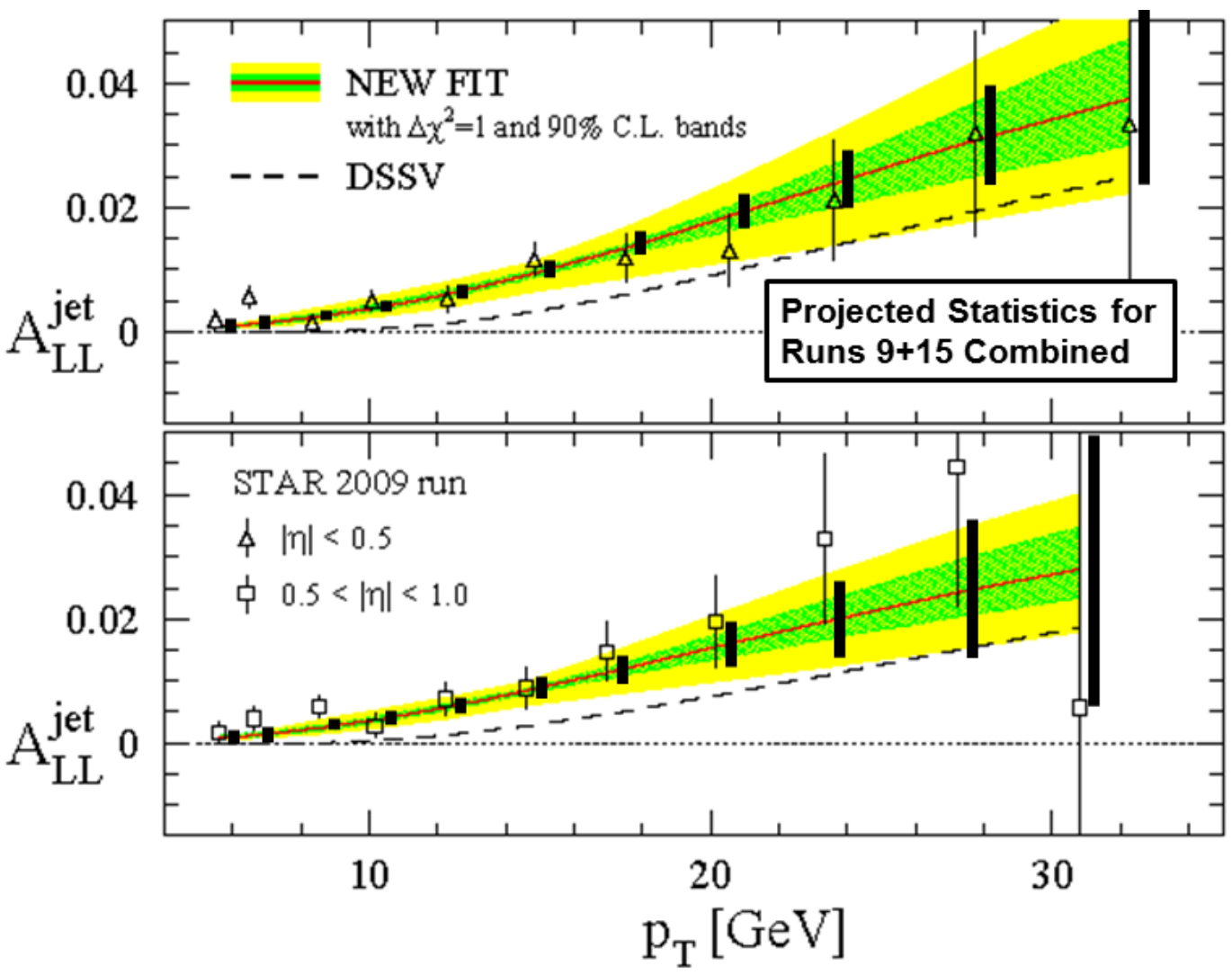

Figure 3-5: The projected statistical precision for $A_{LL}$ vs. $p_T$ for inclusive jets in 200 GeV $p+p$ collisions based on the combined data from the 2009 and 2015 RHIC runs, compared to the uncertainties from the DSSV-2014 fit, which included the 2009 inclusive jet results among the inputs.

Figure 3-4 illustrates the anticipated statistical precision for future inclusive $\pi^0$ $A_{LL}$ measurements at $\sqrt{s}$=510 GeV in the pseudorapidity range $3.1 < \eta < 3.9$ with the PHENIX forward electromagnetic calorimeter (MPC). The estimates are based on the recorded luminosity in Run-2013. Also shown is a theoretical expectation based on the latest DSSV analysis including

the current uncertainties on $\Delta g$. STAR performed a similar measurement utilizing its forward electromagnetic calorimeter (FMS) at $2.5 < \eta < 3.8$ [10]. The significant impact of these new measurements is obvious from comparing the anticipated statistical precision with the current range of theoretical uncertainties, which are mainly driven by our ignorance of $\Delta g$ at momentum fractions in the range from a few times $10^{-3}$ to about 0.01. Data on $A_{LL}$ at forward rapidities will be an ideal addition to the suite of measurements performed so far.

The projected statistical precision for $A_{LL}$ vs. $p_T$ for inclusive jets in 200 GeV $p+p$ collisions based on the combined data from the 2009 and 2015 RHIC runs is shown in Figure 3-5 compared to theoretical expectation based on the latest DSSV-2014 analysis.

To estimate and demonstrate the potential impact of all these preliminary and forthcoming RHIC data for $A_{LL}$ at $\sqrt{s}$ = (200) 510 GeV up to Run-2015 (see Table 3-1) on our knowledge of $\Delta g$, the DSSV group has performed a global analysis with pseudo-data which reflect the anticipated experimental precision. Figure 3-6 shows the running integral $\int_{x_{min}}^{1} dx \Delta g(x, Q^2)$ as a function of $x_{min}$ at $Q^2 = 10$ GeV$^2$ and uncertainty estimates at 90% C.L. based on DSSV framework with and without including the various sets of pseudo-data for PHENIX and STAR. The solid line and the outer band reflect our current knowledge of $\Delta g$ based on published RHIC and other world data similar to what was already shown in Figure 3-3. The inner uncertainty band illustrates the reduction of uncertainties due to the combined set of the projected RHIC measurements discussed above. As can be seen, the preliminary and upcoming RHIC data are expected to reduce the present uncertainties on the truncated integral by about a factor of 2 at $x_{min} = 10^{-3}$ where RHIC data can provide an experimental constraint. For illustration, Figure 3-6 extends down to an $x_{min}$ of $10^{-6}$ well outside the x-region covered by RHIC data. It is interesting to notice that the optimum fit of DSSV prefers an integrated gluon helicity distribution of about **0.36**, i.e., the gluons carry more than 70% of the proton's spin in this scenario at a scale $Q^2 = 10$ GeV$^2$. About half of this value already stems from the $x$-range covered by published RHIC data, as was demonstrated in Figure 3-3.

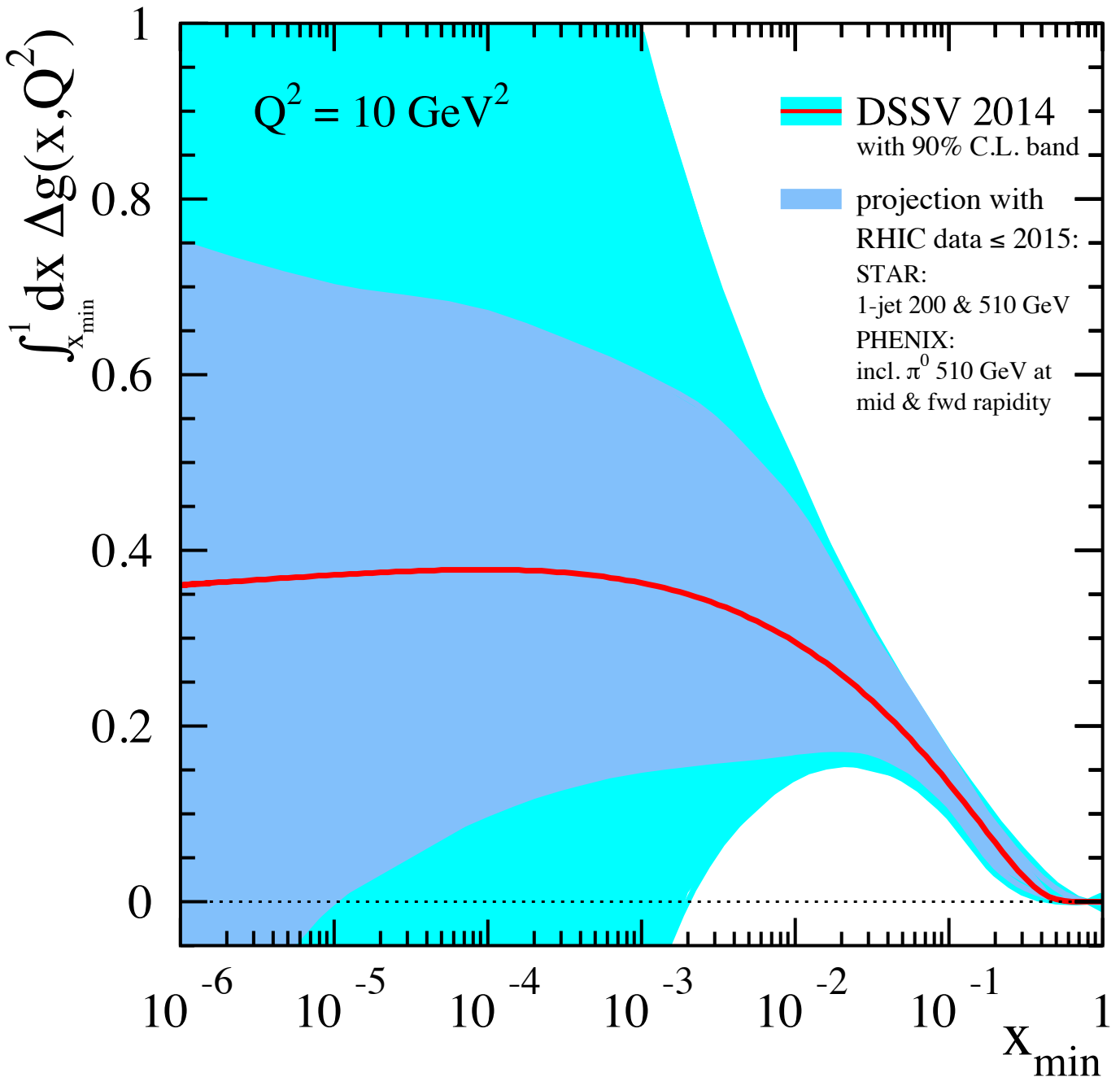

Figure 3-6: The running integral for $\Delta g$ as a function of $x_{min}$ at $Q^2 = 10$ GeV$^2$ as obtained in the DSSV global analysis framework. The inner and outer uncertainty bands at 90% C.L. are estimated with and without including the combined set of projected pseudo-data for preliminary and future RHIC measurements up to Run-2015 (see Table 3-1), respectively.

Figure 3-7 provides a complementary view of the already achieved and still to be expected improvements on our knowledge of $\Delta g$ from the RHIC spin program. Here we look into the maximum allowed variation by the data in a small bin of momentum fraction $x$ and regardless of what happens outside of that interval, rather than combining all contributions from $x_{min}$ to 1 simultaneously as in Figure 3-6. Results are shown for the original DSSV'08 analysis based on PHENIX and STAR data only up to Run-2006, the recent update including the published Run-2009 data,

and a projection including pseudo-data throughout Run-2015. The solid lines represent the respective optimum fits and the bands correspond to uncertainty estimates at 90% C.L. One notices the significant improvements in the uncertainties from the DSSV'08 analysis, which had a rather small $\Delta g$ for the optimum fit back then, to the results including the published 200 GeV Run-2009 data from PHENIX and STAR. In the range $0.05<x<0.3$ the fit now prefers a positive $\Delta g$ within the uncertainty estimates. However, for smaller $x$ values the uncertainties quickly deteriorate. The projected (200) 510 GeV data up to Run-2015 will lead to significant further improvements in the region $0.001<x<0.03$ as is illustrated by the innermost uncertainty bands in Figure 3-7. In the lowest x-bins these improvements are mainly due to the simulated data at forward rapidities while above 0.01 the mid-rapidity data at 510 GeV contribute most.

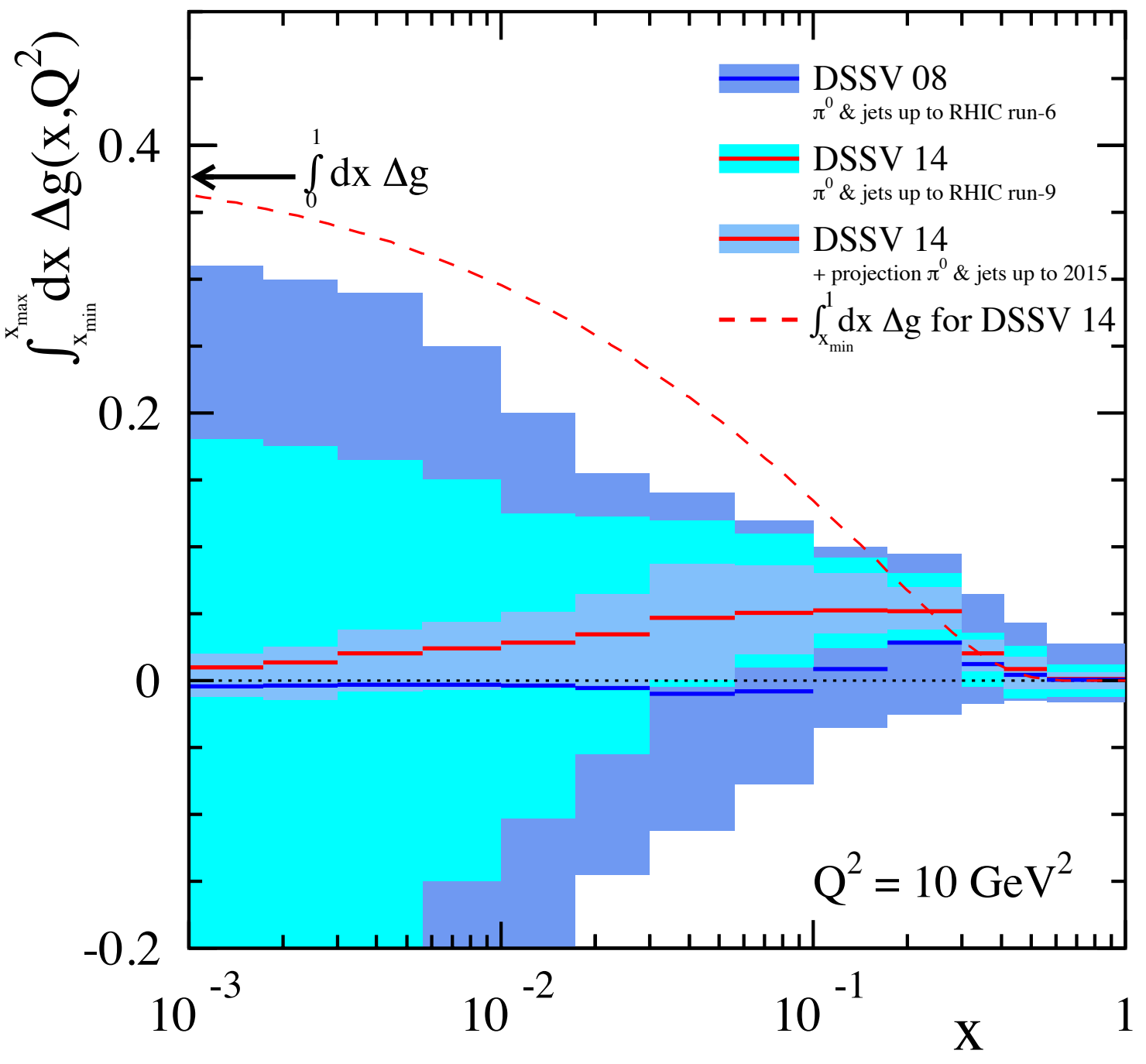

Figure 3-7: Contributions to the integral of $\Delta g(x,Q^2)$ from small bins in $x$ defined by $x_{min}$ and $x_{max}$ at $Q^2 = 10$ GeV$^2$. Results are shown for the original DSSV'08 analysis based on RHIC data only up to Run-2006, the recent update including Run-2009 data from PHENIX and STAR, and a projection including pseudo-data throughout Run-2015. The solid lines represent the respective optimum fits and the bands correspond to uncertainty estimates at 90% C.L. The dashed lines correspond to the truncated integral from $x_{min}$ to 1 as shown in Figure 3-5.

To further improve our understanding of $\Delta g(x)$ and its integral, we plan to make use of correlation measurements such as di-jets and di-hadrons, which give more stringent constraints to the underlying partonic kinematics and thus the functional form of $\Delta g(x)$. It is interesting to note that the functional form of $\Delta g(x)$ provides insight into the dynamical origin of gluons and their polarization inside the proton.

The invariant di-jet mass $M$ is related to the product of the initial partonic $x$ values, whereas the sum $\eta_3 + \eta_4$ of the pseudorapidities of the produced di-jets (labeled as '3' and '4') is related to the ratio of the partonic $x$ values, i.e. $x_1/x_2$. Measurements at both $\sqrt{s}$ = 200 GeV and 500 GeV are preferred to maximize the kinematic reach in $x$. The acceptance of the STAR experiment permits reconstruction of di-jet events with different topological configurations, i.e., different ratios $\eta_3/\eta_4$, ranging from symmetric to asymmetric partonic collisions. The current acceptance of the STAR experiment includes the regions EAST ($-1.0 < \eta < 0.0$) and WEST ($0.0 < \eta < 1.0$) along with some more forward acceptance ($1.09 < \eta < 2.0$ (EEMC)). Figure 3-8 shows the projected statistical uncertainties of $A_{LL}$ for di-jets at $\sqrt{s}$=200 GeV and 500 GeV based on the recorded or to be recorded datasets from 2009 to 2015. The projected uncertainties are compared to the theoretical estimates of the spin asymmetry based on the DSSV-2008 [3] and GRSV-STD [11] sets of the polarized PDFs. As an example, the corresponding $x_1$ and $x_2$ distributions for three different di-jet topologies are shown for $\sqrt{s}$=500 GeV in the lower panels. We note that first preliminary results on pairs of neutral pions from PHENIX at mid-rapidity have been released [12].

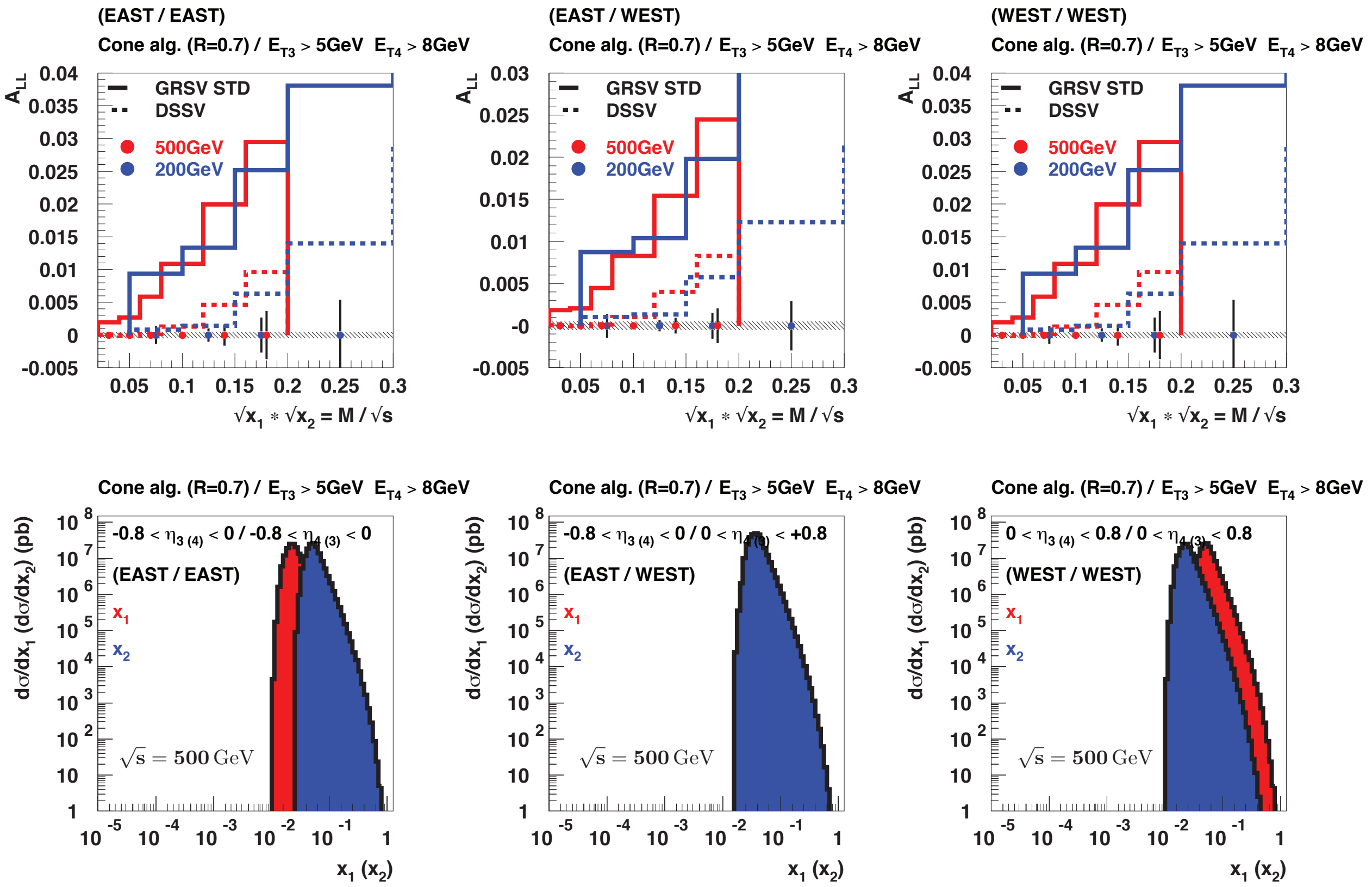

Figure 3-8: Upper panels: Estimates for $A_{LL}$ at NLO accuracy for di-jet production using two sets of polarized PDFs for $\sqrt{s}$=200 GeV (blue) and 510 GeV (red) $p$+$p$ collisions in the current STAR mid-rapidity acceptance region $|\eta| <$ 1.0 together with projected statistical and systematic uncertainties. Lower panels: Relevant $x_1$ and $x_2$ ranges probed by di-jet production in the STAR mid-rapidity acceptance region $|\eta| <$ 1.0.

The above described measurements will impose significant constraints on the polarized gluon distribution beyond what has been achieved so far. However, improving the systematic uncertainties at low $p_T$ arising from various experimental sources remains a challenge that will need to be met in order to access smaller momentum fractions $x$ in the gluon. Theoretical ambiguities both at small $p_T$ and forward rapidities also need to be carefully monitored in an extraction of $\Delta g$. Simultaneous measurements of unpolarized cross sections at the same kinematics – a particular strength of the RHIC program – help to verify the range of validity of the assumed theoretical framework for each measurement. In addition, the complex nature of extracting PDFs by fitting data sets from various experiments and testing the fundamental universality property of helicity PDFs demands further independent measurements with many different final-state particles covering a kinematic range as wide as possible. Therefore, both PHENIX and STAR have been and will continue to investigate, despite their currently limited statistical precision, measurements of $A_{LL}$ for other final states such as charged pions, eta mesons, open charm through single lepton and muon tags, direct photons, and photon/hadron-jet correlations. A detailed discussion of the advantages and disadvantages for each of these channels is beyond the scope of this document.

# 3.2 THE POLARIZED LIGHT SEA QUARK DISTRIBUTIONS

The production of $W^{\pm}$ bosons in longitudinally polarized proton-proton collisions serves as an elegant tool to access the valence and sea quark helicity distributions at a high scale, $Q \sim M_W$, and without the need of FF as in semi-inclusive DIS. The charge of the $W$-boson, or its decay lepton, selects predominantly $u$ and $\bar{d}$ quarks in $W^+$ production and $d$ and $\bar{u}$ quarks in $W^-$ production. Furthermore, due to the parity violating nature of the weak interactions, the $W$ bosons only couple to either left-handed particles or right-handed antiparticles and thus select quarks and antiquarks with helicities parallel or anti-parallel to the longitudinally polarized proton. Consequently, the single spin asymmetry for leptons from $W^+$ decays can then be written at lowest order as

$$A_L^{e^+} = \frac{\int_{\otimes(x_1 x_2)}\{\Delta\bar{d}(x_1)u(x_2)(1+cos\theta)^2 - \Delta u(x_1)\bar{d}(x_2)(1-cos\theta)^2\}}{\int_{\otimes(x_1 x_2)}\{u(x_1)\bar{d}(x_2)(1-cos\theta)^2 + \bar{d}(x_1)u(x_2)(1+cos\theta)^2\}},$$

and similarly for $W^-$ bosons by exchanging up and down quark flavors.

While the valence quark helicity densities are already well known at intermediate $x$ from DIS, the sea quark helicity PDFs are only poorly constrained. The latter are of special interest due to the differing predictions in various models of nucleon structure [see Ref. 13 for a review], which are all able to describe the asymmetry of the unpolarized light quark sea. The existing polarized SIDIS data show a tendency for $\bar{u}$ and $\bar{d}$ having opposite signs while still being consistent with a flavor symmetric light sea within the large uncertainties.

The inaugural run at $\sqrt{s}$ = 500 GeV in 2009 provided the first, but still statistics limited measurement of the longitudinal single-spin asymmetries $A_L$ in $W^{\pm}$ boson production and their subsequent calculable leptonic decay [14]. The rapid analysis of the 2011 ($\sqrt{s}$ = 500 GeV) and the high statistics 2012 ($\sqrt{s}$=510 GeV) longitudinal polarized $p$+$p$ data sets provided the first results for $W^{\pm}$ with substantial impact on the light sea (anti-) quark polarizations.

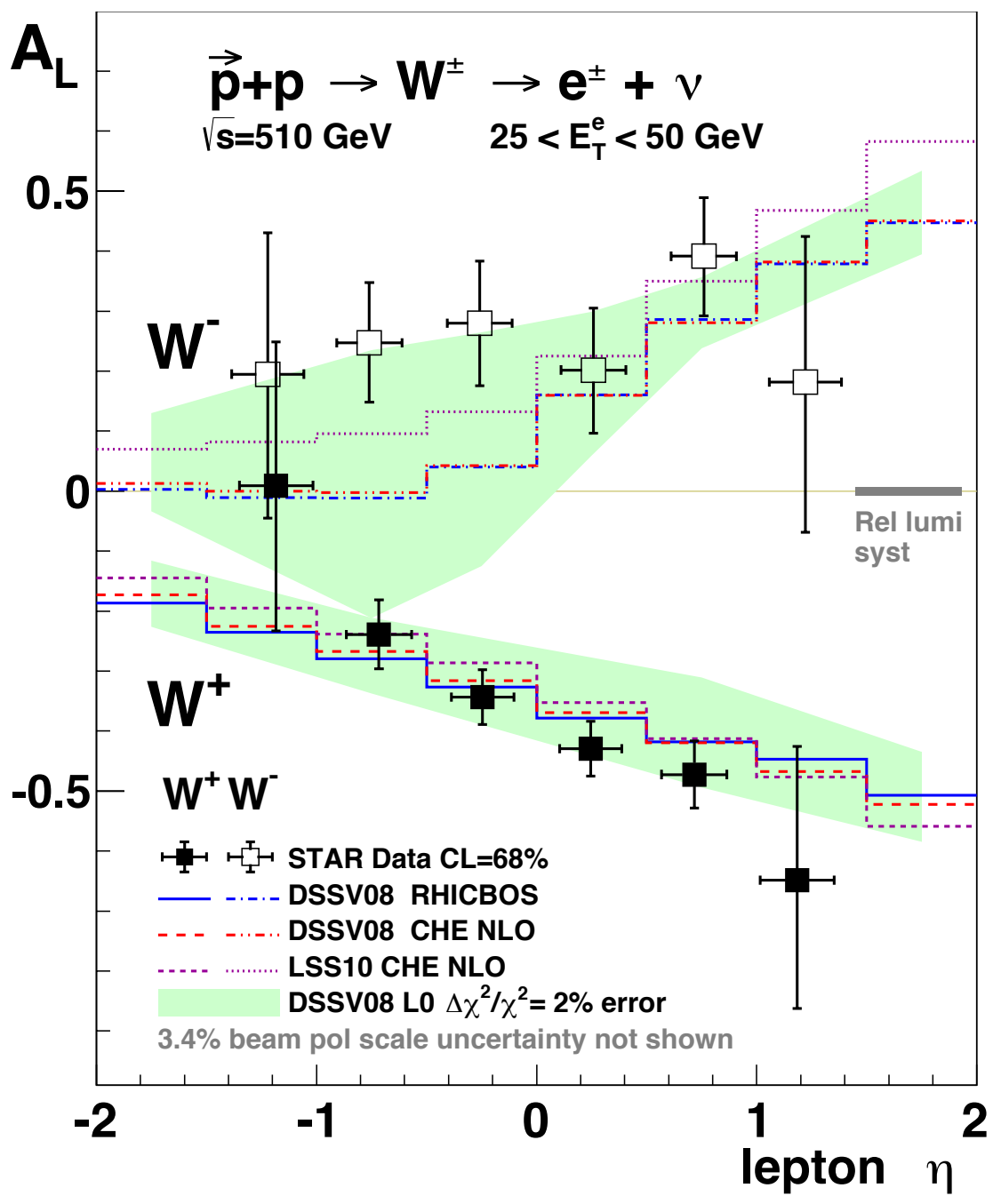

Figure 3-9: Longitudinal single-spin asymmetry $A_L$ for $W^{\pm}$ production as a function of lepton pseudorapidity $\eta_e$ measured by STAR [15] in comparison to theory predictions only based on inclusive and semi-inclusive DIS data.

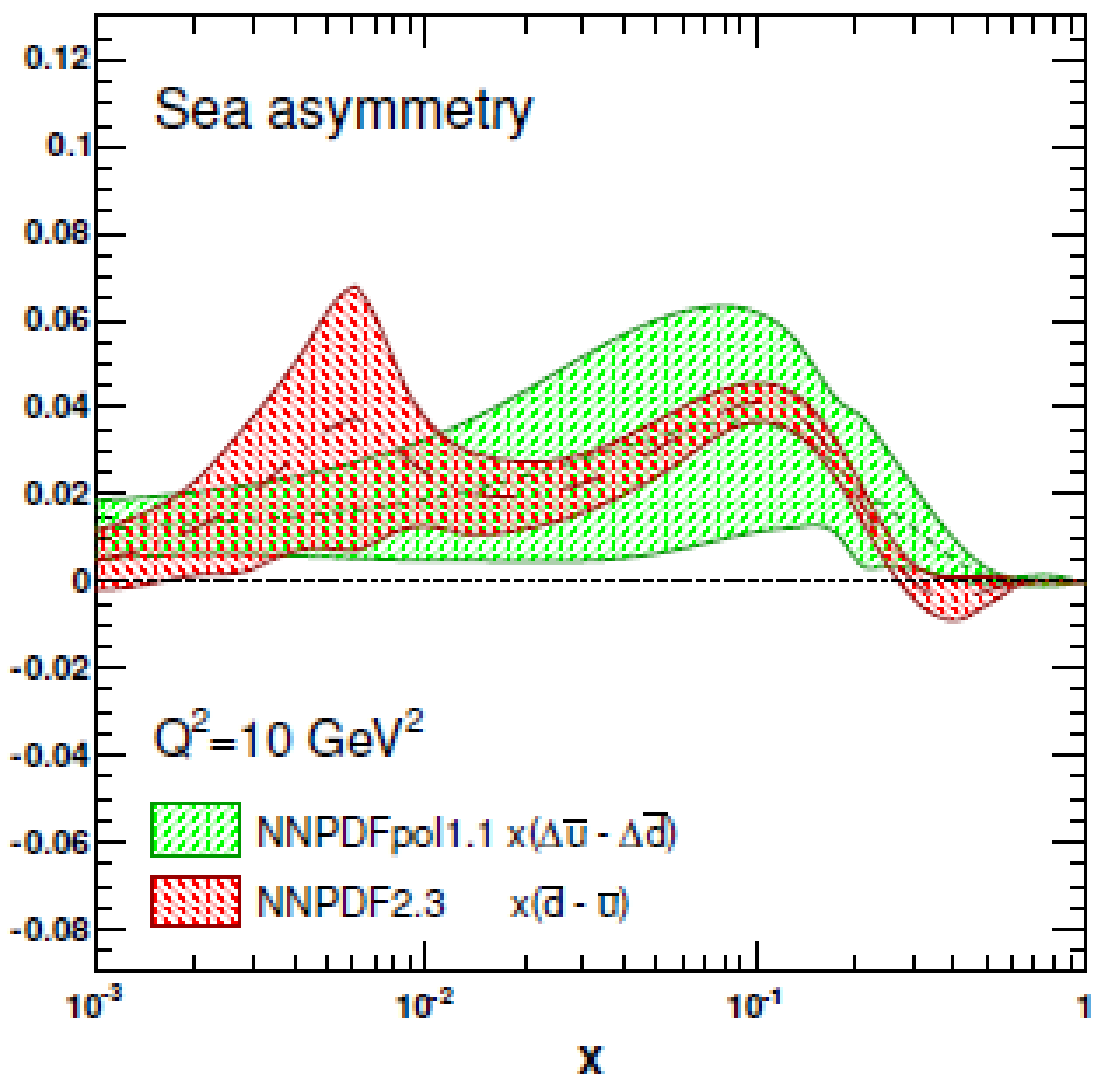

Figure 3-10: Light sea polarized (green) and unpolarized (red) differences between $\bar{u}$ and $\bar{d}$ quarks. The curves are extracted by NNPDF-2.3 for the unpolarized PDFs and by NNPDFpol1.1 for the polarized PDFs, which included the 2012 STAR $W$ single spin asymmetries [15] in their fit [16].

Most interestingly, the 2011+2012 STAR data [15] (see Figure 3-9) at central rapidities prefer $W^-$ decay lepton asymmetries which are systematically larger than the central values of the DSSV08 analysis, directly indicating more positive $\bar{u}$ helicities, while the $W^+$ asymmetries are roughly compatible with the central values of the DSSV08 analysis. As a consequence these results provide the first clear evidence of the flavor symmetry being also broken for the polarized light quark sea with $\Delta\bar{u}$ being positive than $\Delta\bar{d}$ being negative. A positive $\Delta\bar{u} - \Delta\bar{d}$ asymmetry would favor the chiral quark soliton model, the Pauli blocking ansatz and statistical models, while disfavoring many cloud based models.

A first polarized PDF fit by the NNPDF group [9], confirms this finding, although they do not include any SIDIS data at present. Their results show that a symmetric polarized sea in the measured $x$ region can be excluded by more than one sigma, see Figure 3-10.

With the substantially increased statistics from Run-2013 much smaller uncertainties on the asymmetries will be reached for the total data taken between 2011 and 2013, as projected in Figure 3-11. Preliminary results by PHENIX at central and forward rapidities have recently been released, which are consistent with the previous measurements where overlapping in rapidity. The STAR analysis is still ongoing.

With the total 2011 to 2013 data sets analyzed by both experiments the expected uncertainties of the integral on the $\Delta\bar{u}$ helicity in the accessed $x$ range above 0.05 will reach ~1% as shown in Figure 3-13 (left) according to a DSSV analysis including pseudo-data based on the expected uncertainties from both experiments. For the $\Delta\bar{d}$ helicity a precision of around 2% can be expected as is displayed in Figure 3-13 (right). Including pseudo-data based on the projected uncertainties shown in Figure 3-11 into the NNPDF framework the relative uncertainty reduction seen in the DSSV++ fit on the individual $\Delta\bar{u}$ and $\Delta\bar{d}$ parton distribution functions as shown in Figure 3-13 is confirmed. The uncertainty on the flavor asymmetry for the polarized light quark sea $\Delta\bar{u} - \Delta\bar{d}$ is further reduced and a measurement on the 2σ level will be possible (see Figure 3-12). These results demonstrate that the RHIC $W$ program will lead, as soon as all the recorded data are fully analyzed, to a substantial improvement in the understanding of the light sea quark polarizations in the nucleon.

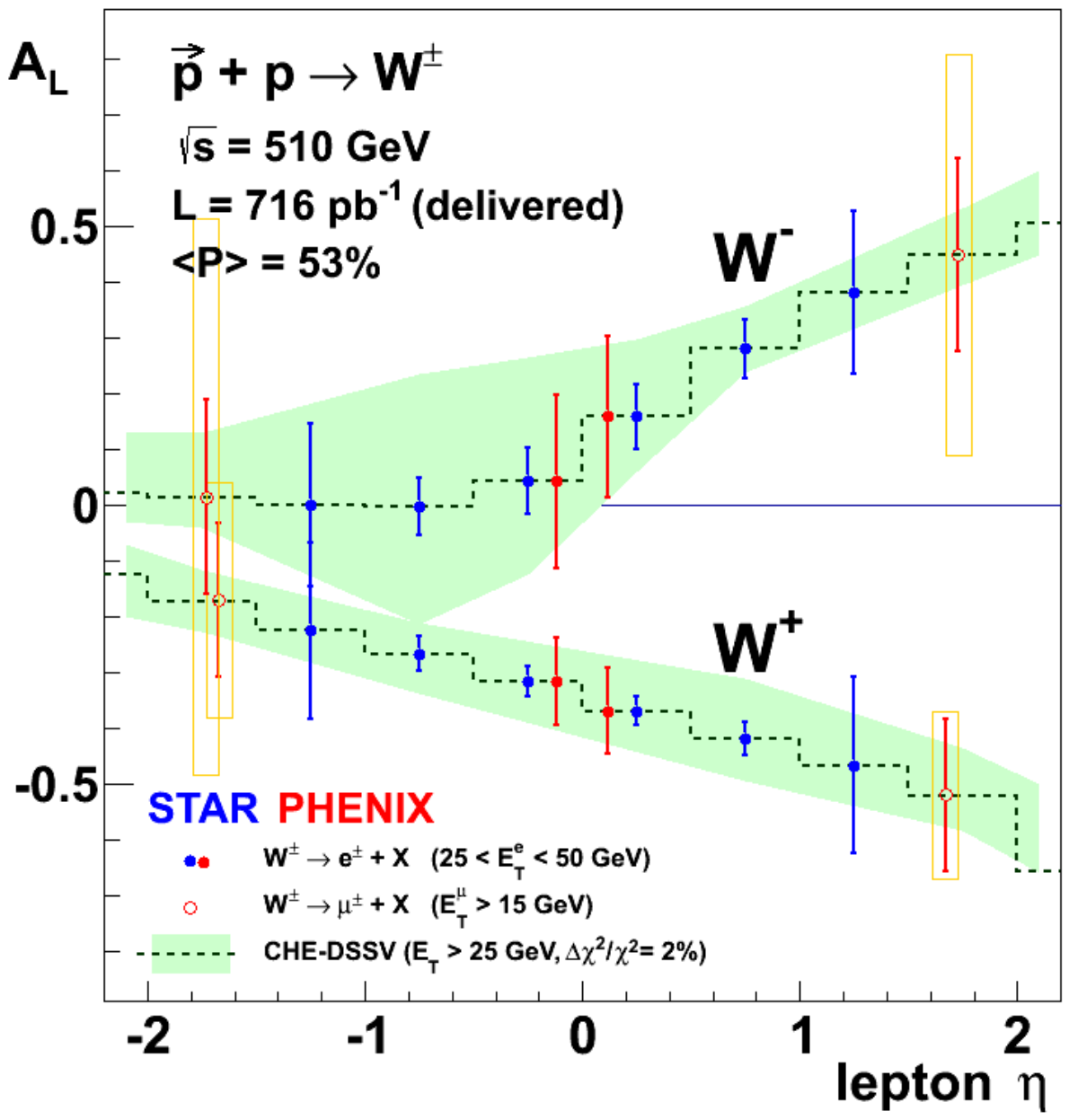

Figure 3-11: Projected uncertainties of the $W$ single spin asymmetries $A_L$ as a function of rapidity.
The total delivered luminosity corresponds to 713 pb$^{-1}$ with an average polarization over the three running periods and both beams of 53%

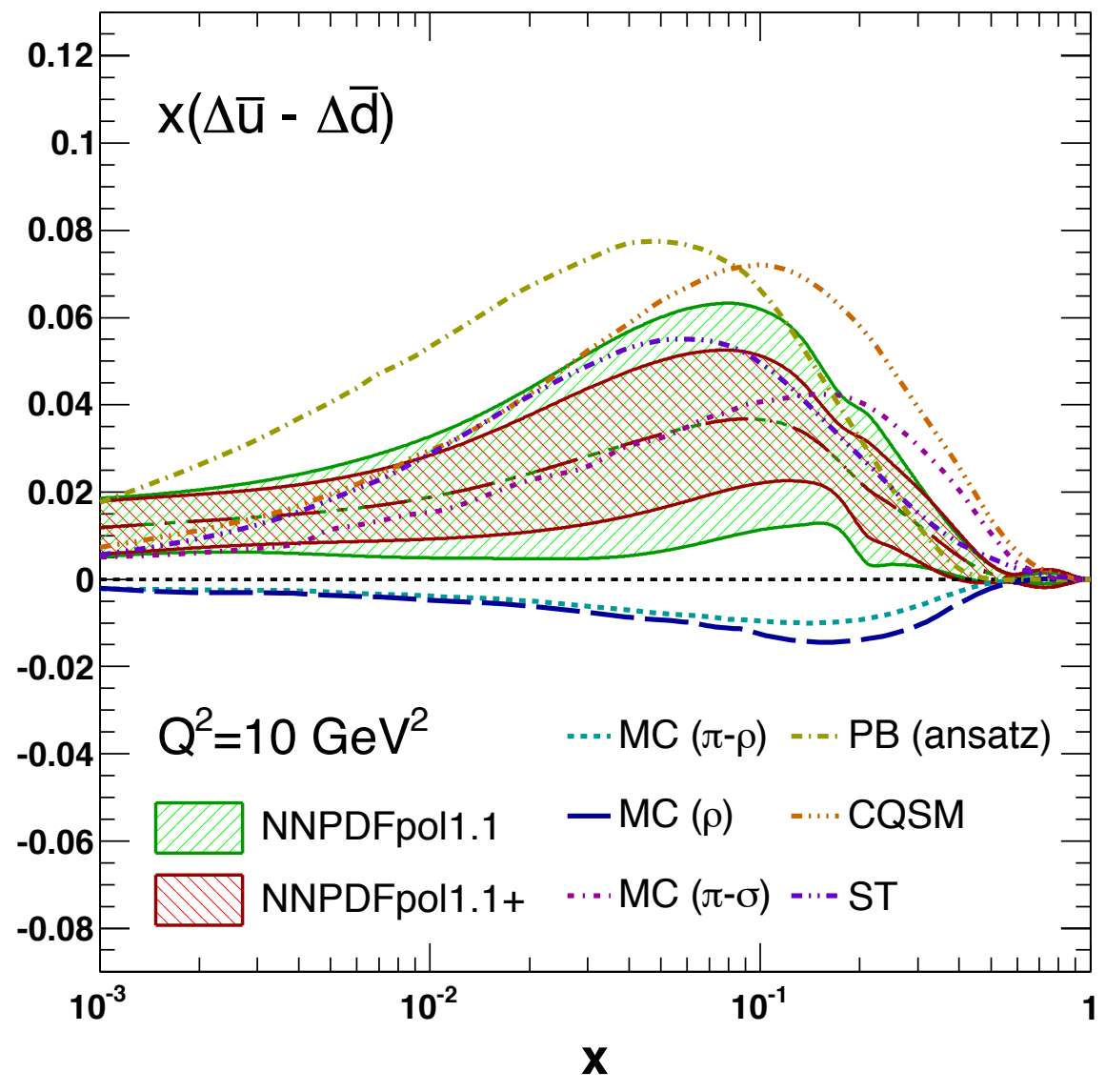

Figure 3-12: The polarized light sea-quark asymmetry $x(\Delta\bar{u} - \Delta\bar{d})$ computed with NNPDFpol1.1 and NNPDFpol1.1+ PDFs, after including the pseudo-data based on the projected uncertainties shown in Figure 3-11, at $Q^2$ = 10 GeV$^2$ compared to various models of nucleon structure (see Ref. [13] for a review).

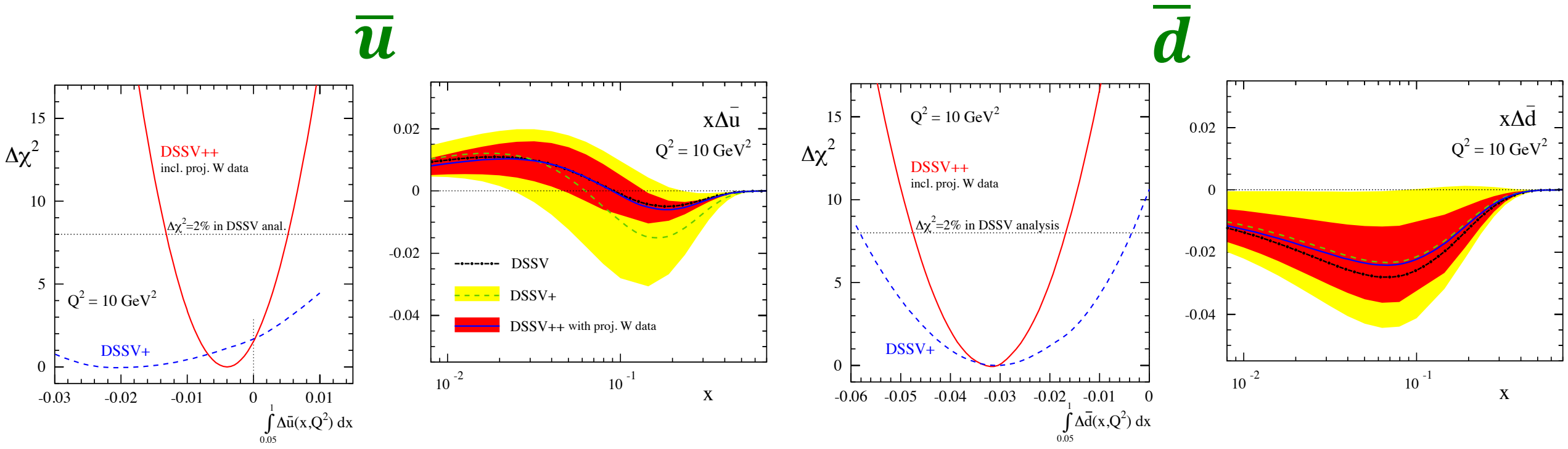

Figure 3-13: $\chi 2$ profiles and x-dependent uncertainty estimates for $\Delta\bar{u}$ (left) and $\Delta\bar{d}$ (right) with (DSSV++) and without (DSSV, DSSV+) including the projections for $W$ boson $A_L$ data shown in Figure 3-11.

As a complementary measurement the unpolarized cross section ratios of positively charged $W$'s over negatively charged $W$'s probe the ratio of the unpolarized light sea quark distributions without the need of nuclear targets and their required corrections. Assuming the strange (anti-) quark contributions are negligible the $W$ cross section ratio can be directly related to the ratio: $R^{W^+/W^-} = \frac{u(x_1)\bar{d}(x_2)+\bar{d}(x_1)u(x_2)}{\bar{u}(x_1)d(x_2)+d(x_1)\bar{u}(x_2)}$.

The STAR experiment has recently shown that in addition to measuring the decay leptons of the $W$-bosons it can also fully reconstruct the kinematics of the $W$-bosons themselves [17]. Figure 3-14 shows the anticipated experimental uncertainties for $R^{W^+/W^-}$ as function of the $W$ boson rapidity for the data recorded for 2011 to 2013. Also shown is the theoretical uncertainty based on the current knowledge of the unpolarized quark distributions, i.e. CT10-NLO [18] calculated using MCFM 6.8 [19]. These data are directly without any assumptions sensitive to the light sea asymmetry in an region of $x > 0.05$, which is relevant to confirm the E866 findings [20] and in particular whether there is a sign change of this ratio at $x$ of 0.25-0.3 currently based on one point being 1.5 statistical standard deviations below zero.

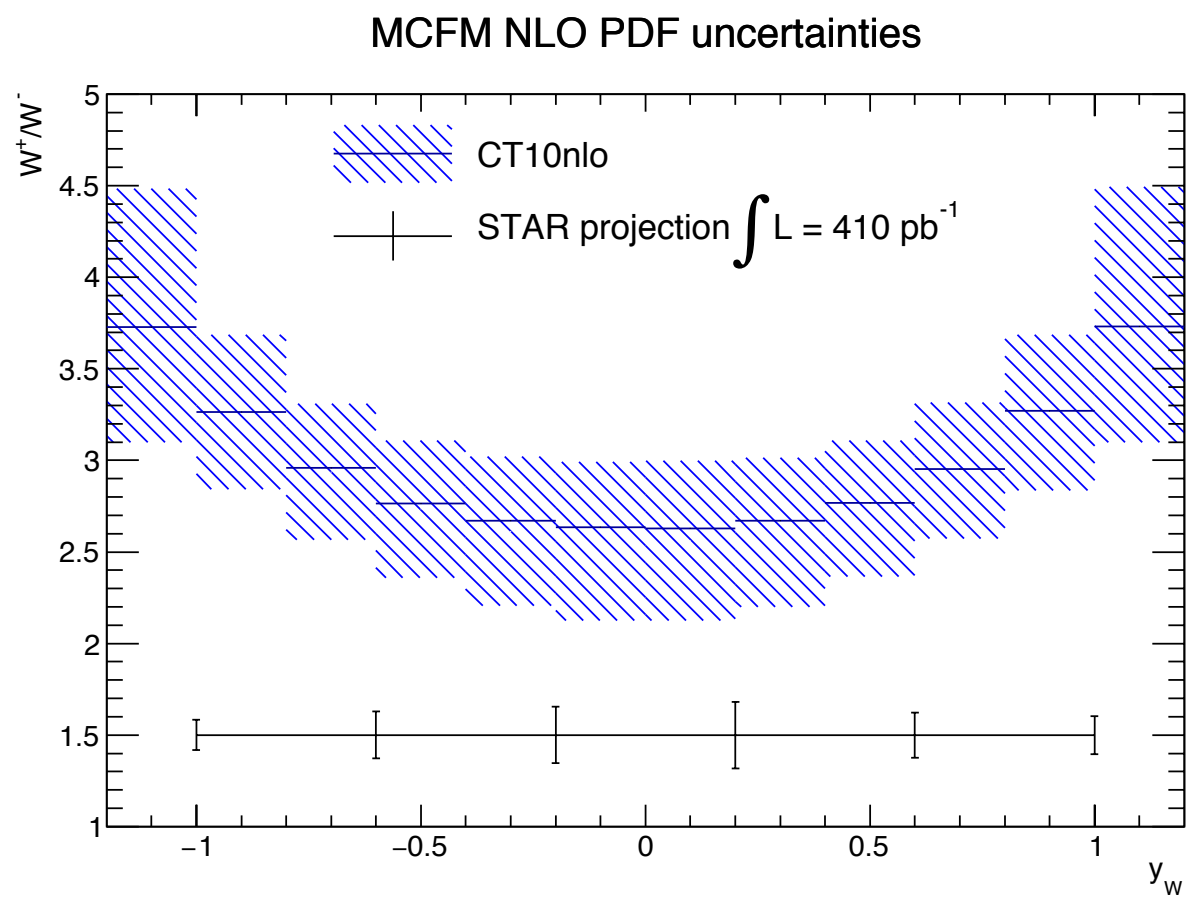

Figure 3-14 Expected uncertainties for the $W$ production charge ratios as a function of the $W$ rapidity as measured by STAR compared to the theoretical uncertainty based on the current knowledge of the unpolarized quark distributions from CT10 [18]. The experimental uncertainties are based on the data recorded 2011 to 2013.

# 4 THE CONFINED MOTION OF PARTONS IN NUCLEONS

A natural next step in the investigation of nucleon structure is an expansion of our current picture of the nucleon by imaging the proton in both momentum and impact parameter space. From TMD parton distributions we can obtain an "image" of the proton in transverse as well as in longitudinal momentum space (2+1 dimensions). At the same time we need to further our understanding of color interactions and how they manifest themselves in different processes. This has attracted renewed interest, both experimentally and theoretically, in transverse single spin asymmetries (SSA) in hadronic processes at high energies, which have a more than 30 year history. Measurements at RHIC have extended the observations from the fixed-target energy range to the collider regime, up to and including the highest center-of-mass energies to date in polarized $p$+$p$ collisions. Figure 4-1 summarizes the measured asymmetries from different RHIC experiments as function of Feynman-$x$ ($x_F \sim x_1$-$x_2$).

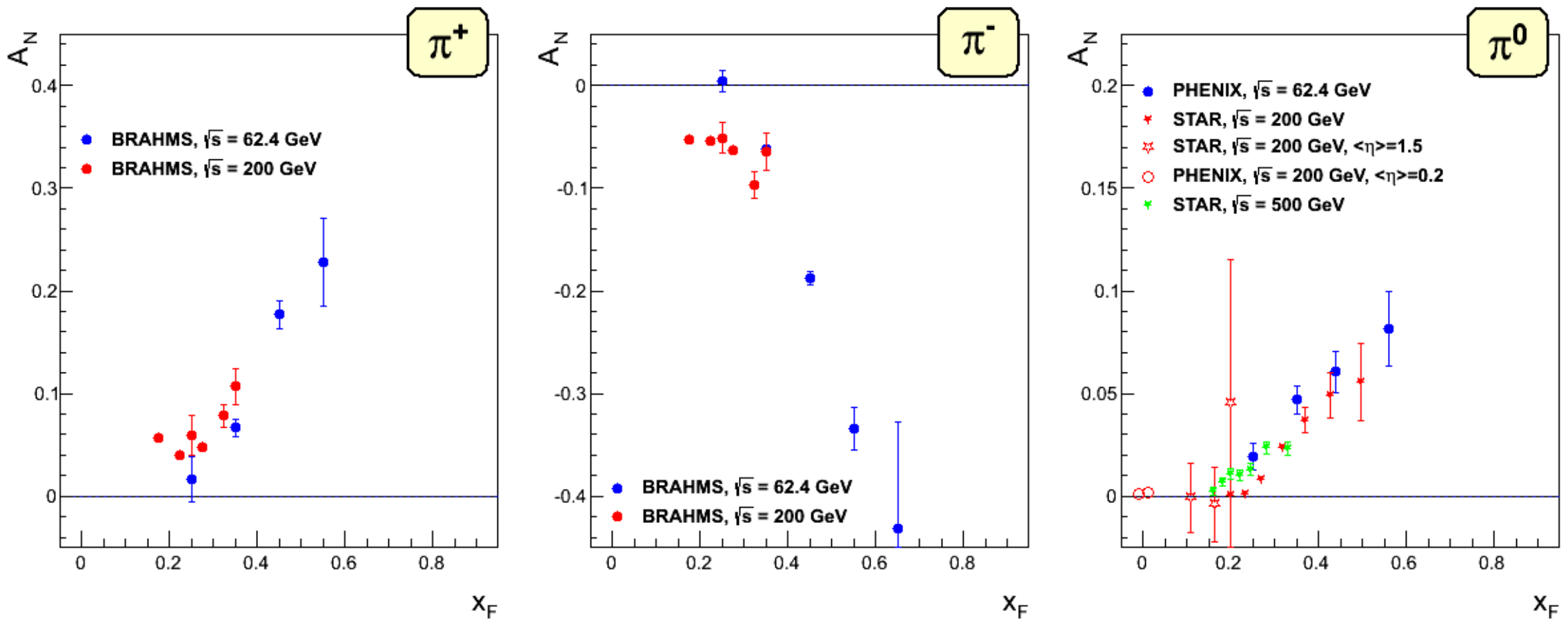

Figure 4-1: Transverse single spin asymmetry measurements for charged and neutral pions at different center-of-mass energies as function of Feynman-$x$.

The surprisingly large asymmetries seen are nearly independent of $\sqrt{s}$ over a very wide range. To understand the observed SSAs one has to go beyond the conventional leading twist collinear parton picture in the hard processes. Two theoretical formalisms have been proposed to explain sizable SSAs in the QCD framework: These are transverse momentum dependent parton distributions and fragmentation functions, such as the Sivers and Collins functions discussed below, and transverse-momentum integrated (collinear) quark-gluon-quark correlations, which are twist-3 distributions in the initial state proton or in the fragmentation process. For many spin asymmetries, several of these functions can contribute and need to be disentangled to understand the experimental observations in detail, in particular the dependence on $p_T$ measured in the final state. The functions express a spin dependence either in the initial state (such as the Sivers distribution or its Twist-3 analog, the Efremov-Teryaev-Qui-Sterman (ETQS) function [21]) or in the final state (via the fragmentation of a polarized quarks, such as the Collins function).

The Sivers function, $f_{1T}^{\perp}$, describes the correlation of the parton transverse momentum with the transverse spin of the nucleon. A non-vanishing $f_{1T}^{\perp}$ means that the transverse parton momentum distribution is azimuthally asymmetric, with the nucleon spin providing a preferred transverse direction. The Sivers function, $f_{1T}^{\perp}$, is correlated with the ETQS functions, $T_{q,F}$, through the following relation:

$T_{q,F}(x,x) = -\int d^2k_\perp \frac{|k_\perp|^2}{M} f_{1T}^{\perp q}(x,k_\perp^2)|_{SIDIS}$ [Eq. 4-1].

In this sense, a measurement constraining the ETQS function indirectly also constrains the Sivers function. We will use this connection repeatedly in the following.

The Collins function, $H_1^{\perp}$, describes a correlation of the transverse spin of a fragmenting quark and the transverse momentum of a hadron singled out among its fragmentation products. A crucial difference between the Sivers and Collins is that the former has a specific process dependence (see Section 4.2). This is a non-trivial theory result, first shown in [22] and extended to *p+p* collisions in [37], which applies to TMD distribution and fragmentation functions in general.

The universality of the Collins FF is of special importance to the *p+p* case where it is always coupled to the chirally odd quark transversity distribution $\delta q(x,Q^2)$, which describes the transverse spin preference of quarks in a transversely polarized proton.

In the last years the focus was on identifying observables that will help to separate the contributions from the initial and final state effects, and will give insight to the transverse spin structure of hadrons. In the following it will be discussed how the current and future data with transverse polarized *p+p* collisions will help to answer the following important questions:

- *Do the large transverse single spin asymmetries survive at high center-of-mass energies?*
- *What is the underlying subprocess responsible for $A_N$?*
- *Is the observed $p_T$-dependence of $A_N$ consistent with theory expectations?*
- *Can the TMD evolution, which is different from the well-known DGLAP evolution, be seen in the RHIC data?*
- *Can the magnitude to which factorization is broken in p+p collisions be determined?*

| Year | √s (GeV) | Recorded Luminosity for transversely polarized *p+p* STAR | Recorded Luminosity for transversely polarized *p+p* PHENIX | <P> in % |
|---|---|---|---|---|
| 2006 | 62.4 | 0.2 $pb^{-1}$ | 0.02 $pb^{-1}$ | 48 |
| | 200 | 8.5 $pb^{-1}$ | 2.7 $pb^{-1}$ | 57 |
| 2008 | 200 | 7.8 $pb^{-1}$ | 5.2 $pb^{-1}$ | 45 |
| 2011 | 500 | 25 $pb^{-1}$ | --- | 53/54 |
| 2012 | 200 | 22 $pb^{-1}$ | 9.7 $pb^{-1}$ | 61/58 |
| 2015 | 200 | 50 $pb^{-1}$ | 40 $pb^{-1}$ | 60 |
| 2016 | 500 | 400 $pb^{-1}$* | not defined | 55 |

Table 4-1: Luminosity recorded by STAR in the past transverse polarized *p+p* runs from 2006 onward. The PHENIX numbers are for |vtx| < 30cm. The bottom row reflects the STAR and PHENIX beam use request for 2015. (* delivered luminosity)

Our current understanding is/will be based on the already taken or soon to be taken data sets listed in Table 4-1.

To disentangle the different subprocesses it is important to identify less inclusive measurements (which are particularly sensitive to certain processes). Table 4-2 identifies observables that allow separating the contributions from polarization effects in initial and final states, and will give insight to the transverse spin structure of hadrons. At this point we should emphasise that most observables in *p+p* collisions can only be related to the transverse spin structure of hadrons through the Twist-3 formalism, where only one hard scale is required. This is typically the $p_T$ of a produced particle or jet, which at RHIC is sufficiently large in much of the phase space. By contrast, the TMD framework requires two hard scales, $p_T$ and $Q$ with $p_T << Q$. Di-jets, azimuthal dependences of hadrons within a jet, *W, Z,* or Drell-Yan production are observables in *p+p* collisions providing two such scales. Moreover, TMD factorization in *p+p* collisions may be broken for processes with observed hadrons or jets in the final state [23]. A better understanding of this issue remains an outstanding task.

| Initial State | Final State |
|---|---|
| $A_N$ as function of rapidity, $E_T$, $p_T$ and $x_F$ for inclusive jets, direct photons and charmed mesons<br><br>$A_N$ as a function of rapidity, $p_T$ for $W^{\pm}$, $Z^0$ and DY | $A_{UT}$ as a function of the azimuthal dependence of the correlated hadron pair on the spin of the parent quark (transversity x interference fragmentation function)<br><br>Azimuthal dependences of hadrons within a jet (transversity x Collins fragmentation function)<br><br>$A_N$ as function of rapidity, $p_T$ and $x_F$ for inclusive identified hadrons (transversity x Twist-3 fragmentation function) |

Table 4-2: Observables to separate the contributions from initial and final states to the transverse single spin asymmetries. Two-scale processes are indicated in blue and one-scale ones in black.

# 4.1 TRANSVERSTITY AND THE COLLINS AND INTERFERENCE FRAGMENTATION FUNCTIONS

As described above, for a complete picture of nucleon spin structure at leading twist one must consider not only unpolarized and helicity distributions, but also those involving transverse polarization, such as the transversity distribution, [24, 25, 26]. The transversity distribution can be interpreted as the net transverse polarization of quarks within a transversely polarized proton [25]. Transversity is difficult to access due to its chiral-odd nature, requiring the coupling of this distribution to another chiral-odd distribution. Semi-inclusive deep inelastic scattering (SIDIS) experiments have successfully probed transversity through two channels: asymmetric distributions of single pions, coupling transversity to the transverse-momentum-dependent Collins FF [27], and asymmetric distributions of di-hadrons, coupling transversity to the so-called "interference fragmentation function" (IFF) [28] in the framework of collinear factorization. Taking advantage of universality and robust proofs of TMD factorization for SIDIS, recent results [29,30,31,32] have been combined with $e^+e^-$ measurements [33,34] isolating the Collins and IFFs for the first global analyses to extract simultaneously the transversity distribution and polarized FF [35, 36]. In spite of this wealth of data, the kinematic reach of existing SIDIS experiments, where the range of Bjorken-$x$ values does not reach beyond $x \lesssim 0.3$, limits the current extractions of transversity.

Following the decomposition as described in [37,38, 39] the Collins effect times the quark transversity distribution and the IFF times the quark transversity distribution may be accessed through single spin asymmetries of the azimuthal distributions of hadrons inside a high energy jet and the azimuthal asymmetries of opposite signed pion pairs respectively. Figure 4-2 and Figure 4-3 show the first observations of nonzero Collins [40] and di-hadron asymmetries [41] in $p$+$p$ collisions at $\sqrt{s}$ = 200 and 500 GeV by STAR. These results are from transversely polarized data taken in 2006, 2011, and 2012, and demonstrate that transversity is accessible in polarized proton collisions at RHIC. STAR finds that the azimuthal asymmetry of pions in polarized jet production depends strongly on $j_T$, the momentum of the pion transverse to the jet thrust axis. The upper panel of Figure 4-2 shows that large asymmetries are seen when a wide range of $j_T$ values ($j_T \sim \Delta R \times p_{T,jet}$) are accepted, whereas the lower panel shows that much smaller asymmetries are found when the measurement is restricted to larger values of $j_T$. In both cases, the 200 and 500 GeV measurements are consistent. A comparison of the transversity signals extracted from the Collins effect and IFF measurements will explore questions about universality and factorization breaking, while comparisons of measurements at 200 and 500 GeV will provide experimental constraints on evolution effects.

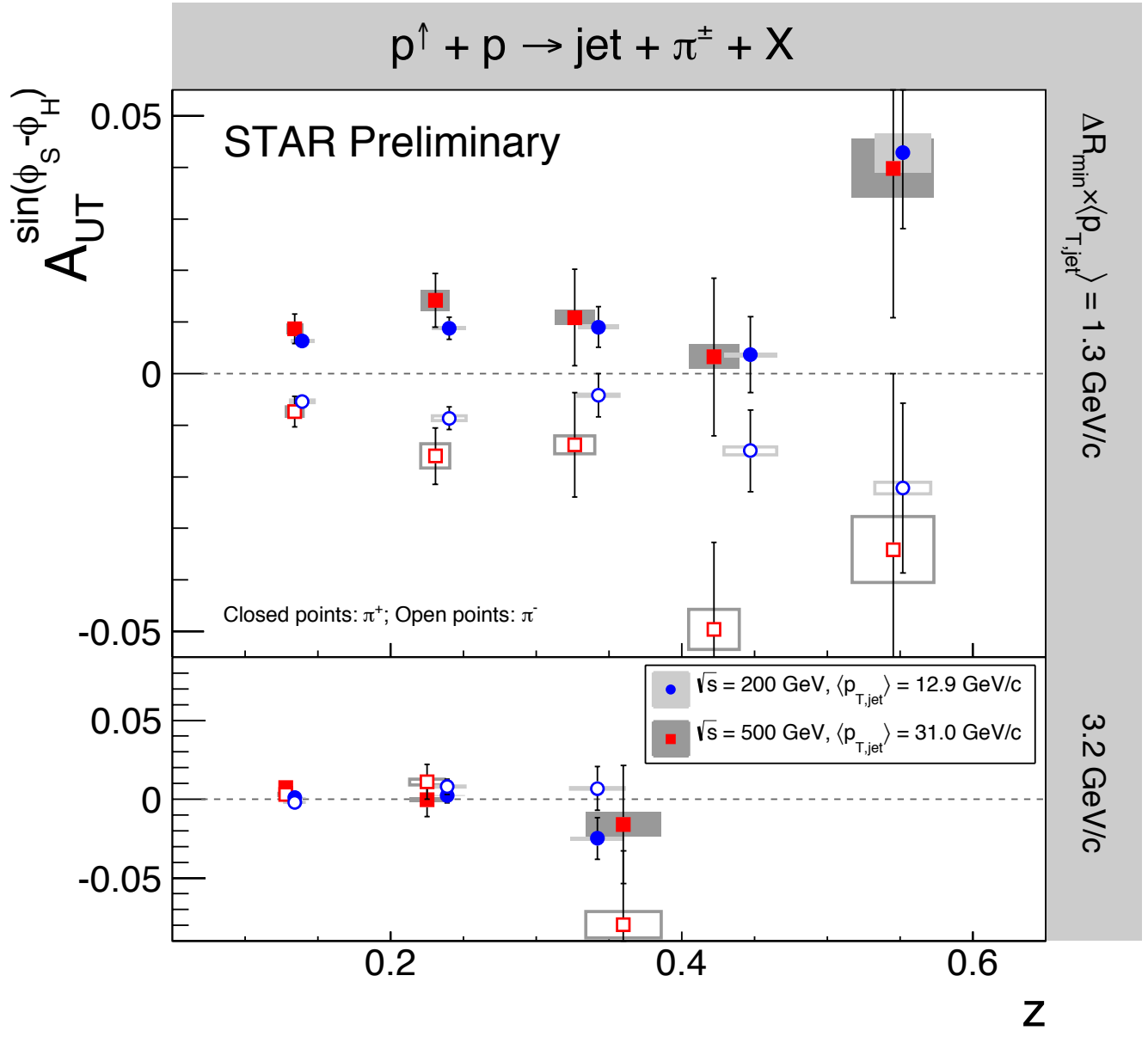

Figure 4-2: $A_{UT}^{\sin(\phi_S-\phi_h)}$ vs. $z$ for charged pions in jets at 0 <$\eta$< 1 from $p$+$p$ collisions at $\sqrt{s}$ = 200 GeV and 500 GeV by STAR. The $p_{T,jet}$ ranges have been chosen to sample the same parton $x$ values for both beam energies. The angular cuts, characterized by the minimum distance of the charged pion from the jet thrust axis, have been chosen to sample the same $j_T$–values ($j_T \sim \Delta R \times p_{T,jet}$). These data show for the first time a non-zero asymmetry in $p$+$p$ collisions sensitive to transversity x Collins FF.

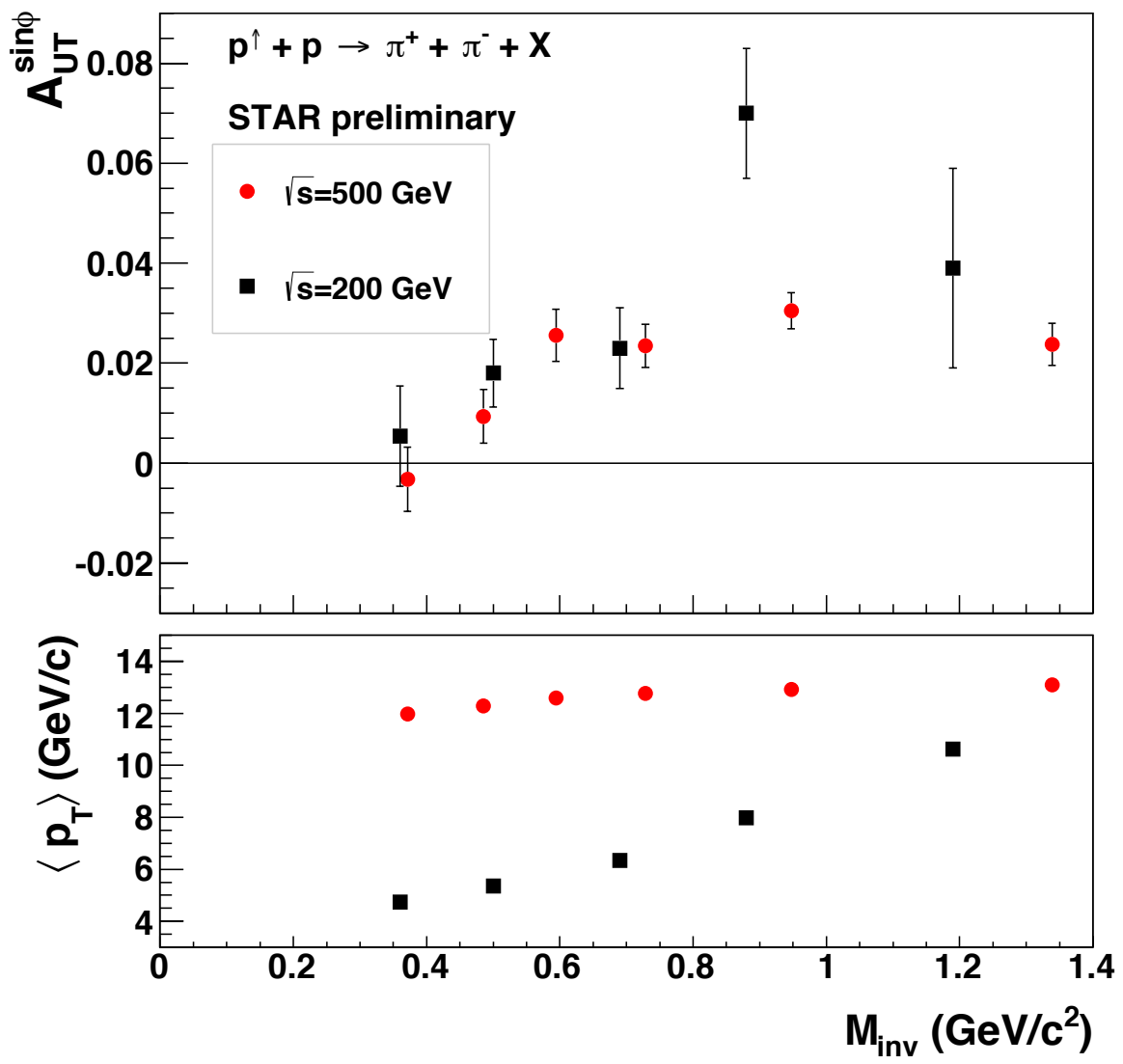

Figure 4-3: $A_{UT}^{sin\phi}$ as a function of $M_{\pi+\pi-}$ (upper panel) and corresponding $p_{T_{\pi+\pi-}}$ (lower panel). A clear enhancement of the signal around the ρ-mass region is observed both at √s=200 GeV and 500 GeV by STAR for -1 < $\eta$ < 1. These data show for the first time a nonzero asymmetry in $p$+$p$ collisions sensitive to transversity x IFF.

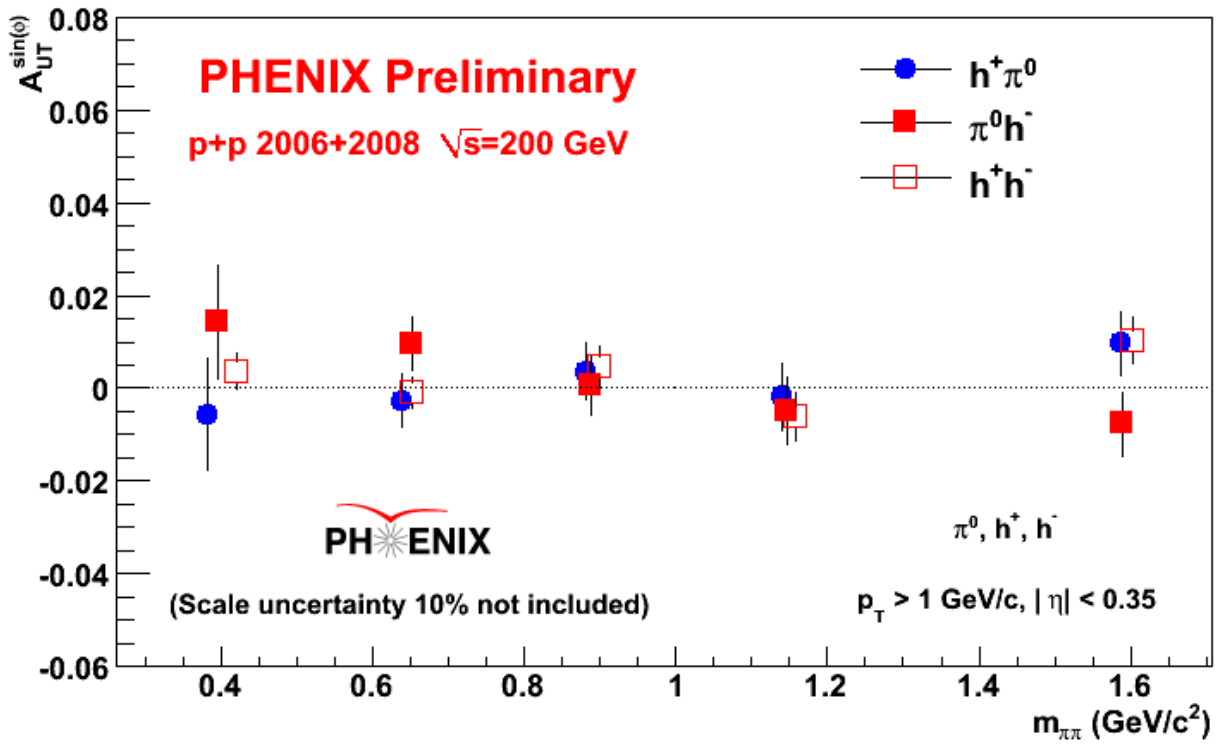

Figure 4-4: $A_{UT}^{sin\phi}$ as a function of the invariant mass for different hadron combinations at √s=200 GeV by PHENIX for -0.35 < $\eta$ < 0.35.

By accessing the Collins asymmetry through the distribution of pions within a jet, one may also extract the $k_T$ dependence of transversity, giving insight into the multidimensional dependence of the distribution. Alternative azimuthal distributions provide sensitivity to physics beyond transversity including gluon linear polarization [38], for which STAR is providing the first ever-experimental limits. STAR will have three times as much data at 200 GeV than shown in Figure 4-2 after the 2015 RHIC run, and has proposed to record over an order of magnitude more data at 500 GeV in 2016. This will enable far more detailed, multi-dimensional examination of the Collins and IFF asymmetries.

Figure 4-4 shows the di-hadron asymmetry in preliminary data from $p^\uparrow + p \rightarrow \pi^+ + \pi^- + X$ at $|\eta| < 0.35$ at $\sqrt{s} = 200$ GeV measured by the PHENIX detector. Contrary to the STAR result these asymmetries are consistent with zero. This behavior is attributed to the limited kinematic coverage in pseudorapidity and $p_T$, which pushes the sampled $x$ range to lower values at which the

magnitude of the transversity distribution drops rapidly.

Figure 4-5 shows a nonzero Collins asymmetry $p^{\uparrow} + p \rightarrow jet + \pi^0 + X$ measured at $2.8 < \eta < 4.0$ [42]. The jet is reconstructed only from the calorimetric energy of the jet deposited in the STAR FMS, an electromagnetic calorimeter at rapidity $2.8 < \eta < 4.0$. The kinematics of this measurement is identical to the one of the large forward asymmetries $A_N$ as depicted in Figure 4-1 for which the Collins mechanism was one of the possible explanations to cause the large asymmetries. This measurement clearly shows that the Collins mechanism alone cannot explain the observed inclusive $\pi^0$-asymmetries.

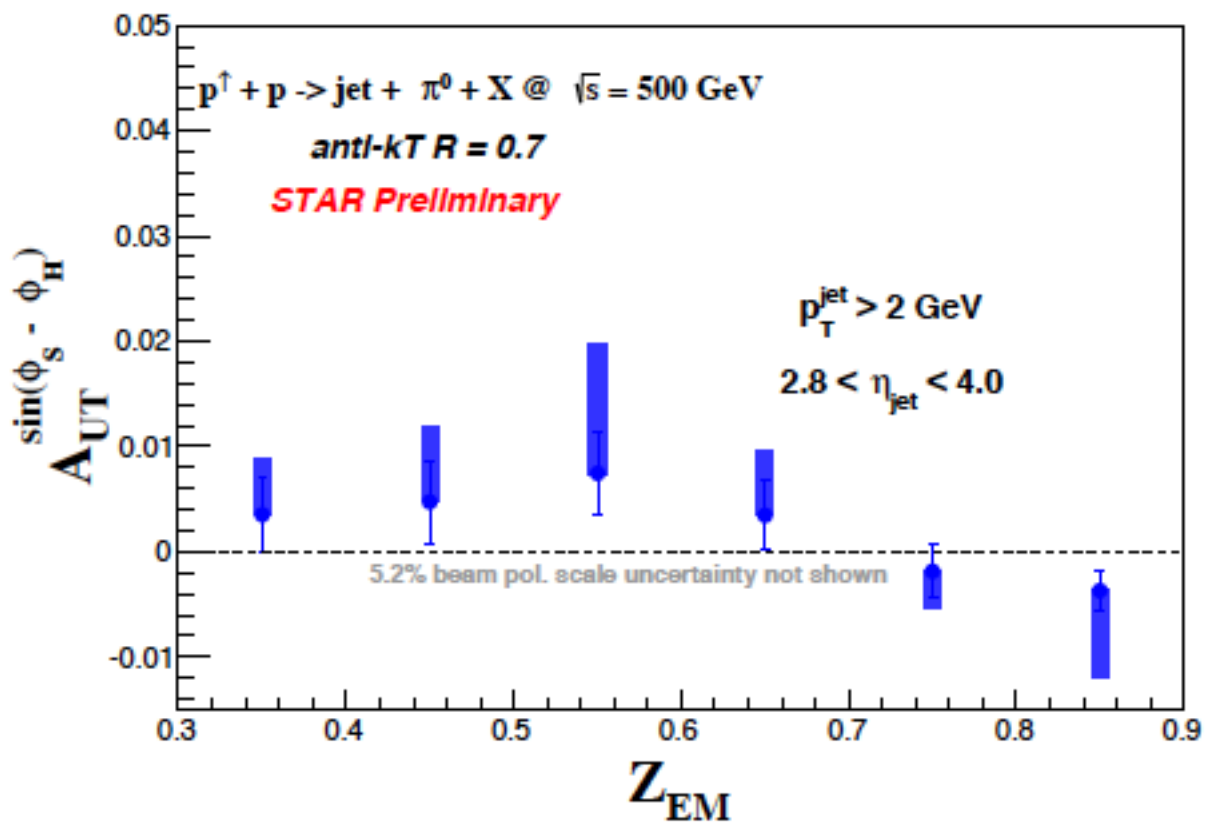

Figure 4-5: The azimuthal asymmetry $A_{UT}^{sin\phi_S - \phi_h}$ of a $\pi^0$ inside a calorimetric jet measured by the STAR FMS at $2.8<\eta< 4.0$ as function of $z_{EM}$, the fractional momentum of the outgoing parton/jet carried by the outgoing hadron.

While the measurements of transversity through the Collins FF need TMD factorization to hold in *p+p* scattering, di-hadron asymmetries utilize collinear factorization. Thus, not only can more precise measurements of these effects in *p+p* improve our knowledge of transversity, such measurements are/will be invaluable to test the longstanding theoretical questions, such as the magnitude of any existing TMD factorization breaking. Extractions at RHIC kinematics also allow the possibility for understanding the TMD evolution of the Collins FF (e.g. Ref. [43]) by comparing to those extractions from SIDIS and $e^+e^-$ data. Probing transversity in *p+p* collisions also provides better access to the d-quark transversity than is available in SIDIS, due to the fact that there is no charge weighting in the hard scattering QCD 2→2 process in *p+p* collisions. Further *p+p* collisions allow access to Transversity for $x > 0.3$ where it is currently very poorly known. Extending the measurements of transversity to the high $x$ region provides access to the tensor charge $\int_0^1 (\delta q^a(x) - \delta \bar{q}^a(x))dx = \delta q^a$ [44] a quantity essential to understand the nucleon structure at leading twist and calculable in lattice calculations.

# 4.2 POLARIZATION EFFECTS IN THE PROTON: SIVERS AND TWIST-3

In the past the primary contributions of PHENIX and STAR to transverse spin physics have been through the study of forward neutral pion production in $p$+$p$ collisions (see, for example, ref. [45,46]). This effort has been extended to include the first measurements at $\sqrt{s}$ = 200 GeV of the transverse spin asymmetry $A_N$ for the $\eta$ meson [47]. The STAR Run-2011 data taken with transverse polarization at $\sqrt{s}$ = 500 GeV have revealed several surprising results.

Figure 4-6 shows the transverse single spin asymmetry $A_N$ for "electromagnetic" jets (i.e. jets with its energy only measured in a electromagnetic calorimeter) detected in the FMS at $2.5 < h < 4.0$ as a function of the jet $p_T$ for different photon multiplicities and ranges in jet energy. It can be clearly seen that with increasing number of photons in the "electromagnetic jet" (increasing "jettiness" of the event) the asymmetry becomes smaller. Jets with an isolated $\pi^0$ have the largest asymmetry consistent with the asymmetry in inclusive $\pi^0$ events, as seen from the right-most panel in Figure 4-1. For all jet energies and photon multiplicities in the jet, the asymmetries are basically flat as a function of jet $p_T$, a feature also already seen for inclusive $\pi^0$ asymmetries. Recently, it has been proposed that in the collinear, Twist-3 factorization approach a significant portion of the sizable inclusive pion asymmetries seen at forward pseudorapidity is due to Twist-3 FFs coupled to transversity [48]. This calculation is the first one which showed, similar to the experiment [49], a flat $p_T$-dependence for $A_N$. The ability for this approach to describe adequately the effects seen at SIDIS and at RHIC is a potentially significant breakthrough in the longstanding mystery surrounding the nonzero inclusive asymmetries at forward pseudorapidity (e.g. Ref. [50]). For these reasons, the most desirable kinematic region for future study at RHIC is in the region of $\eta > 2$.

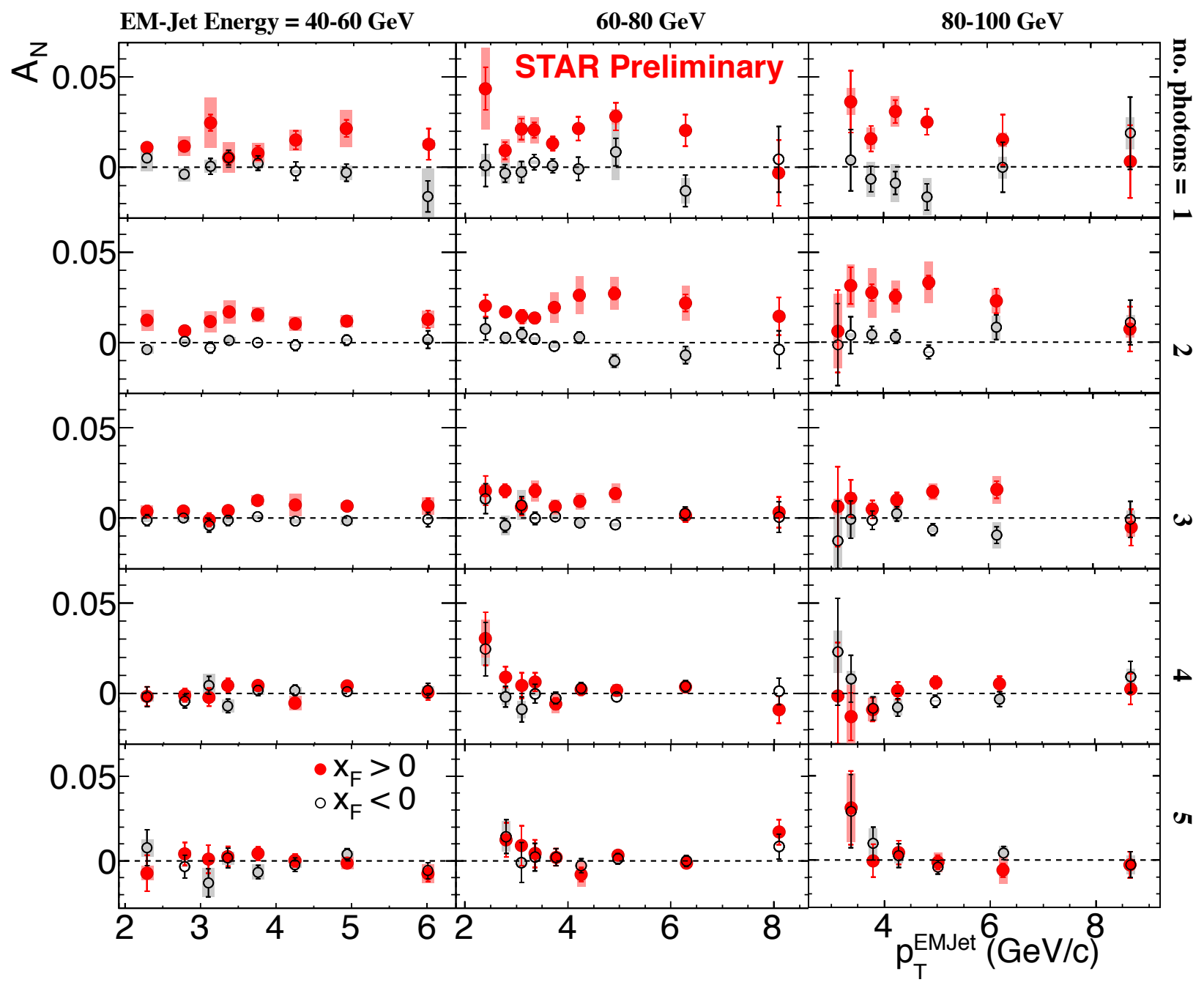

Figure 4-6: The transverse single spin asymmetry $A_N$ for "electromagnetic" jets detected in the FMS ($2.5 < \eta < 4.0$) as function of the jet $p_T$ and the photon multiplicity in the jet in bins of the jet energy. This behavior raises serious questions regarding how much of the large forward $\pi^0$ asymmetries are due to the same underlying dynamics as jet production namely 2→2 parton scattering processes.

To further study these effects the transverse single spin asymmetry $A_N$ of these electromagnetic jets has also been studied in correlation with an away side jet in the rapidity range $-1 < \eta < 2$. Requiring an additional correlated away-side jet the asymmetry for isolated forward $\pi^0$ mesons becomes smaller. For further details see reference [51]. Both these observations raise serious questions regarding how much of the large forward $\pi^0$ asymmetries are due to the same underlying dynamics as jet production namely 2→2 parton scattering processes. The STAR results are in agreement with preliminary results from the $A_N$DY collaboration at RHIC, which meas-ured $A_N$ for inclusive jets at $\sqrt{s}$ = 500 GeV on the order of ~5x10$^{-3}$ [52]. In [53] it is argued that this behavior is consistent with the fact that the Twist-3 parton correlation functions for *u* and *d* valence quarks cancel, because their behavior follows the one obtained for the Sivers function from fits to the SIDIS data, which show the *u* and *d* quark Sivers function to have opposite sign but equal magnitude. The same cancelation occurs calculating $A_N$ for $\pi^0$ at forward rapidities. But even more importantly it has been shown that the sign of the asymmetry calculated in the Twist-3 formalism does not match the one measured for $\pi^0$ $A_N$ [54].

All these observations indicate that the underlying subprocess causing a significant fraction of the large transverse single spin asymmetries in the forward direction is of diffractive nature. The Roman Pot PHASE-II* upgrade at STAR [55,56] will allow to make a measurement of $A_N$ $\pi^0$ for single and double diffractive events by tagging one or both protons in the Roman Pots. A discovery of large transverse single spin asymmetries in diffractive processes would open a new avenue to study the nature of the pomeron exchange in *p+p* collisions. To measure this asymmetry is a goal for the part of Run-2015 with transverse polarized *p+p* collisions.

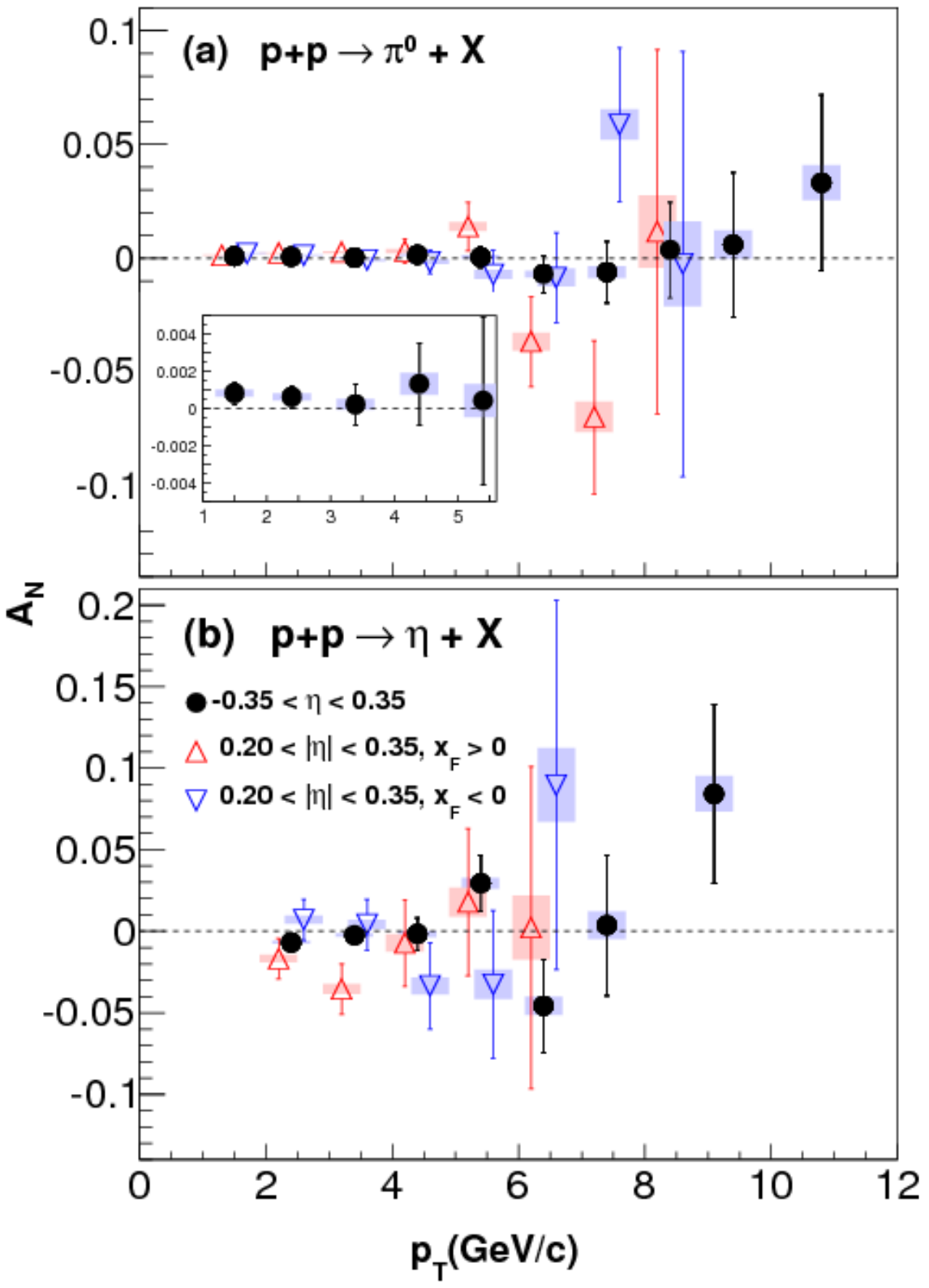

Figure 4-7: The AN measured at midrapidity ($|\eta|$ < 0.35), as function of $p_T$ for $\pi^0$ (a) and $\eta$ (b) mesons.Triangles are slightly forward/backward going sub-samples of the full data set (circles). These are shifted in $p_T$ for better visibility. An additional uncertainty from the beam polarization is not included.

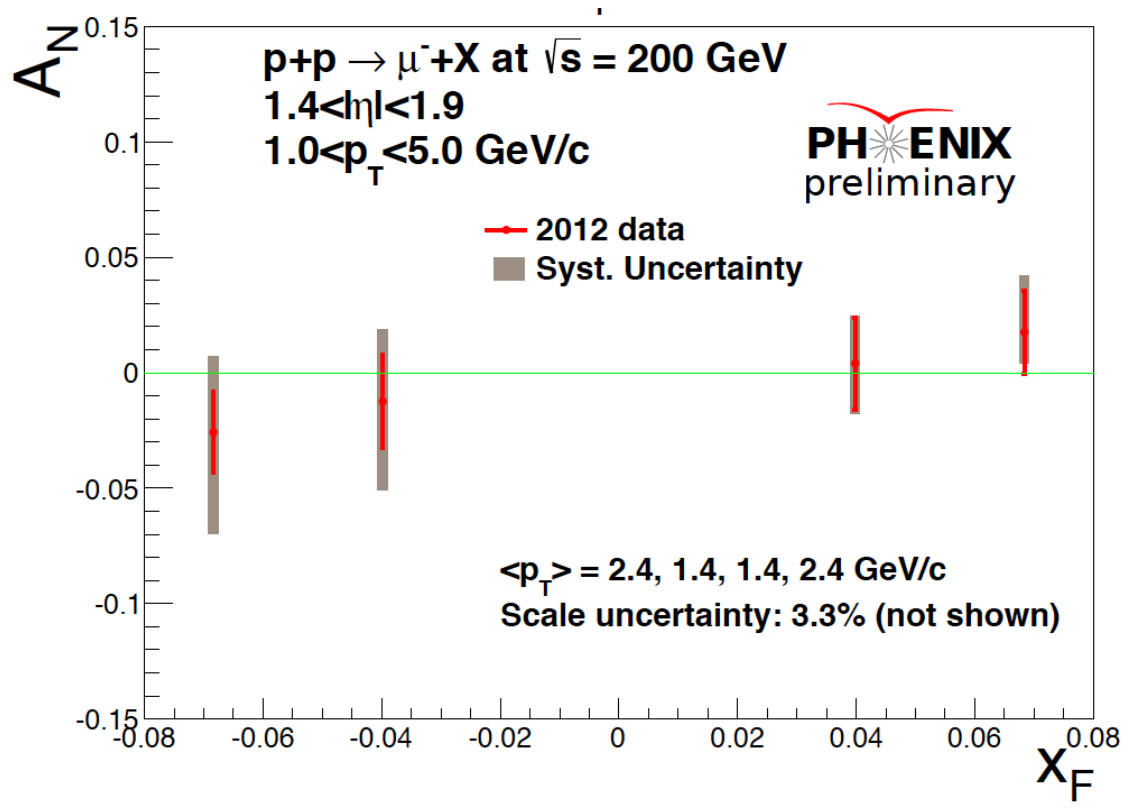

Figure 4-8: Single spin asymmetries of single muons measured at forward rapidities (1.4<η<1.9). The results are corrected for background contributions from punch-through hadrons. Systematic uncertainties are dominated by the uncertainty in the background fraction.

Figure 4-7 shows $A_N$ for $\pi^0$ (a) and $\eta$ (b) mesons at midrapidity, no significant deviation from zero can be seen in the results within the statistical uncertainties in the covered transverse momentum range. Fragmentation effects would likely dominate any difference in the two-meson asymmetries. Either these are small or suppressed by the contributing transversity distribution in the covered kinematic range.

With the dominant underlying subprocesses being *qg* and *gg* scattering (see Figure 2-3 left) these asymmetries, like earlier low statistics results [57], can be used to constrain the gluon Sivers function [58] dominantly through the Twist-3 formalism. Figure 4-8 shows $A_N$ for inclusive muons measured at 1.4 <$|\eta|$<1.9 with the PHENIX muon arms. Theses inclusive muons come dominantly from open charm and J/Ψ mesons, which are only produced through gluon-gluon

fusion and therefore give a direct sensitivity to the integral over $x$ to the gluon Sivers function in the Twist-3 formalism through the analog relation as in Eq. 4-1.

An important aspect of the Sivers effect that has emerged from theoretical study is its process dependence. In SIDIS, the quark Sivers function includes the physics of a final state effect from the exchange of (any number of) gluons between the struck quark and the remnants of the target nucleon. On the other hand, for the virtual photon production in the Drell-Yan process, the Sivers asymmetry appears as an initial state interaction effect. As a consequence, the quark Sivers functions are of opposite sign in these two processes and this non-universality is a fundamental prediction from the color gauge invariance of QCD. The experimental test of this sign change is one of the open questions in hadronic physics (NSAC performance measure HP13) and will deeply test our understanding of QCD factorization. The COMPASS experiment at CERN is pursuing this sign change through DY using a pion beam in the years 2015 and 2016 and new initiatives have been proposed e.g. at FNAL.

While the required luminosities and background suppressions for a meaningful measurement of SSA in Drell-Yan production are challenging, other channels can be exploited in $p+p$ collisions, which are of the same sensitivity to the predicted sign change. These include in the TMD formalism the measurement of SSA of $W^{\pm}$ and $Z$ bosons and in the Twist-3 formalism the SSA for prompt photons and inclusive jets. These are either already accessible with the existing PHENIX and STAR detectors or need only modest upgrades and/or continued polarized beam operations.

Figure 4-9 shows the predicted $A_N$ for DY (left) [67] and $W^-$ [63] (right) **before any TMD evolution is taken into account**. Lately, there have been several theoretical predictions for the transverse single spin asymmetries for DY, $W^{\pm}$ and $Z^0$ bosons including TMD-evolution, for examples see [59,60,61] and references therein. In all cases the asymmetries have been significantly reduced. The TMD evolution equations contain in addition to terms, which can be calculated in QCD, also non-pertubative terms, whose parameters need to be obtained from fits to data. Unfortunately there is not yet a consensus as to how to obtain and handle the non-pertubative input in the TMD evolution, for details see [62]. This complication leads to large uncertainties in the prediction for the DY, $W^{\pm}$ and $Z^0$ SSA, which can only be addressed by future measurements.

The transversely polarized data set in Run-2011 at $\sqrt{s}$ = 500 GeV allowed STAR to reconstruct the transverse single spin asymmetries for $A_N$ for $W^{\pm}$ and $Z^0$ bosons. Especially the measurement of the $A_N$ for $W^{\pm}$ bosons is challenging where contrary to the longitudinally polarized case, it is required to completely reconstruct the $W$ bosons as the kinematic dependences of $A_N$ can not easily be resolved through the high $p_T$ decay lepton, for details see [63,64].

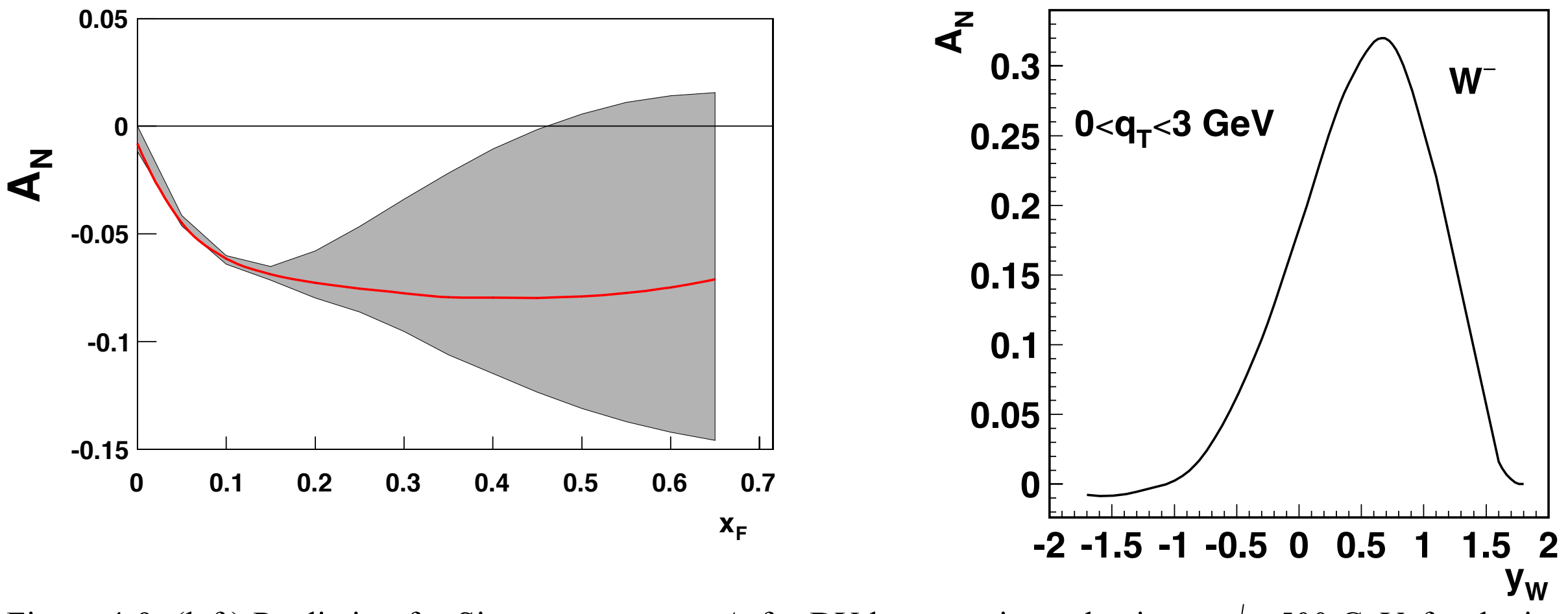

Figure 4-9: (left) Prediction for Sivers asymmetry $A_N$ for DY lepton pair production at $\sqrt{s}$=500 GeV, for the invariant mass 4<$Q$<8 GeV and transverse momenta 0 < $q_T$ < 1 GeV [67]. (right) $A_N$ as a function of $W^-$ boson rapidity at $\sqrt{s}$=500 GeV [63]. **Both predictions are before any TMD evolution is applied.**

Due to the large STAR acceptance it is possible to reconstruct the $W$ boson kinematics from the recoil jet, a technique used at D0, CDF and the LHC experiments to reconstruct the $W$ boson kinematics. Figure 4-10 (upper left) shows the transverse single spin asymmetries for $A_N$ for $W^{\pm}$ as a function of the $W$ boson rapidity $y$. The asymmetries have also been reconstructed as a function of the $p_T$ of the $W$ boson. For the $Z^0$ boson (Figure 4-10 (lower left)) the asymmetry could only be reconstructed in one $y$-bin with the current limited statistics (25 pb$^{-1}$). Details for this analysis can be found in [17]. The analysis represents an important proof of principle similar to the first Run-2009 $W^{\pm}$ $A_L$ measurements.

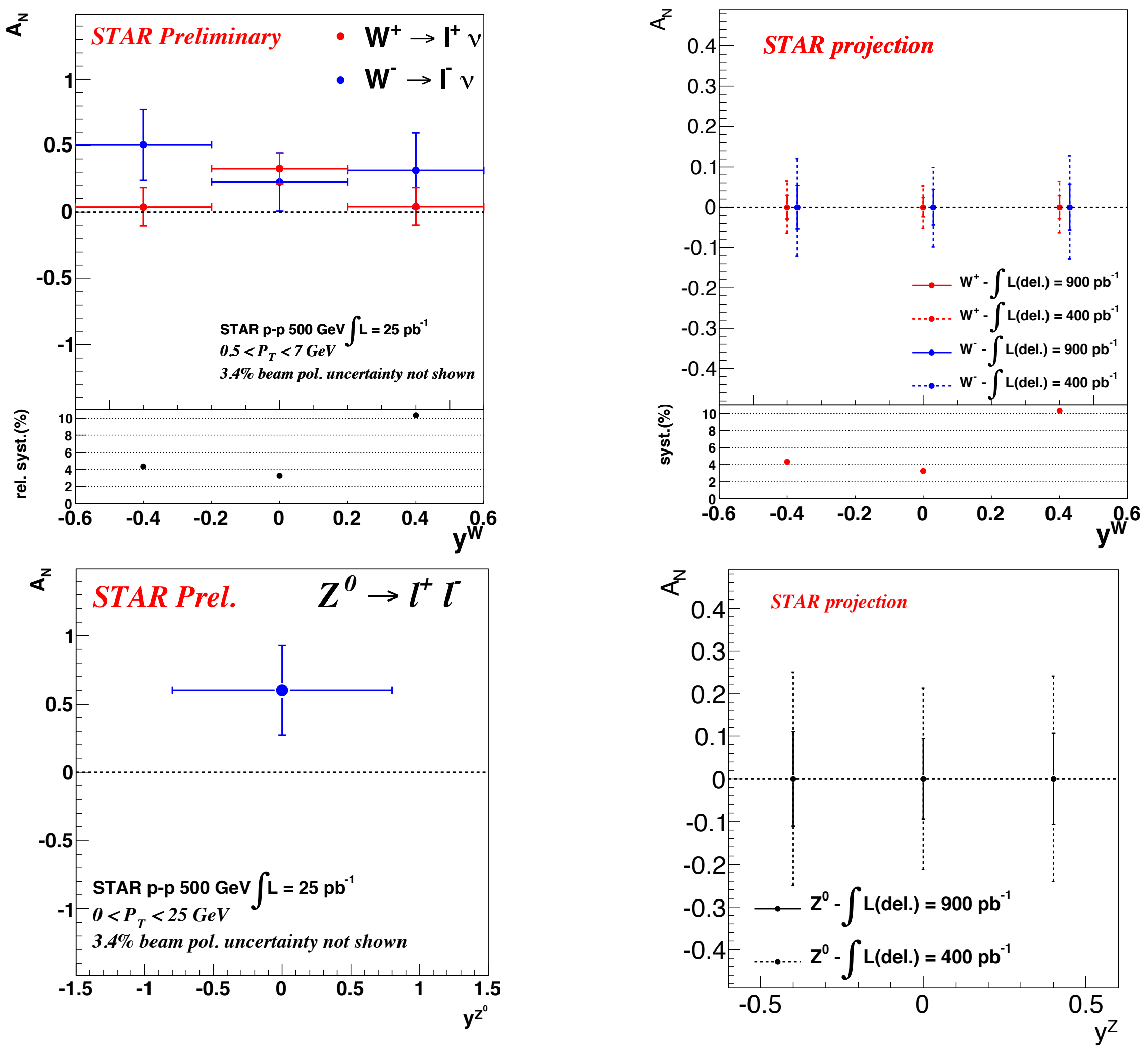

Figure 4-10: left column: The transverse single spin asymmetries $A_N$ for $W^{\pm}$ and $Z$-bosons as function of the $W(Z)$ boson rapidity $y$. right column: The projected uncertainties for transverse single spin asymmetries of $W^{\pm}$ and $Z^0$ bosons as a function of rapidity for a delivered integrated luminosity of 400 (900) pb$^{-1}$ and an average beam polarization of 55%.

$W^{\pm}$ bosons production provides an ideal tool to study the spin-flavor structure of sea quarks inside the proton. Such a measurement of the transverse single spin asymmetry will provide the world wide first constraint on the sea quark Sivers function in a $x$-range where the measured asymmetry in the $\bar{u}$ and $\bar{d}$ unpolarized sea quark distribution functions, as measured by E866 [65], can only be explained by strong non-pQCD contributions. At the same time, this measurement is also able to access the sign change of the Sivers function, if the effect due to TMD evolution on the asymmetries in in the order of a factor 5 reduction.

Figure 4-10 (right) shows the projected uncertainties for transverse single spin asymmetries of $W^{\pm}$ and $Z^0$ bosons as function of rapidity for a delivered integrated luminosity of 400 (900) pb$^{-1}$ and an average beam polarization of 55%. The 400 (900) pb$^{-1}$ correspond to 7 (14) week RHIC-

Run utilizing the concept of dynamic $\beta^*$ squeeze through the duration of a RHIC fill. The dynamic $\beta^*$ squeeze provides a factor 2 increase of the luminosity in a fill, compared to Run-2013, as the luminosity profile through the fill is kept flat. Such a run can be realized as early as in 2016.

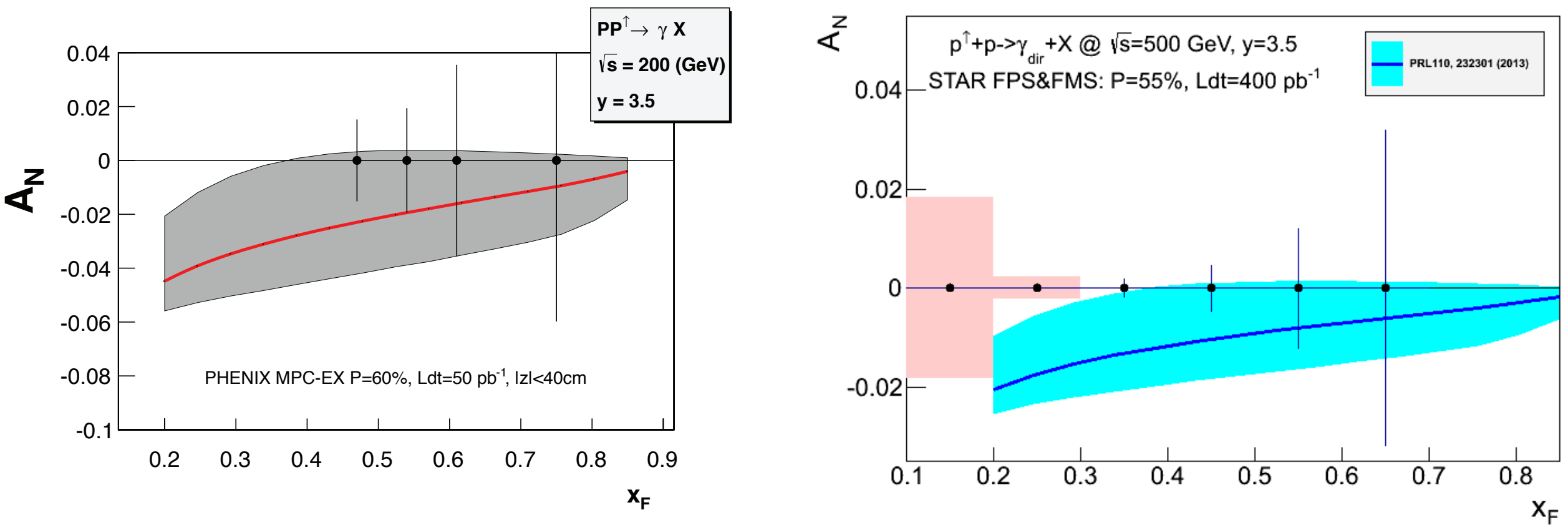

Figure 4-11: Statistical and systematic uncertainties for the direct photon $A_N$ after background subtraction compared to theoretical predictions from Ref. [67] for $\sqrt{s}$ = 200 GeV as measured by PHENIX (left) and for $\sqrt{s}$ = 500 GeV as measured by STAR. If the correlation between the Twist-3 correlation functions and the Sivers function as described in [Eq.4-1] would be violated the asymmetries would have the same magnitude but being positive.

Transverse single spin asymmetries in direct photon production provide a different path to access this sign change through the formalism utilizing the Twist-3 parton correlation functions. For the 2015 polarized $p$+$p$ run both PHENIX and STAR will install preshowers in front of their forward electromagnetic calorimeters the MPC and the FMS, respectively [66, 55]. These upgrades will enable a measurement of the SSA for direct photons. Figure 4-11 (left) shows the statistical and systematic uncertainties for the direct photon $A_N$ as to be obtained by PHENIX in Run-2015 at $\sqrt{s}$ = 200 GeV, similar uncertainties will be obtained by STAR. The asymmetry can be measured up to $x_F \sim 0.7$ where the $\pi^0$ asymmetries are largest (see Figure 4-1). The curve represents a calculation based on Twist-3 parton correlation functions, $T_{q,F}$, constrained by the Sivers function obtained from a fit to the world SIDIS data [67]. Figure 4-11 (right) shows the same simulation and uncertainties for a direct photon SSA at $\sqrt{s}$ = 500 GeV a measurement earliest possible in 2016. The theoretical asymmetries are reduced by a factor 2 due to evolution effects at $\sqrt{s}$ = 500 GeV. For these asymmetries no cancelation as for jets, due to the opposite sign but similar magnitude of the $u$ and $d$ quark Twist-3 correlation functions is expected due to the electromagnetic nature of the process the individual parton densities are weighted with the respective quark charge $e_q^2$.

The ultimate test for the TMD factorization, evolution and the relation between the Sivers function and the Twist-3 correlation function (see Eq. 4.1) would be to measure $A_N$ for $W^{\pm}$, $Z^0$ boson, DY production and direct photons. To obtain a significant measurement of $A_N$ for DY production, the DY leptons need to be detected at rapidities 2 to 4 for a lepton pair mass above 4 GeV. This is a highly nontrivial measurement, as backgrounds mainly due to QCD 2→2 processes need to be suppressed by a factor ~$10^6$. A first attempt to measure $A_N$ for DY could be done during the 2016 run, using in the case of STAR the FMS, preshower and a newly installed postshower to obtain the needed background suppression. In the case of PHENIX the $A_N$ for DY should be measured at $1.2 < \eta < 2.2$ and for a lepton pair mass from 2 to 4 GeV to have a sizable asymmetry. Detailed measurements of $A_N$ for DY will only be possible with forward upgrades for both PHENIX and STAR.

# 4.3 PHYSICS WITH TRANSVERSE POLARIZED p+A COLLISIONS

### *Single Transverse Spin Asymmetry in Polarized Proton-Nucleus Collisions:*

As a result of exciting recent theoretical developments, the scattering of a polarized proton on an unpolarized nuclear target appears to have the potential to extend and deepen our understanding of QCD. In the frame where the nucleus is relativistic, its wave function consists of densely packed quarks and gluons, which constantly split and merge with each other. At high enough energies the density of the gluons is so high that the saturation regime is reached, characterized by strong gluon fields and scattering cross sections close to the unitarity bound. The saturated wave function is often referred to as the Color Glass Condensate (CGC) (for details see [68]).

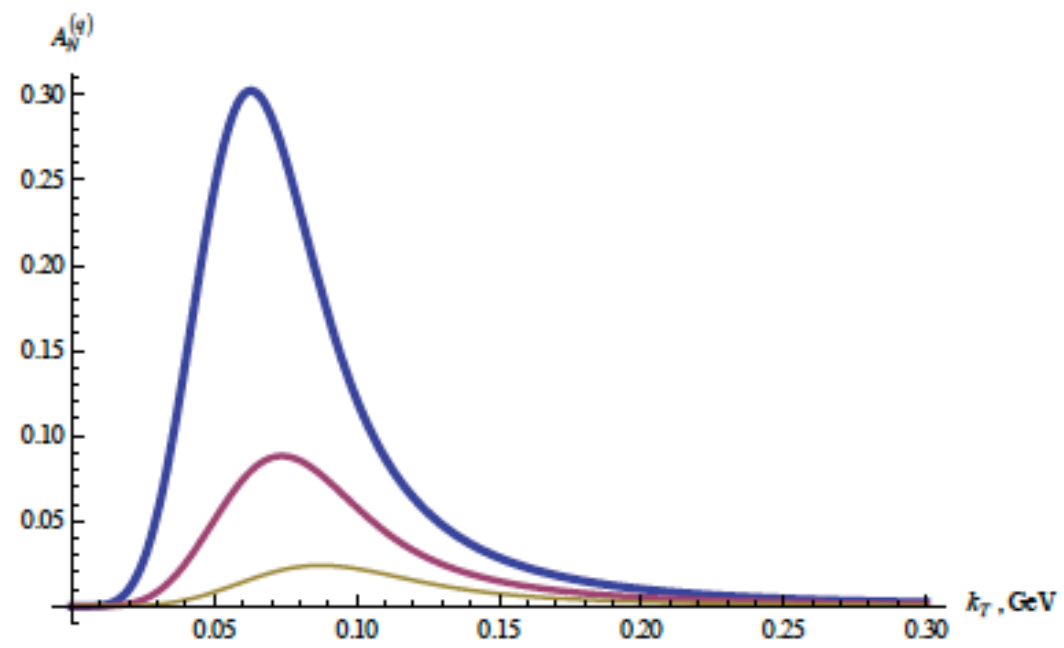

Figure 4-12: Quark SSA from Eq. (81) in [69] plotted as a function of $k_T$ for different values of the target radius: $R$ = 1 fm (blue curve), $R$ = 1.4 fm (red curve), and $R$ = 2 fm (gold curve) for $\alpha$ = 0.7.

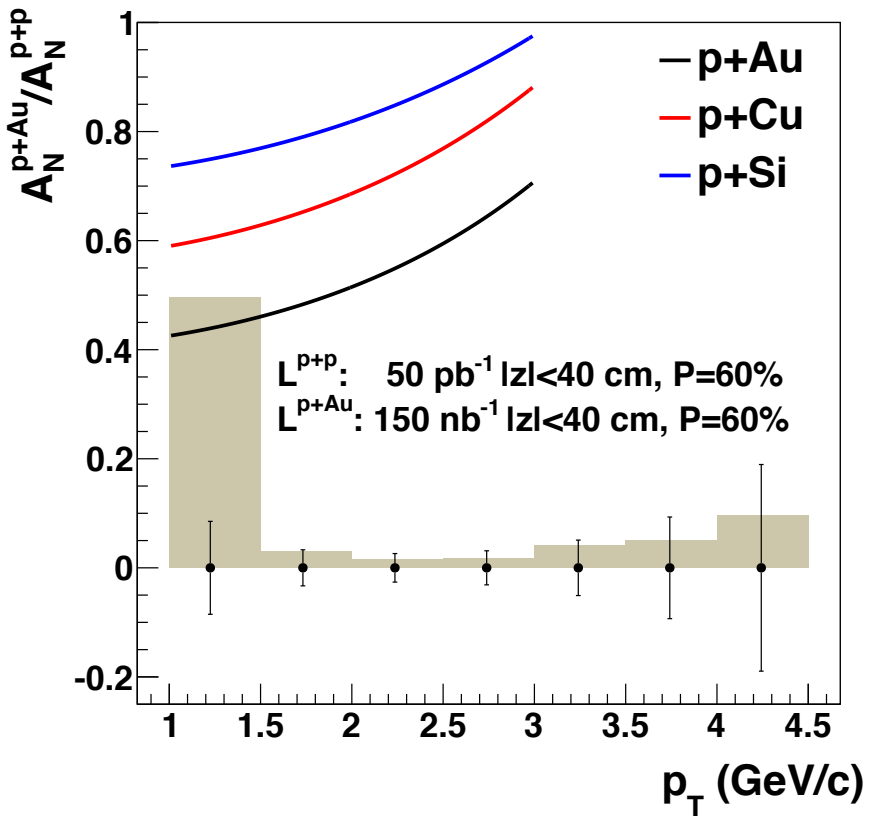

Figure 4-13: The projected statistical and systematic uncertainties for the ratio of $A_N^{pA}/A_N^{pp}$ measured for $\pi^0$'s in the PHENIX MPC-EX for the 2015 requested transverse $p+p$ and $p+A$ running [66]. The colored curves follow Eq. 17 in Ref. [73] assuming $Q_s^p$ = 1 GeV (solid) with $Q_s^A = A^{1/3}\, Q_s^p$

Scattering a polarized probe on this saturated nuclear wave function may provide a unique way of probing the gluon and quark TMDs. In particular, the single transverse spin asymmetry $A_N$ may provide access to the elusive nuclear Weizsaecker-Williams (WW) gluon distribution function [70], which is a solid prediction of the CGC formalism [71] but is very difficult to measure experimentally. The nuclear effects on $A_N$ may shed important light on the strong interaction dynamics in nuclear collisions. While the theoretical approaches based on CGC physics predict that hadronic $A_N$ should decrease with increasing size of the nuclear target [72,73,74] (see Figure 4-12), some approaches based on perturbative QCD factorization predict that $A_N$ would stay approximately the same for all nuclear targets [75]. The asymmetry $A_N$ for prompt photons is equally important to measure. The contribution to the photon $A_N$ from the Sivers effect [76] is expected to be nonzero, while the contributions of the Collins effect [77] and of the CGC-specific odderon-mediated contributions [78] to the photon $A_N$ are expected to be suppressed [73,79]. Clearly experimental data on polarized proton-nucleus collisions are needed in order to distinguish different mechanisms for generating the single spin asymmetry $A_N$ in nuclear and hadronic collisions.

Figure 4-13 shows the statistical und systematic uncertainties for the ratio of $A_N$ $\pi^0$ for the 2015 RHIC transverse polarized $p+p$ and $p+Au$, respectively, as measured by PHENIX. The curves represent the theoretical prediction [73] for the suppression of SSA in the nuclear medium. This measurement will not only allow us to get a handle on the saturation scale, but will also help to understand the underlying subprocess leading the large forward SSA in transverse polarized $p+p$. To distinguish further between the different theoretical models predicting a suppression of $A_N$ in $p+A$ it will be essential to measure simultaneously $A_N$ for direct photons and $\pi^0$. The same measurements can be performed by STAR (for details see [55]).

### *Access to the generalized parton distribution $E_g$:*

"How are the quarks and gluons, and their spins distributed in space and momentum inside the nucleon?"

"What is the role of orbital motion of sea quarks and gluons in building the nucleon spin?"

These are key questions, which need to be answered to understand overall nucleon properties like the spin structure of the proton. The formalism of generalized parton distributions provides a theoretical framework, which allows some answers to the above questions [80]. Exclusive reactions in DIS, i.e., deeply virtual Compton scattering, have been mainly used to constrain GPDs. RHIC with its possibility to collide transversely polarized protons with heavy nuclei has worldwide the unique opportunity to measure $A_N$ for exclusive $J/\psi$ in ultra-peripheral $p^{\uparrow}$+Au collisions (UPC) [81]. A nonzero asymmetry would be the first signature of a nonzero GPD $E$ for gluons, which is sensitive to spin-orbit correlations and is intimately connected with the orbital angular momentum carried by partons in the nucleon and thus with the proton spin puzzle. To measure $A_N$ for exclusive $J/\psi$ in ultra-peripheral $p^{\uparrow}$+Au collisions provides an advantage in rate as the emission of the virtual photon from the gold nucleus is enhanced by $Z^2$ compared to ultra-peripheral $p^{\uparrow}$+p collisions. Detecting the scattered polarized proton in "Roman Pots" (RP) and vetoing the break-up of the gold nucleus can ensure exclusivity of the process. The event generator SARTRE [82], which also describes well the STAR results for $\rho^0$ results for UPC in Au+Au collisions, has been used to simulate exclusive $J/\psi$ -production in $p^{\uparrow}$+Au UPC (see Figure 4-14 for the two contributing processes).

To select the $J/\psi$ with the photon generated from the Au-beam cuts in the RP as installed for PHASE-II* need to be applied, at 200 GeV the RP PHASE-II* system has a $t$-acceptance from -0.016 GeV2 to -0.2 GeV$^2$ shows the probed $x$-$Q^2$ plane (left), if the photon is emitted by the gold beam and the rapidity distribution of the $J/\psi$ -meson for both processes generating the "UPC-photon" after applying all the cuts to reduce background from the "p-shine" UPC-events [55]. The final number of $J/\psi$'s for a recorded luminosity of 300 nb$^{-1}$ is ~7k, which is enough to have a look to the $A_N$ as function of $t$. This measurement can be further improved with a high statistics $p^{\uparrow}$+A run in the later part of the decade and a high statistics $p^{\uparrow}$+p run in 2011.

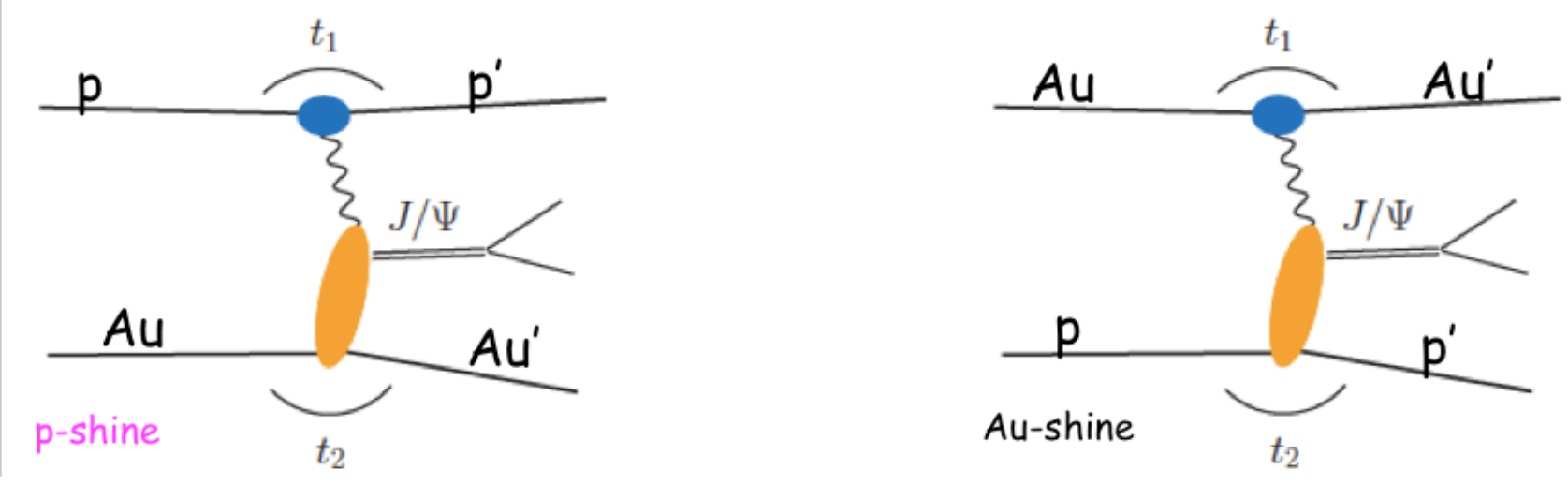

Figure 4-14: Possible processes to generate exclusive $J/\psi$ in $p^{\uparrow}$+Au UPC in one case the photon is emitted by the proton, "p-shine" (left) in the other by the gold, "Au-shine" (right).

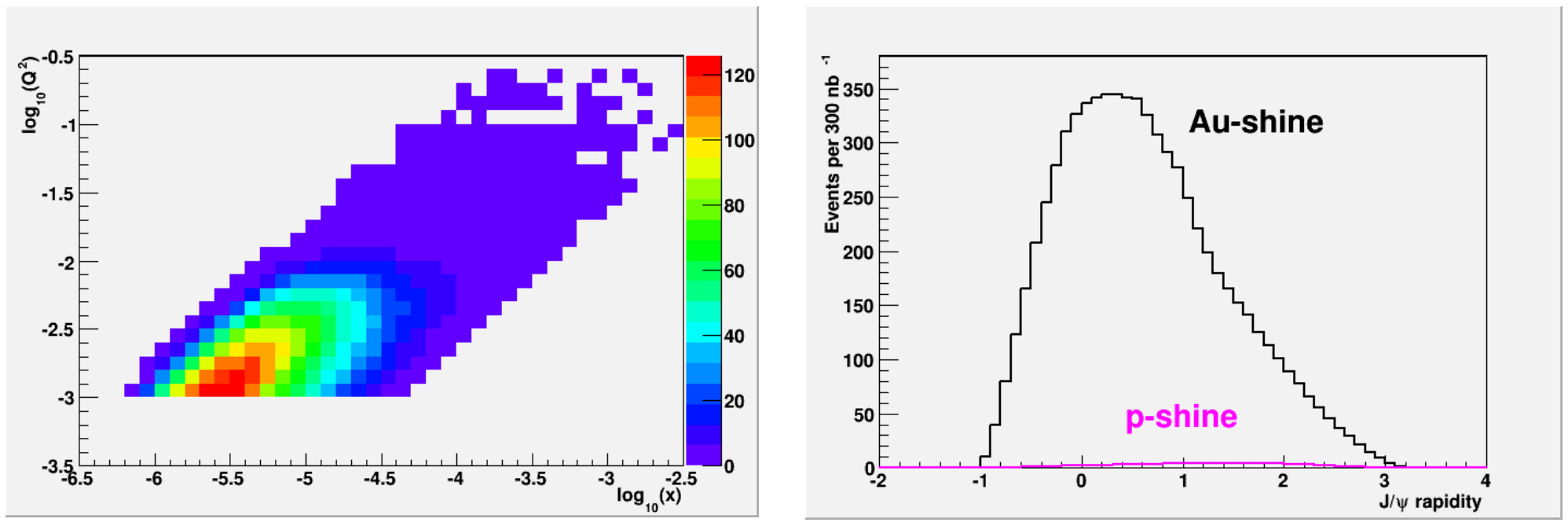

Figure 4-15: (left) The probed $x$-$Q^2$ distribution for Au-shine events. (right) The $J/\psi$ rapidity distribution for both events types after applying all cuts.

# 5 UNIQUE FUTURE OPPORTUNITIES WITH POLARIZED PROTONS AT RHIC

## 5.1 DETECTOR UPGRADES

In response to a charge from the BNL Associate Lab Director Berndt Mueller, the STAR and PHENIX Collaborations documented their plans for future $p$+$p$ and $p$+$A$ running at RHIC in 2021 and beyond [83, 56]. The time period covered by the charge coincides with scheduled $p$+$p$ and $p$+$A$ running at $\sqrt{s}$ = 200 GeV as part of the proposed sPHENIX [84] run plan and assumes modest increases in RHIC performance [85]. A summary of the detailed white papers from PHENIX and STAR is given in the following sections. While the white papers are consistent with the current plan for RHIC running in the next decade, both collaborations also present opportunities that could be exploited with additional polarized proton running at 510 GeV.

### 5.1.1 The fsPHENIX Upgrade

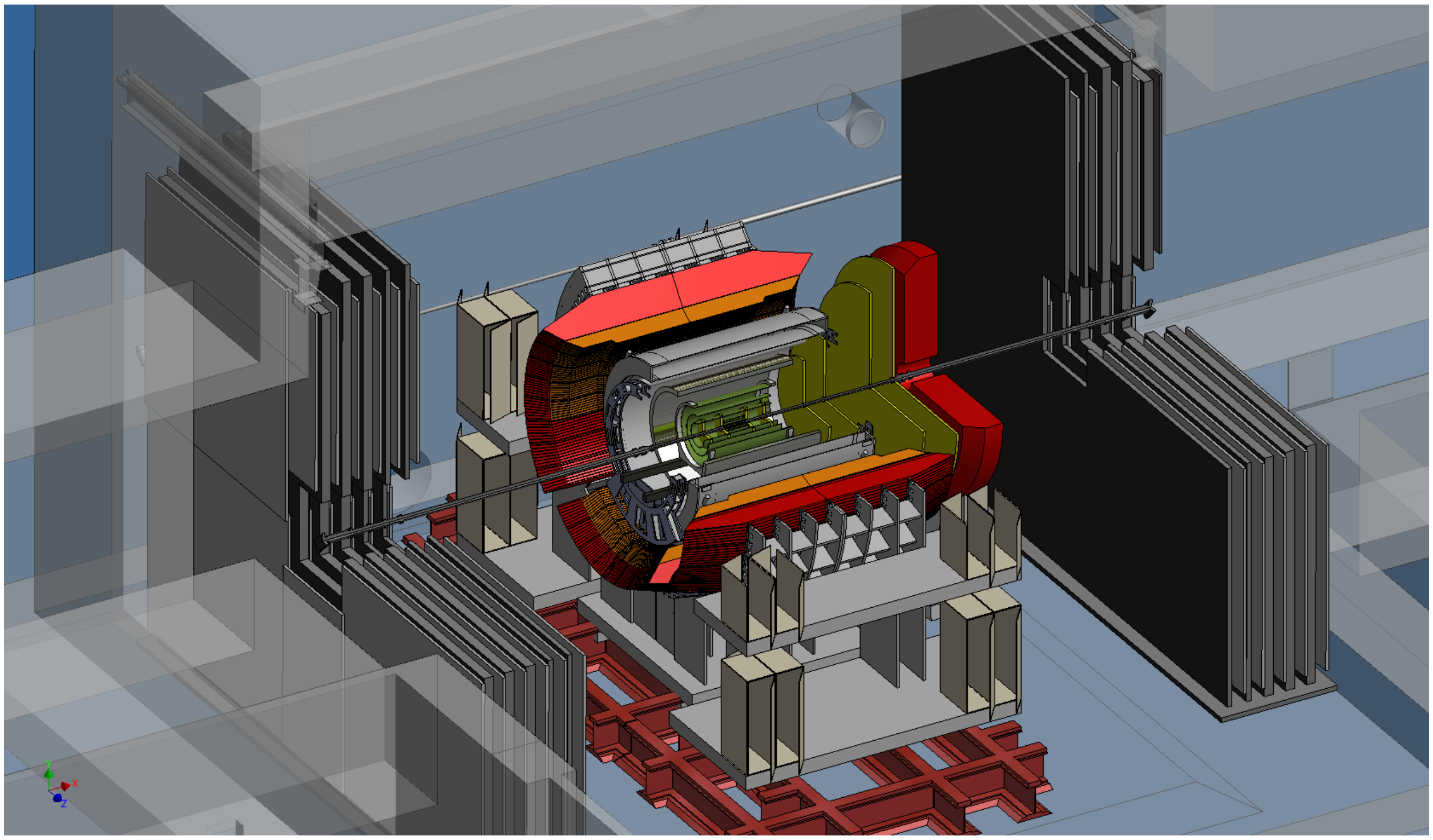

Figure 5-1: Engineering rendering of the sPHENIX (central barrel) and fsPHENIX (forward spectrometer) apparatus. The magnet flux return at both ends of the sPHENIX barrel is not shown in this rendering. See text for details of the fsPHENIX spectrometer arm.

The PHENIX collaboration has previously presented plans for the evolution of PHENIX to a next-generation heavy-ion detector [86], and ultimately to an EIC detector [87]. The time period covered by the charge from the BNL ALD  is intermediate to these existing proposals and would coincide with scheduled $p$+$p$ and $p$+$A$ running at $\sqrt{s}$ = 200 GeV as part of the proposed sPHENIX run plan.

This fsPHENIX ("forward sPHENIX") physics program focuses on physics observables at forward angles with additional detectors augmenting the sPHENIX upgrade. With the possibility that in the transition to the EIC the RHIC collider may not maintain the ability to provide hadron collisions, opportunities for significant new discoveries would be lost if the existing investment in RHIC is not fully exploited with measurements in the forward region. The fsPHENIX physics program centers on a comprehensive set of measurements of jet production in transversely spin-polarized *p+p* and *p+A* collisions, exploiting the unique capability of the RHIC collider to provide beams of protons with high polarization in addition to a variety of unpolarized nuclear beams. In *p+p* collisions we propose a program of intra- and inter-jet measurements designed to separate and elucidate the sources of the large single spin asymmetries measured previously at RHIC for single hadrons. Similar measurements in spin-polarized *p+A* collisions make use of transverse single-spin asymmetries as a probe of the parton densities in nuclei at small-*x*. This forward jet physics program can be done with the running already requested for sPHENIX in 2021-2022, and in fact enhances the sPHENIX heavy ion jet program with increased dijet coverage.

In addition, the planned sPHENIX running in *p+p* and *p+A* collisions at at $\sqrt{s} = 200$ GeV will also yield a measurement of the single spin asymmetry for Drell-Yan virtual photons (2.0<M<2.5 GeV, via the decay to two muons) complementary to the measurement will be made earlier by COMPASS-II. To full realize the physics potential in Drell-Yan, a dedicated running period at 510 GeV will be required in order to yield new, high-statistics Drell-Yan measurements with extended reach into the valence region of the polarized nucleon. By measuring both low-mass and high-mass Drell-Yan muon pairs this dataset will be able to explore and constrain the evolution of TMD functions in the polarized nucleon.

In order to optimize the use of available resources, the fsPHENIX detector, as shown in Figure 5-1, is developed around the proposed sPHENIX central detector and the re-use of existing PHENIX detector systems (such as the muon identifier and the forward vertex detector), as well as elements of a future EIC detector forward hadron arm. The majority of the cost of the fsPHENIX detector, estimated to be $12M including overhead and contingency, but not labor, can be viewed as a down payment on an EIC detector that will be needed during the EIC era. About 90% of the cost of the fsPHENIX detector is shared with an EIC detector. In addition to providing an important set of new physics measurements that may not be possible in the EIC era, an early investment in fsPHENIX would help provide day-1 readiness of the EIC detector.

The conceptual design of the fsPHENIX appratus is as follows. A magnetic field for particle tracking and charge identification is provided by shaping the sPHENIX superconducting solenoid field with a steel piston located around the beam line and an iron return yoke designed as part of the hadronic calorimeter. High resolution tracking near the interaction point is obtained from a reconfigured version of the existing FVTX detector. Three new GEM stations provide intermediate tracking and excellent momentum determination for charged particles.

A hadronic calorimeter measures total jet energy, position and size. The GEM tracker and hadronic calorimeter are identical in design to those proposed for the EIC detector forward hadron arm.

The existing PHENIX North Muon Identifier (MuID) system, covering 1.2<η<2.4, is used for muon identification. We propose to extend the pseudorapidity region for muon identification from η=2.4 to η=4, by building a "miniMuID" with a design similar to the existing PHENIX MuID.

The new detector subsystems to be built as part of fsPHENIX and a future EIC detector include:

- A forward hadronic calorimeter
- Three stations of GEM tracking chambers
- The "miniMuID", for muon identification at high pseudorapidity
- A magnetic field shaper for the forward region

The baseline fsPHENIX design could be upgraded with a RICH detector (the one designed for the EIC detector built around sPHENIX perfectly fits the outlined physics goals) to provide charged kaon and pion identification. Such and addition would allow exciting new physics measurements, such as the Collins asymmetry for identified particles in jets.

## *Jet Asymmetries*

The measurement of the inclusive jet single spin asymmetry accesses the correlation between the spin of the proton and transverse motion of its partonic constituents. An open, and very interesting question is whether the partial cancellation of the contribution of up and down quarks is responsible for the small size of the inclusive jet asymmetries measured by $A_N$DY.

By selecting jets with positively charged hadrons or negatively charged hadrons with jet fractional energies $z$>0.5 one can bias the jet's partonic process fractions towards larger up or down quark fractions. Based on the parton information from PYTHIA simulations, one can find that this selection improves the u quark purity at large $z$ to ~80% and the d quark purity to ~30% of all jets, see Figure 5-2. Under the positive charge sign requirement, the up quark contribution reaches high purities easily across all of the inspected $x_F$ values. The down quark contribution is enhanced over the natural abundance by nearly a factor of 3 in the negatively charged leading hadron sample, but does not rise above the up quark fraction for two reasons: (1) the smaller initial fraction provided by nature and (2) a weaker response to the charge sign requirement due to the down quark's smaller 1/3 electrical charge.

Jets in the fsPHENIX acceptance will also arise from other sources, namely gluons but also a smaller portion from proton beam remnants in the forward-most parts of the detector. These other sources are small at large $x_F$, contributing below ~20% to the jet sample. At smaller $x_F$ the contribution from gluons rises as should be expected from changes in the parton distribution function. Beam remnants have been simulated using a PYTHIA $k_T$=0.36 GeV/c setting appropriate for fragmentation scale processes and matching BRAHMS proton/pion ratios at forward rapidity. Other contributions were calculated using hard scattering settings within Tune A.

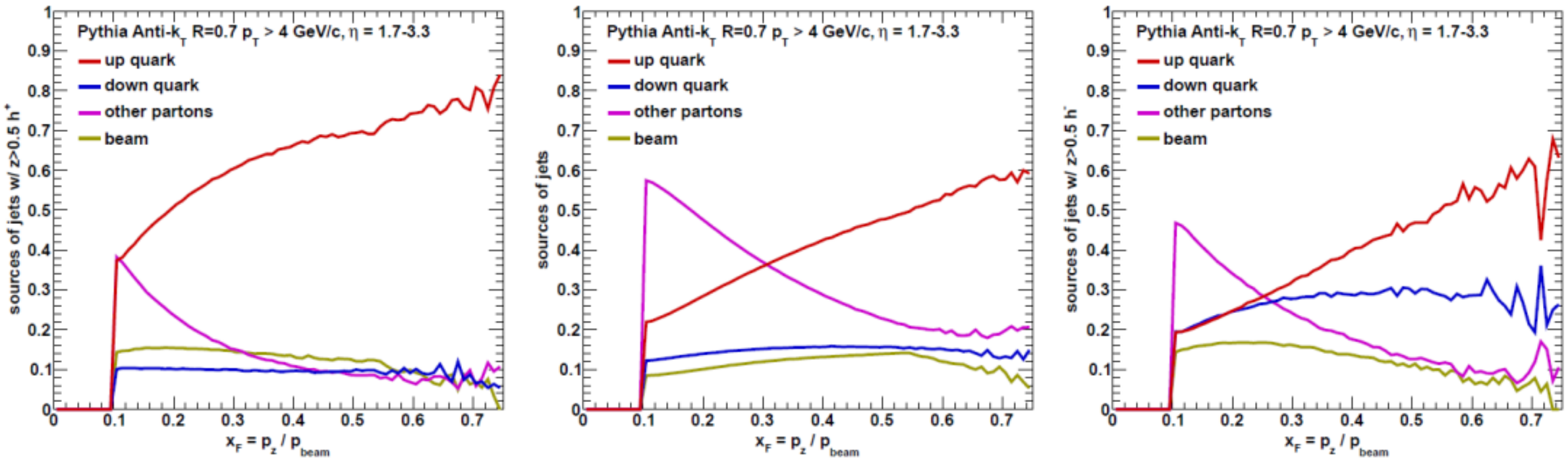

Figure 5-2: Up (red), down (blue) quark and other purities (magenta and brown) for jets in the rapidity range 1.7-3.3 and transverse momenta above 4 GeV as a function of the reconstructed jet $x_F$ when requiring a $h^+$ (left panel) with z > 0.5, no hadron selection (middle panel), or $h^-$ (right panel) with z > 0.5.

Using the leading hadron sign discrimination we can we can benchmark the ability of fsPHENIX to extract extract the up and down quark spin asymmetries using using a simple illustrative model comprised of the jet jet purities shown in Figure 5-2, the efficiencies of our our jet charge sign cuts, and two models proposed for for the up and down quark canceling needed to reproduce the inclusive jet spin asymmetries measured measured by $A_N$DY with the jet statistics collected in a in a baseline year of $p$+$p$ collisions with an integrated luminosity of 97 $pb^{-1}$. The theoretical inputs are taken from the error bands and linearly extrapolated to cover the full $x_F$ range. The $A_N$ values for up and down quarks are applied for the jet fractions within the positive charge selected, the natural abundance, and the negative charge selected sample separately. Other sources of jets were taken to contribute no asymmetry. The beam polarization was taken to be 60%. The resulting raw measurement projections are shown in the left column of

Figure 5-3. Statistical uncertainties are applied to these projections after accounting is made for the efficiencies of the leading charge cuts, determined using a GEANT4 simulation. The uncertainties are larger for the negatively charged hadron cuts where the efficiencies are lower. The inclusive jet sample is corrected for the beam polarization and appears in the final form in the right column. A simple two parameter fit is applied to the three raw measurements with our model of the jet purities to extract the projected statistical uncertainties on the theoretical inputs, also shown in the right column. The precision of the up and down quark $A_N$ are similar after extraction as all three inputs from the left column were used.

The results of this illustrative example demonstrate how the sign of the $A_N$ cancellation in the inclusive jet sample will be immediately apparent already in the raw measurements. Our final ability to distinguish between the up and down quark contributions to the inclusive $A_N$ will depend on how large that cancellation is in nature. The two theoretical descriptions have different scales for this cancellation, but both are readily distinguished by the expected statistical precision of the jet sample to be collected. In both cases, the statistical precision will improve on the theoretical model uncertainty from fitting other world data.

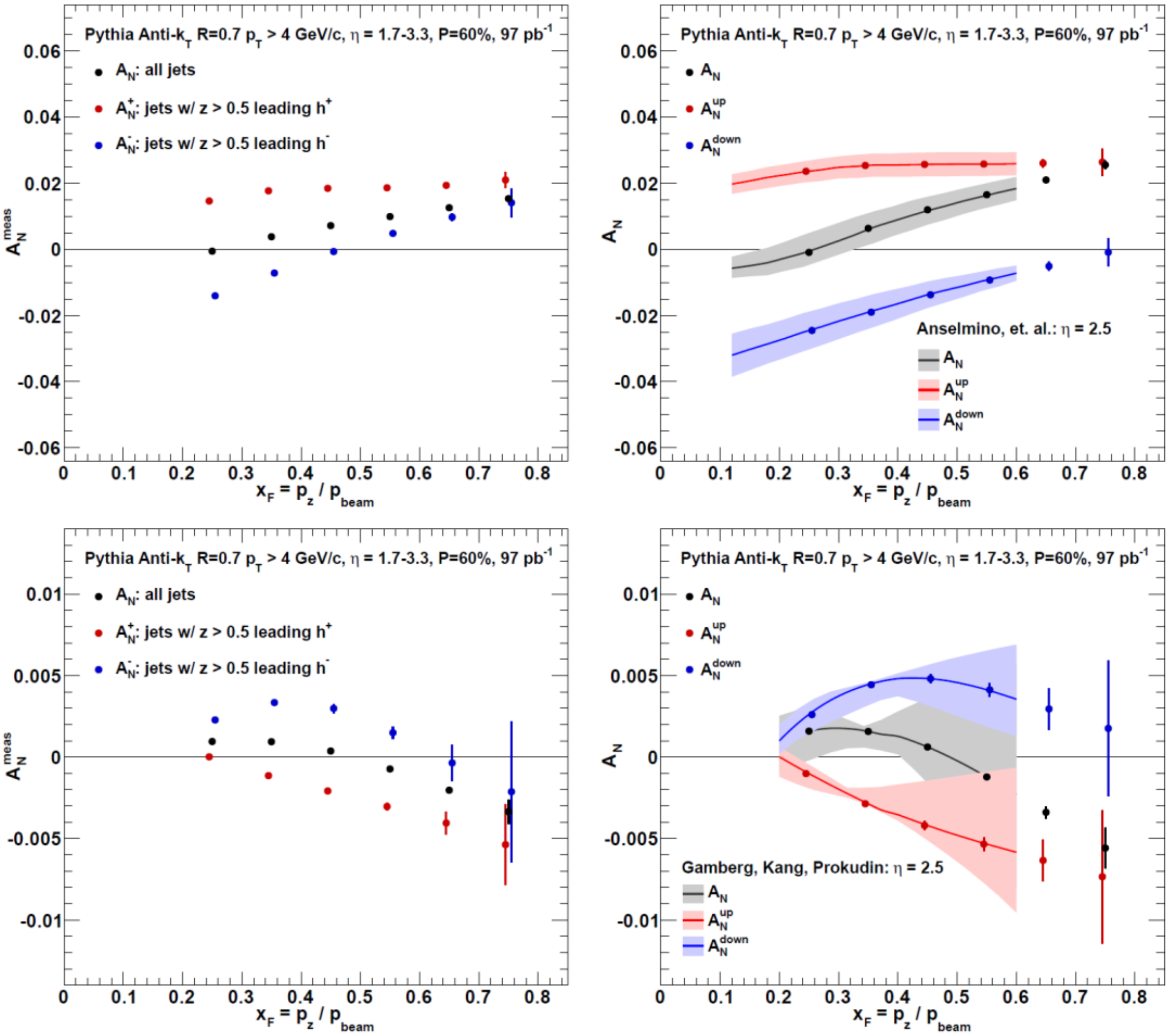

Figure 5-3: Projected statistical precision for jet $A_N$ measurements (left column) and extracted theory constraints (right column) for theoretical inputs from Anselmino, et. al. (top row) and Gamberg, Kang, & Prokudin (bottom row) using jets with $p_T > 4$ GeV/c in the pseudorapidity range 1.7-3.3. Shaded bands depict existing theoretical uncertainties assuming a fit to world data that involves no spin-dependent fragmentation. Error bars show the expected statistical uncertainties from 97 $pb^{-1}$ of *p+p* at 200 GeV.

The polarized Drell-Yan measurement with fsPHENIX is a competitive program with fixed-target experiments that will run prior to fsPHENIX. The goal of fsPHENIX is to investigate the Sivers transverse-spin asymmetry in the Drell-Yan process in order to compare it with that from the DIS process and to test the "modified universality" of the Sivers TMD. This has become one of the top priorities for the world-wide hadronic physics community. fsPHENIX will provide complementary information to that which will be obtained by fixed-target experiments by opening a new kinematic regime. A comparison of the asymmetries at fixed-target energies and collider energies also provides a unique test of TMD evolution.

Table 5-1 shows a comparison of the polarized Drell-Yan measurement in COMPASS-II and fsPHENIX. Information on COMPASS-II is obtained from the COMPASS-II proposal [88]. For fsPHENIX we use information from a GEANT4 Monte Carlo of the detector, as well as a 10 week RHIC run at 200 GeV and a 15 week RHIC run at 510 GeV. What may seem surprising at first is that with limited running the significance of the fsPHENIX Figure of Merit (FoM) results for high-mass pairs at 200 GeV is comparable to COMPASS-II. This is in large part due to the use of a polarized $NH_3$ target in COMPASS-II, which has a significant depolarization factor ($f$=0.22), while the RHIC beam polarization is relatively high ($P$=0.6). At 510 GeV the statistical power of fsPHENIX is dramatically improved due to the higher cross section and longer running time.

| | COMPASS-II | fsPHENIX 200 GeV | fsPHENIX 510 GeV |
|---|---|---|---|
| $L_{avg}$(cm$^{-2}$s$^{-1}$) | $1.18\times10^{32}$ | $0.76\times10^{32}$ | $6.48\times10^{32}$ |
| Average $L$ /week | 14.3 pb$^{-1}$/week | 18.7 pb$^{-1}$/week | 128 pb$^{-1}$/week |
| Accelerator eff. | 0.8 | (included above) | (included above) |
| Detector up-time | 0.85 | 0.6 | 0.6 |
| Vertex cut | n/a | 0.62 | 0.62 |
| Sampled $L$ /week | 9.7 pb$^{-1}$/week | 6.9 pb$^{-1}$/week | 47.6 pb$^{-1}$/week |
| week/year | 20 | 10 | 15 |
| Sampled $L$ /year | 194 pb$^{-1}$/year | 69 pb$^{-1}$/year | 714 pb$^{-1}$/year |
| Dimuon trigger eff. | 0.81 | 0.81 | 0.81 |
| **High mass: 4 GeV/$c^2$ < $M$ < 9 GeV/$c^2$** | | | |
| Reconstruction eff. | 0.8 | 0.312 | 0.305 |
| Offline $L$ /year | 126 pb$^{-1}$/year | 17.5 pb$^{-1}$/year | 177 pb$^{-1}$/year |
| Cross section $\sigma$ | 1291 pb | 1199 pb | 2542 pb |
| Acceptance $\Omega$ | 0.35 | 0.14 | 0.19 |
| $\sigma\cdot\Omega$ | 452 pb | 171 pb | 478 pb |
| $K$ factor (assumption) | 2 | 1.38 | 1.38 |
| Dimuon/year $L\cdot\sigma\cdot\Omega\cdot K$ | 115000/year | 4150/year | 117000/year |
| $FoM$/year | 2230/year | 747/year | 14600/year |
| $\delta A_T^{\sin\phi_S} = 1/\sqrt{FoM}$ | 0.021 | 0.037 | 0.0083 |
| **Low mass: 2 GeV/$c^2$ < $M$ < 2.5 GeV/$c^2$** | | | |
| Reconstruction eff. | 0.8 | 0.285 | 0.272 |
| Offline $L$ /year | 126 pb$^{-1}$/year | 16.0 pb$^{-1}$/year | 157 pb$^{-1}$/year |
| Cross section $\sigma$ | 6231 pb | 2811 pb | 4630pb |
| Acceptance $\Omega$ | 0.43 | 0.22 | 0.21 |
| $\sigma\cdot\Omega$ | 2679 pb | 610 pb | 955 pb |
| $K$ factor (assumption) | 2 | 1.38 | 1.38 |
| Dimuon/year $L\cdot\sigma\cdot\Omega\cdot K$ | 674000/year | 13500/year | 207000/year |
| $FoM$/year | 13200/year | 2430/year | 25900/year |
| $\delta A_T^{\sin\phi_S} = 1/\sqrt{FoM}$ | 0.0087 | 0.020 | 0.0062 |

Table 5-1: Comparison of Drell-Yan measurements by COMPASS-II and fsPHENIX (1.2 < η < 4), where FoM is $N/2\cdot(f|S_T|)^2$ (COMPASS-II, f=0.22, |S_T|=0.9) or $N/2\cdot P^2$ (fsPHENIX, P=0.6 at √s=200 GeV and P=0.5 at √s=510 GeV).

In order to compare the kinematic coverage of COMPASS-II and fsPHENIX, we performed a simple PYTHIA simulation of the Drell-Yan signal with simple geometrical acceptance and without magnetic field. For COMPASS-II, we assumed a detector acceptance determined by the main spectrometer magnet, and a beam momentum of 190 GeV/c. For fsPHENIX, we applied a pseudorapidity cut $1.2 < \eta < 4$, and studied collision energies of 200 GeV and 510 GeV. Figure 5-4 shows the FoM as a function of the partonic fraction $x$ probed in the polarized proton. This clearly shows that the fsPHENIX measurements has the ability to probe to higher in the proton, where extractions of the Sivers function from SIDIS data indicate the Sivers function is larger. A combination of the COMPASS-II and fsPHENIX Drell-Yan measurements will allow the extraction of the Sivers function over a wide range in $x$, while a comparison of low-mass and high-mass pair asymmetries will permit a study of the evolution of the asymmetries in $Q^2$.

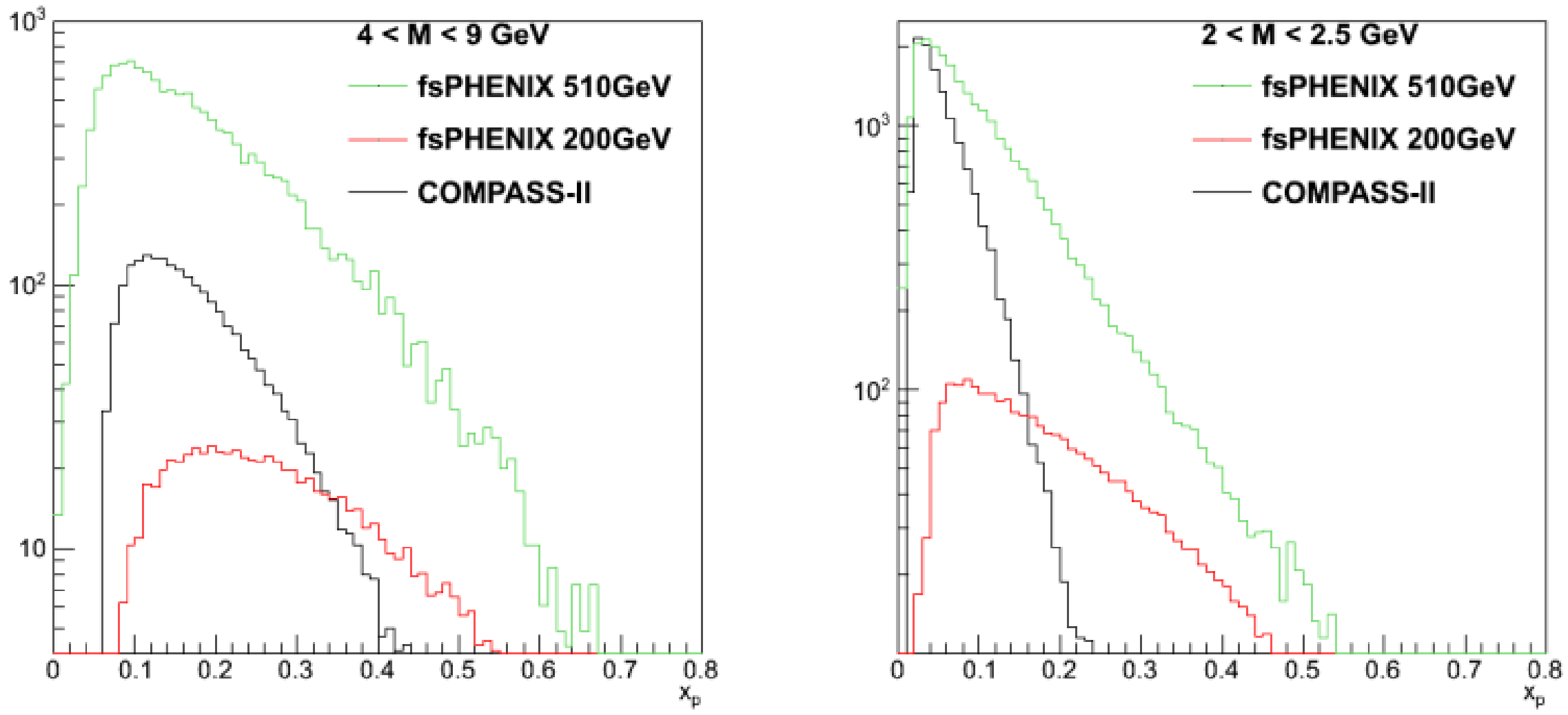

Figure 5-4: Left: A comparison of the FoM of COMPASS-II and fsPHENIX at high mass 4 GeV/$c^2$ < M <9 GeV/$c^2$. The black line shows FoM of COMPASS-II, the red line shows FoM of fsPHENIX at $\sqrt{s}$=200 GeV and the green line shows FoM of fsPHENIX at $\sqrt{s}$=510 GeV. Right: A similar comparison of FoM of COMPASS-II and fsPHENIX at low mass 2 GeV/$c^2$ < M < 2.5 GeV/$c^2$.

## 5.1.2 STAR Upgrades

### *Forward Calorimeter System*

The STAR forward upgrade is mainly driven by the desire to explore QCD physics in the very high or low region of Bjorken *x*. Previous STAR efforts using the FPD and FMS detectors, in particular the refurbished FMS with pre-shower detector upgrade in Runs 2015-2016, have demonstrated that there are outstanding QCD physics opportunities in the forward region as we have outlined in the previous sections. In order to go much beyond what STAR would achieve with the improved FMS detector, STAR proposes a forward detector upgrade with superior detection capability for neutral pions, photons, electrons, jets and leading hadrons covering a pseudorapidity region of 2.5-4.5.

At the core of the forward detector upgrade is the Forward Calorimeter System (FCS). The design of the FCS is driven mainly by detector performance, the integration into the STAR system and the cost optimization. Whenever possible we also minimize the number of mechanical components and the construction and operation resources needed from the collaboration in order to carry out the forward upgrade project under the expected constraints for budget and other resources.

The FCS consists of a Spaghetti ElectroMagnetic Calorimeter (SPACal) followed by a Lead and Scintillating Plate sampling Hadronic Calorimeter (HCal). The SPACal is made of Tungsten powder and scintillating fibers as such it has achieved one of the highest densities and among the most compact calorimeters. The Moliere radius of the SPACal is about 2.3 cm and we have chosen the size of each module to be 2.5×2.5×17 cm$^3$ corresponding to 23 $X_0$ in length. The HCal is made of Lead and Scintillator tiles with a tower size of 10×10×81 cm$^3$ corresponding to 4 interaction length. Our goal is to have a fully compensated calorimeter system. The proposed FCS has 120×80 SPACal towers and 30×20 HCal towers covering an area of 3×2 m$^2$. We are currently investigating the possibility to locate the FCS closer to the interaction vertex reducing the number of towers necessary for the same pseudorapidity coverage.

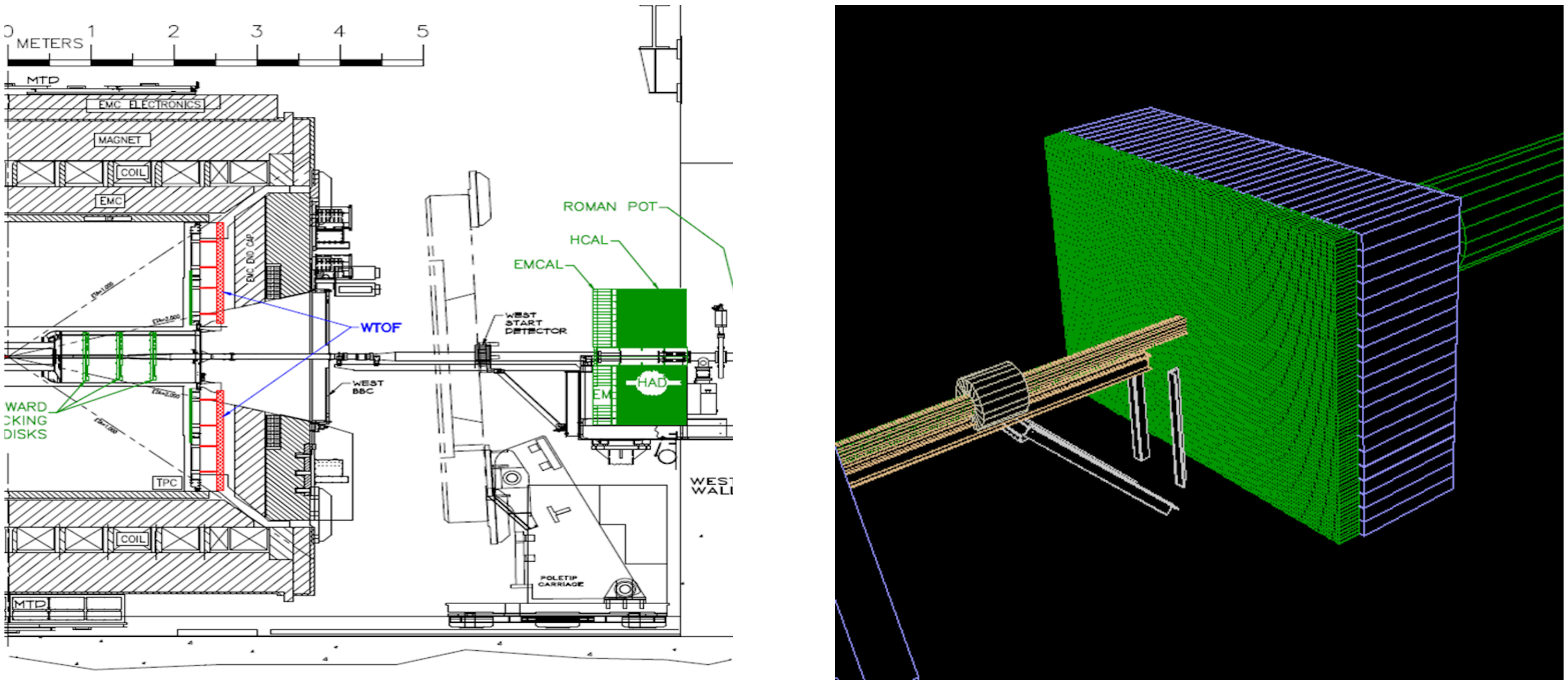

Figure 5-5: Location of the FCS at the West side of the STAR Detector system and a GEANT model of the FCS in the STAR simulation software

Figure 5-5 shows the location of the proposed FCS at the West side of the STAR detector system and a schematic description of the FCS in the STAR Monte Carlo simulation software. The read-out for the SPACal will be placed in the front so that there will be no significant dead gaps between the SPACal and the HCal. Wavelength shifting slats are used to collect light from the HCAL scintillating plates to be detected by photon sensors at the end of the HCal. Multiple Silicon PMTs will be used to read out each SPACal and HCal module, 4 for SPACal and 8 for HCal, respectively.

The SPACal construction technique using Tungsten powders and scintillating fibers and the compact read-out scheme using SiPMTs have been the subject of a detector R&D project by a team of STAR groups from UCLA, TAMU, PSU, IU and BNL under the support of the BNL/JLab EIC generic detector R&D program. In 2012 we constructed several prototype SPACal modules and carried out a beam test at FNAL to prove the validity of the basic concept for the Tungsten powder and scintillating fiber construction technique. In 2013-2014 with additional support from STAR for the calorimeter R&D, we refined the SPACal construction technology and built wedge-shaped SPACal modules for EIC barrel applications and normal SPACal modules for STAR FCS applictions. We also developed a novel construction technique for HCal by stacking Lead and Scintillator plate in-situ. Students and post-docs just before the test run constructed an array of 4×4 prototype HCal modules at the FNAL test beam site. We envision that a full HCal detector can be assembled at the STAR experimental hall within a few months during the summer shutdown period.

Figure 5-6 shows a newly constructed array of 4×4 HCal modules at the FNAL test beam facility. The right panel shows the energy resolutions for the FCS SPACal and HCal detectors as a function of the beam energy. SiPMTs were used for the read-out of both calorimeter detectors. The measured energy resolution as a function of beam energy is consistent with our Monte Carlo simulations for the detector performance. The measured values have been used in our physics simulations for the proposed forward detector upgrade. We have established a viable detector construction technology and proved the feasibility of SiPMT read-out scheme for the proposed STAR FCS. The performance of the prototype FCS testing at FNAL in March 2014 demonstrated that the proposed FCS detector will meet the STAR physics requirements.

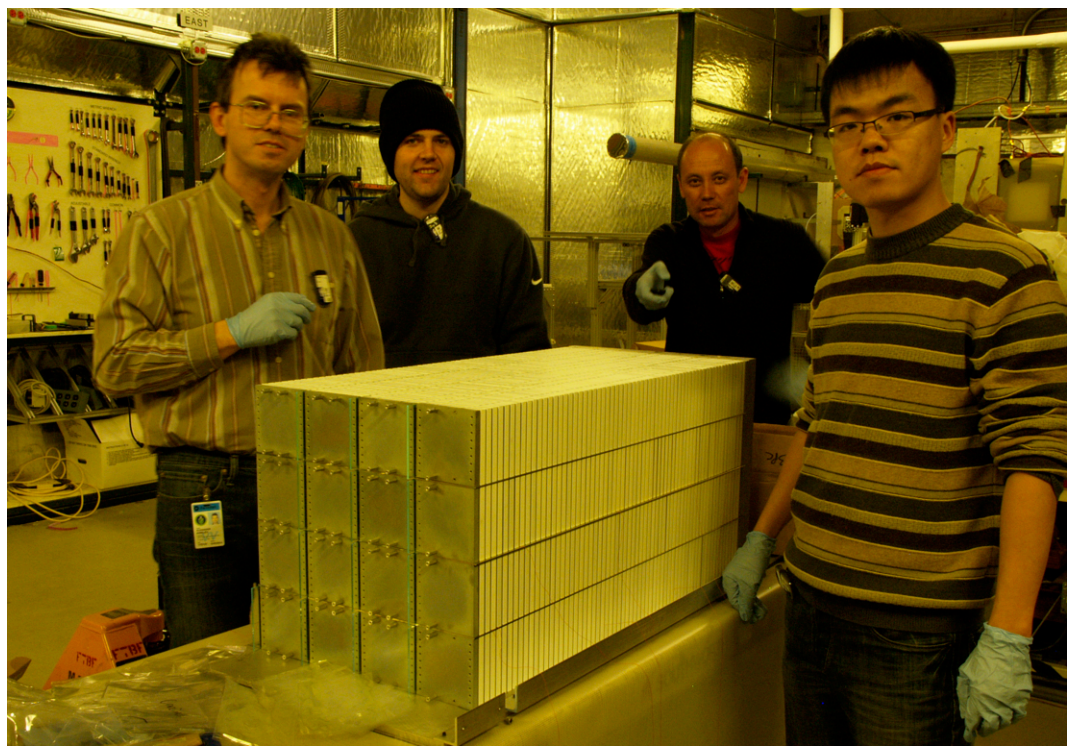

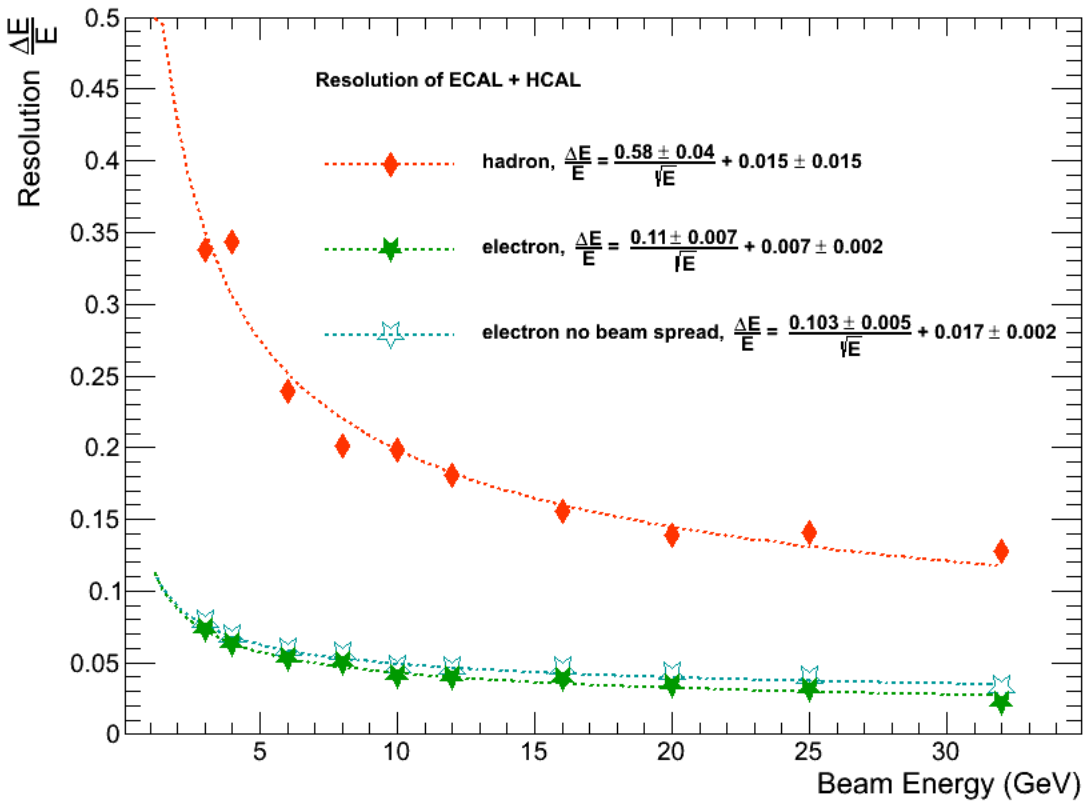

Figure 5-6: Prototype of HCAL calorimeter being assembled at FNAL for the test run. Energy resolution of the FCS for electrons and hadrons measured at test run at FNAL in 2014

## Forward Tracking System

In addition to the FCS, a Forward Tracking System (FTS) is also under consideration for the STAR forward upgrade project. Such a FTS has to cope with the STAR 0.5 T Solenoid magnet field to discriminate charge sign for transverse asymmetry studies and those of electrons and positrons for Drell-Yan measurements. It needs to find primary vertices for tracks and point them towards the calorimeters in order to suppress pile-up events in the anticipated high luminosity collisions, or to select particles from Lambda decays. It should also help with electron and photon identification by providing momentum and track veto information. In order to keep multiple scattering and photon conversion background under control, the material budget of the FTS has to be small. These requirements present a major challenge for detector design in terms of position resolution, fast readout, high efficiency and low material budget.

STAR has considered two possible detector technology choices: the Silicon detector technology and Gas Electron Multiplier (GEM) technology. STAR has gained considerable experience in both technologies from the FGT (Forward GEM Tracker) construction and the Intermediate Silicon Tracker (IST) construction in recent years. Several groups are pursuing GEM-based detector R&D under the auspices of generic EIC R&D program. Further evaluation of the GEM tracker option based on the STAR FGT experience will be needed if STAR will pursue such a technology.

Silicon detectors have been widely used in high-energy experiments for tracking in the forward direction. For example, Silicon strip detectors have been successfully used at many experiments: the Dzero experiment at the Tevatron, CMS and LHCb at the LHC, and PHENIX at the RHIC. More recent designs incorporate hybrid Silicon pixel detectors, which resulted in the improvement of position resolutions and removal of ghost hits, but unfortunately they also significantly increased the cost and material budget. According to preliminary Monte Carlo simulations, charge sign discrimination power and momentum resolution for the FTS in the STAR Solenoid magnet depends mostly on phi resolution, and is insensitive to the r-position resolution. Therefore a Silicon mini-strip detector design would be more appropriate than a pixel design. We are evaluating a design that consists of three to four disks at z locations at about 70 to 140 cm. Each disk has wedges covering the full 2π range in $\phi$ and 2.5-4 in $\eta$. The wedge will use Silicon mini-strip sensors read out from the larger radius of the sensors. Compared to the configuration of reading out from the edges along the radial direction, the material budget in the detector acceptance will be smaller since the frontend readout chips, power and signal buses and cooling lines can be placed outside of the detector acceptance.

STAR will continue to evaluate these technology options for the FTS design. More R&D efforts are needed to demonstrate the technical feasibility of these options through Monte Carlo simulations and detector prototyping. We hope to narrow the choice when we develop the technical proposal for the STAR forward upgrade in the coming a few years.

## Physics Opportunities with the STAR forward upgrade

### *Kinematics of inclusive forward jets in p+p with the proposed forward upgrade*

Both the measurement of the helicity structure and the transverse spin structure of the nucleon use reconstructed jets and di-jets to narrow the phase space of partonic kinematics. Since jets serve as proxies for the scattered partons, reconstructed jets allow the selection of events with a specific weighting of the fractional momenta of the parent protons carried by the scattering partons, assuming a 2-to-2 process. Here we call the fractional momentum carried by the parton coming from the beam along the *z*-axis (towards the proposed forward upgrade instrumentation) $x_1$, and the fractional momentum carried by the other parton $x_2$.

For measurements of $A_{LL}$ it is important to select events where one $x$ is determined with good accuracy within the kinematic region in which one wants to measure $\Delta g(x)$ and the other $x$ is in a region where the helicity distribution is well known, i.e. in the region of medium to high values of $x$. The transverse spin structure of the nucleon on the other hand is usually accessed using transverse single spin asymmetries and semi-inclusive measurements. This means one measures azimuthal asymmetries of the final state, where the distribution function of interest couples to a spin dependent FF that serves as a polarimeter. Consequently we studied how in single- and di-jet events, the jet pseudorapidity $\eta$ and $p_T$ are related to the underlying partonic variables $x_1$ and $x_2$. We also studied the matching between reconstructed jets and scattered partons and the resolutions with which the parton axis can be reconstructed from the reconstructed detector jets. The latter is important to evaluate how well azimuthal asymmetries around the outgoing parton axis will be reconstructed by looking at asymmetries of reconstructed particles around the reconstructed jet axis.

The simulations are based on PYTHIA Tune A at $\sqrt{s}$=500 GeV and a minimum partonic $p_T$ of 3 GeV. Fast detector simulations have been used to account for the resolutions of the STAR barrel and the forward upgrade detectors (for details see [56]). Jets are reconstructed with an anti-$k_T$ algorithm with a radius of 0.7. An association between reconstructed jets and scattered partons is defined to be a distance in $\eta$-$\phi$ space of less than 0.5. In the following, reconstructed jets are referred to as "detector jets" and jets found using stable, final state particles "particle jets."

Figure 5-7 (left) shows the regions of $x$ that can be accessed by jets in the forward region. A minimum jet $p_T$ of 3 GeV/*c* was chosen to ensure that the momentum transfer is sufficiently high for pQCD calculations to be valid. At high $x$, values of $x$~0.6 should be reachable. This compares well with the current limit of SIDIS measurements, $x$~0.3, and encompasses the region in $x$ that dominates the tensor charge. To investigate the possibility of selecting specific $x$ regions, in particular high $x$, the dependence of $x$ on the jet $p_T$ and pseudorapidity was studied. Figure 5-7 (right) shows $x_1$ as a function of jet $p_T$.

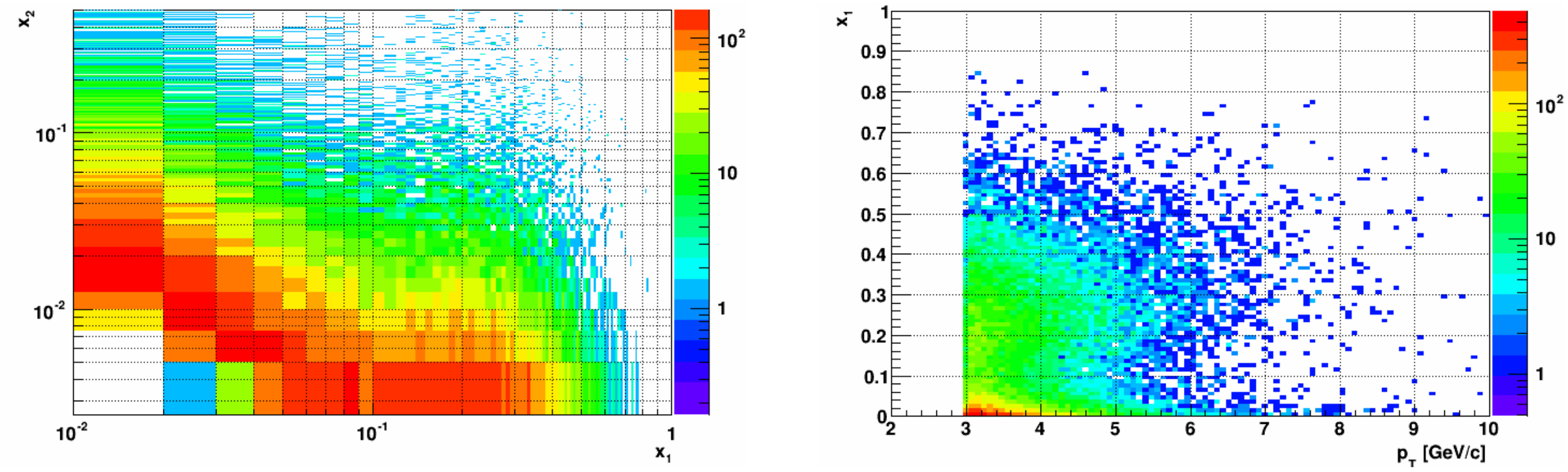

Figure 5-7: (left) Distribution of the partonic variables $x_1$ and $x_2$ for events with a jet with $p_T > 3$ GeV/$c$ and $2.8 < \eta < 3.5$. $x_1$ values of around 0.6 can be reached whereas $x_2$ goes as low as $7\times10^{-3}$.
(right) x1 versus jet $p_T$. As expected, there is a correlation between the x accessed and the pT of the jet. However, there is an underlying band of low x1 values. This can be improved by further restricting the η range of the jet. Here $2.8 < \eta < 3.8$.

For measurements of azimuthal asymmetries of jets or hadrons within a jet to probe the transverse spin structure of the nucleon it is important to reconstruct reliably the outgoing parton direction. Therefore, the matching of reconstructed jets to scattered partons was studied (Figure 5-8). In general, matching and parton axis smearing improves with $p_T$, which may be connected to the jet multiplicity that rises with transverse momentum.

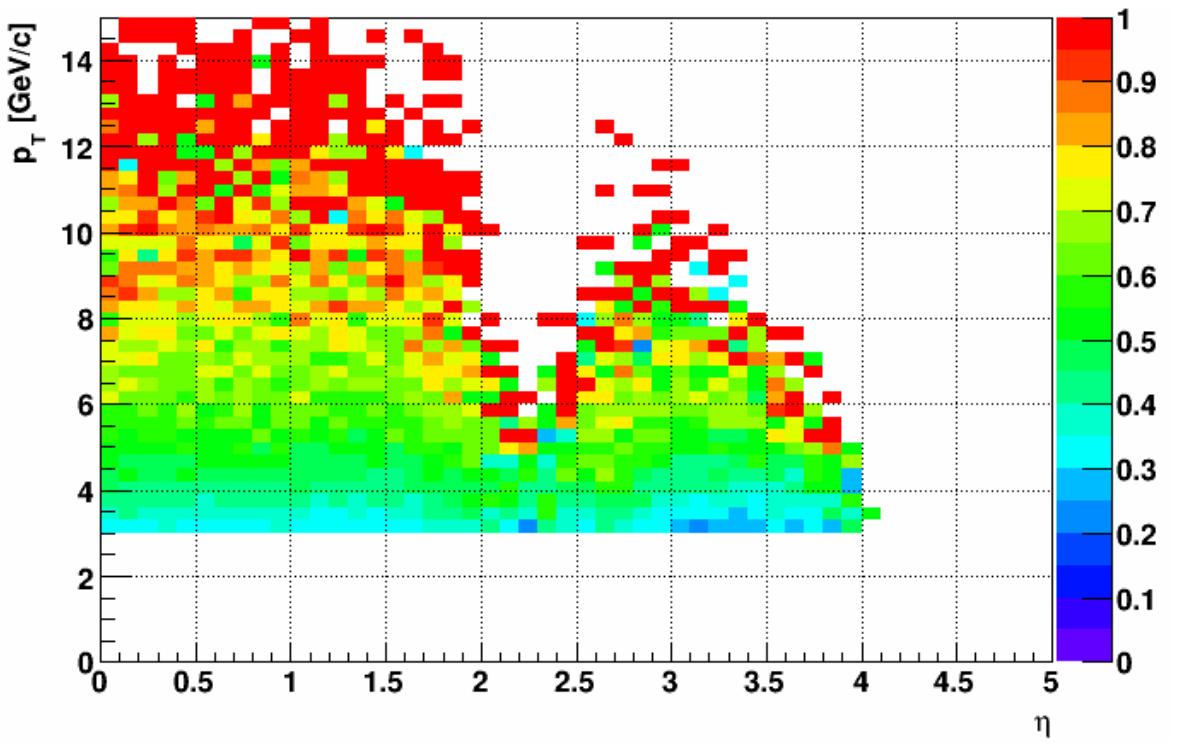

Figure 5-8: Matching Fraction between detector jets and partons. The matching fraction at low pT is only around 50%, but grows to over 90% for high pT. Unfortunately, the statistics at high pT in the forward region is small.

## *Confined Motion of Partons in Nucleons:*

The planned STAR upgrades for the second half of this decade include expansion of the TPC tracking capability by about one half of a unit of pseudorapidity as well as charged-particle tracking capability and hadronic calorimetry to the forward subsystems, spanning the range $2.8 < \eta < 3.7$. Tracking upgrades are critically necessary for Collins and di-hadron measurements that require robust charge-sign discrimination.

In Figure 5-9 we show the expected Collins asymmetries for $p^{\uparrow} + p \rightarrow jet + \pi^{\pm} + X$ at $2.8 < \eta < 3.7$ and $\sqrt{s} = 500$ GeV. Jets are required to have a minimum $p_T$ of 3 GeV/$c$. The 2008 transversity and Collins FF parameterization by the Torino group [35] has been inserted into a leading-order PYTHIA simulation using CDF Tune A. Jets are reconstructed utilizing an anti-$k_T$ algorithm, and the asymmetries are calculated relative to the associated hard-scattered parton. The projections assumed 1 fb$^{-1}$ of luminosity with 60% beam polarization. Asymmetries of nearly ~2% are expected for both flavors of pions. In Figure 5-10 we show a comparison of di-hadron asymmetries at the "detector" level, with the fast simulation detector smearing, to those at the "particle" level, before simulated detector smearing. Based on the simulation, the effects of kinematic smearing to the asymmetries are expected to be quite small. This suggests that within the same subsystem, one can simultaneously measure in a robust fashion the Collins asymmetry within the TMD framework and the di-hadron asymmetry within the collinear framework. These measurements are critical for extending current understanding of transversity and questions concerning TMD evolution, factorization breaking, and universality.

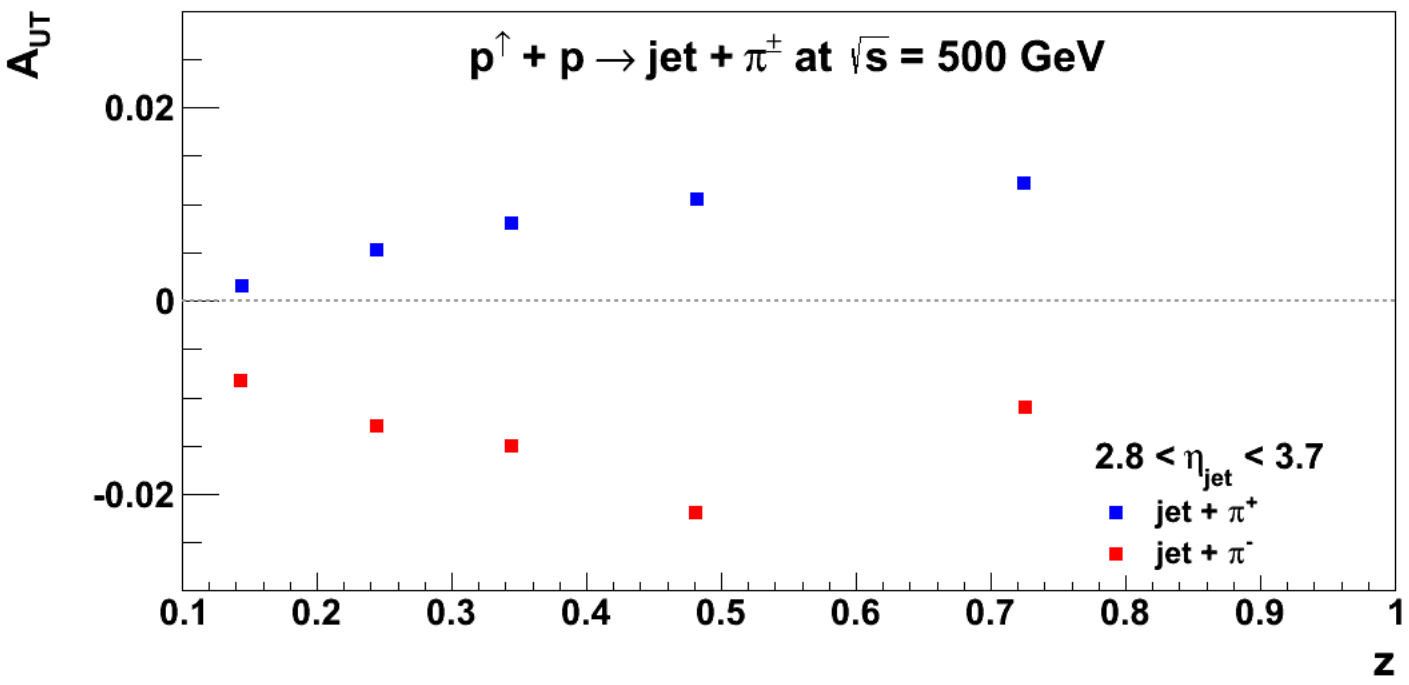

Figure 5-9: Expected Collins asymmetries assuming the Torino parameterization [35] within a leading-order PYTHIA Monte Carlo for charged pions within jets produced with $2.8 < \eta < 3.7$ and $p_T > 3$ GeV/$c$. The expectations assume 1fb$^{-1}$ of integrated luminosity, and statistical uncertainties are smaller than the size of the points. Jets are reconstructed utilizing an anti-$k_T$ algorithm, and the asymmetries are calculated relative to the axis of the hard scattered parton.

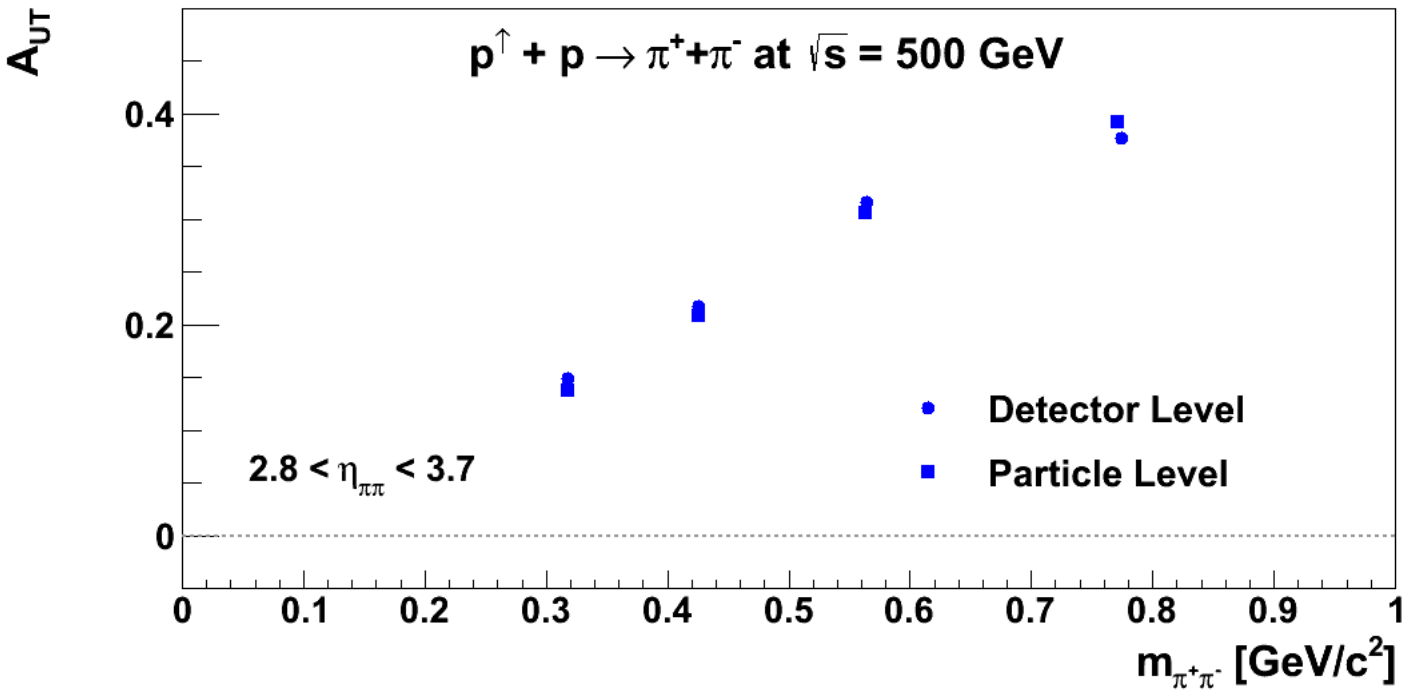

Figure 5-10: Comparison of IFF asymmetries at the "detector" level and at the "particle" level for charged pions produced within $2.8 < \eta < 3.7$. Asymmetries are shown as a function of di-hadron invariant mass, and assuming a parameterization inspired from FF measurements at Belle [34]. The projections assume 1fb$^{-1}$ of integrated luminosity, and statistical uncertainties are smaller than the size of the points.

In Figure 5-11 we show the expected Sivers asymmetries [89] for $p^{\uparrow} + p \rightarrow jet + X$ at $2.8 < \eta < 3.7$ and $\sqrt{s} = 500$ GeV. Jets are reconstructed in the same manner as discussed above for the Collins asymmetries, and the Torino parameterization is assumed for the Sivers function [35]. Since the inclusive jet asymmetry provides only a single hard scale, namely, jet $p_T$, the Twist-3 framework is most naturally suited for theoretical interpretation. However, the current estimates give a sense for the size of such effects. One can see that for 1 fb-1 statistics may be sufficient to observe a nonzero asymmetry. However, the effects are expected to be quite small, at an order less than 1%. The magnitude of this projection is qualitatively similar to existing inclusive jet asymmetries at forward pseudorapidity [52].

Recent theoretical work [53] has found that by taking into account initial-state and final-state interactions between the hard scattered parton and the polarized remnant, extractions of the Sivers function from SIDIS data [29,30] are consistent with existing inclusive jet data from $p$+$p$ scattering [52]. The extracted Sivers functions were used to derive the Twist-3 function $T_{q,F}(x,x)$ [90] that was then used to compute the corresponding inclusive jet asymmetry for $p$+$p$ scattering. The prediction compares favorable to the measured asymmetry, indicating a process-dependence to the Sivers effect. Due to the small size of the apparent inclusive jet asymmetries more precise measurements are needed.

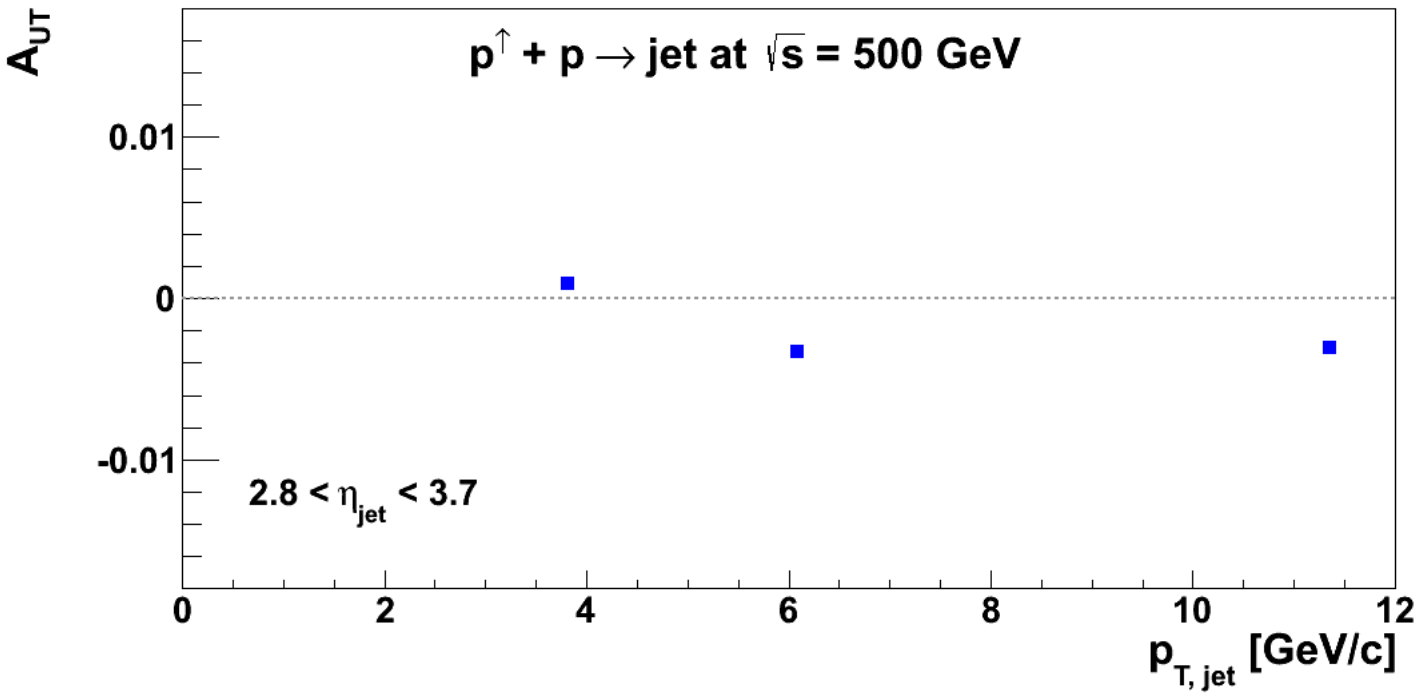

Figure 5-11: Expected Sivers asymmetries based on the Torino parameterization [35] within a leading-order PYTHIA Monte Carlo for jets produced with $2.8 < \eta < 3.7$ and $p_T > 3$ GeV/$c$. The expectations assume 1 fb$^{-1}$ of integrated luminosity, and statistical uncertainties are smaller than the size of the points. Jets are reconstructed utilizing an anti-$k_T$ algorithm, and the asymmetries are calculated relative to the axis of the hard scattered parton.

In addition to the inclusive jet measurements outlined above, di-jet measurement allow further probes of the transverse momentum dependent structure of the nucleon. Here the relative transverse momentum between the jets, $k_T$, and gives the additional soft scale needed for the TMD framework. In addition, accessing functions like Sivers [89] and Boer-Mulders [91] in $p$+$p$ collisions allows one to explore additional asymmetries that may result from the "color-entanglement" in $p$+$p$, which also leads to the breakdown of factorization theorems [92].

As already discussed in section 3.1 the measurements of beam-spin dependence in di-jet production will allow better constraints of the underlying event kinematics to constrain the shape of the gluon polarization. It has been shown that the current acceptance of the STAR experiment permits reconstruction of di-jet events with different topological configurations, i.e. different $\eta_3/\eta_4$ combinations. It is in particular the access to the large $\eta_{3\,/}\,\eta_4$ region which allows to probe gluons in QCD processes at very small $x$-values. The proposed new instrumentation for the forward region would cover a nominal range in $\eta$ of $2.5 < \eta < 4.0$ and therefore enhance asymmetric ($x_1 < x_2$ or $x_1 > x_2$) partonic collisions.

Figure 5-12 shoes the asymmetries $A_{LL}$ as a function of the invariant mass $M/\sqrt{s}$ for four topological di-jet configurations involving at least the forward system labeled as FCS in combination with either EAST, WEST, EEMC and FCS. It is in particular the EEMC / FCS and FCS / FCS configurations which would one allow to probe $x$ values as low as $10^{-3}$. The theory curves at NLO level have been evaluated for DSSV-2008 [3] and GRSV-STD [11]. The systematic uncertainty, which is assumed to be driven by the relative luminosity uncertainty of $\delta R = 5{\cdot}10^{-4}$, is clearly dominating over the size of the statistical uncertainties. Any future measurements in this topological configuration of very forward measurements would clearly benefit from improved relative luminosity measurements.

Figure 5-13 shows the $x$ coverage for four topological di-jet configurations. The proposed forward di-jet production measurements, shown in Figure 5-12 in combination with measurements of the current STAR acceptance region (see Figure 3-8) would allow to probe spin phenomena of gluons well below the current accessible region of $0.05 < x < 1.0$. Such measurements would provide critical initial insight into the nature of the proton spin. The proposed program offers unique and timely opportunities to advance the understanding of gluon polarization in the polarized proton, prior to a future Electron-Ion Collider that with sufficient kinematic reach will probe the $x$-dependence of the gluon polarization to well below $10^{-3}$ with high precision [93].

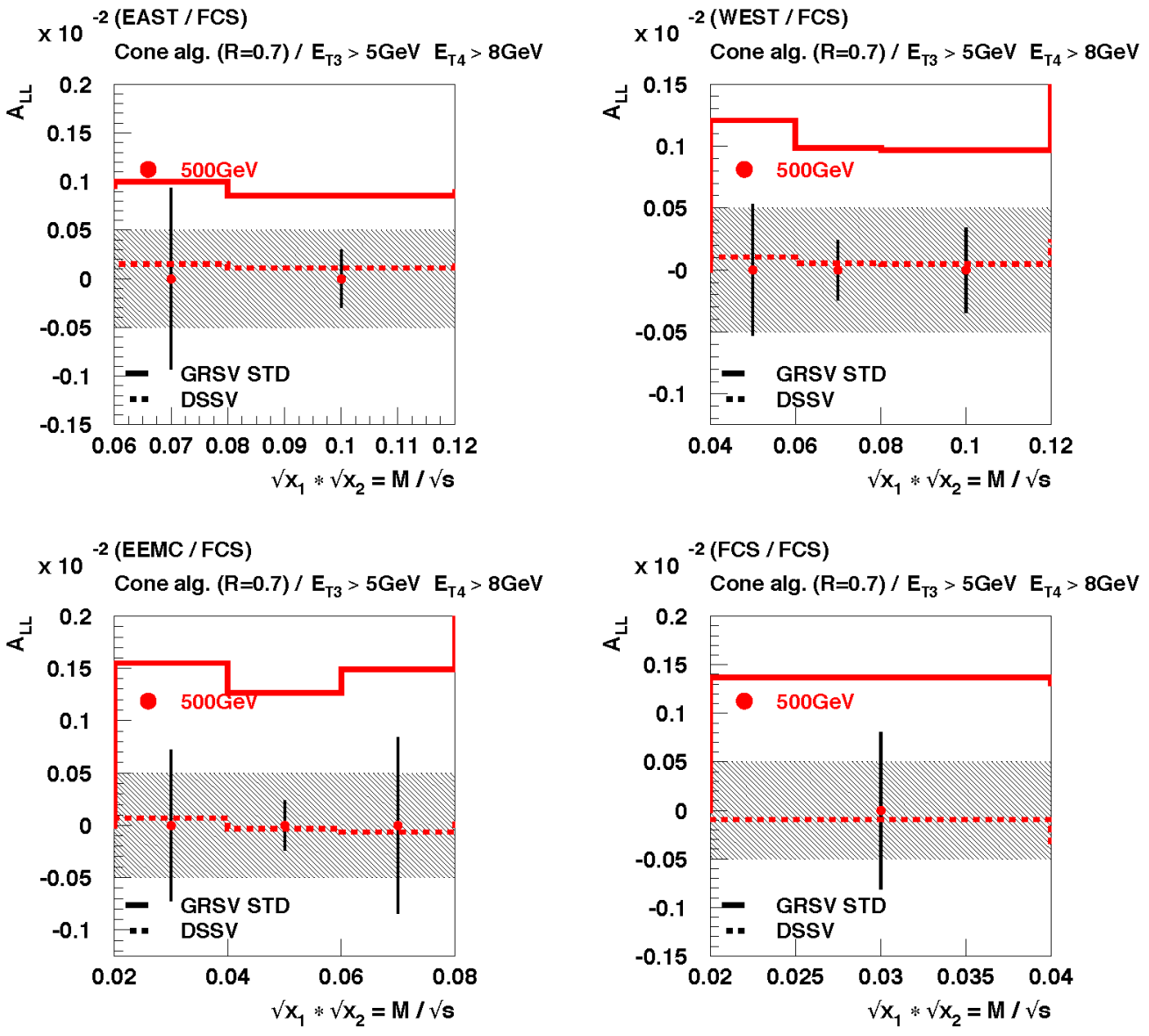

Figure 5-12: $A_{LL}$ NLO calculations as a function of $M/\sqrt{s}$ for $2.8 < \eta < 3.7$ together with projected statistical and systematic uncertainties. An uncertainty 5·10-*4* has been assumed for the systematic uncertainty due to relative luminosity. A beam polarization of 60% and a total delivered luminosity of $1\,\mathrm{fb}^{-1}$ have been assumed with a ratio of 2/3 for the ratio of recorded to delivered luminosity.

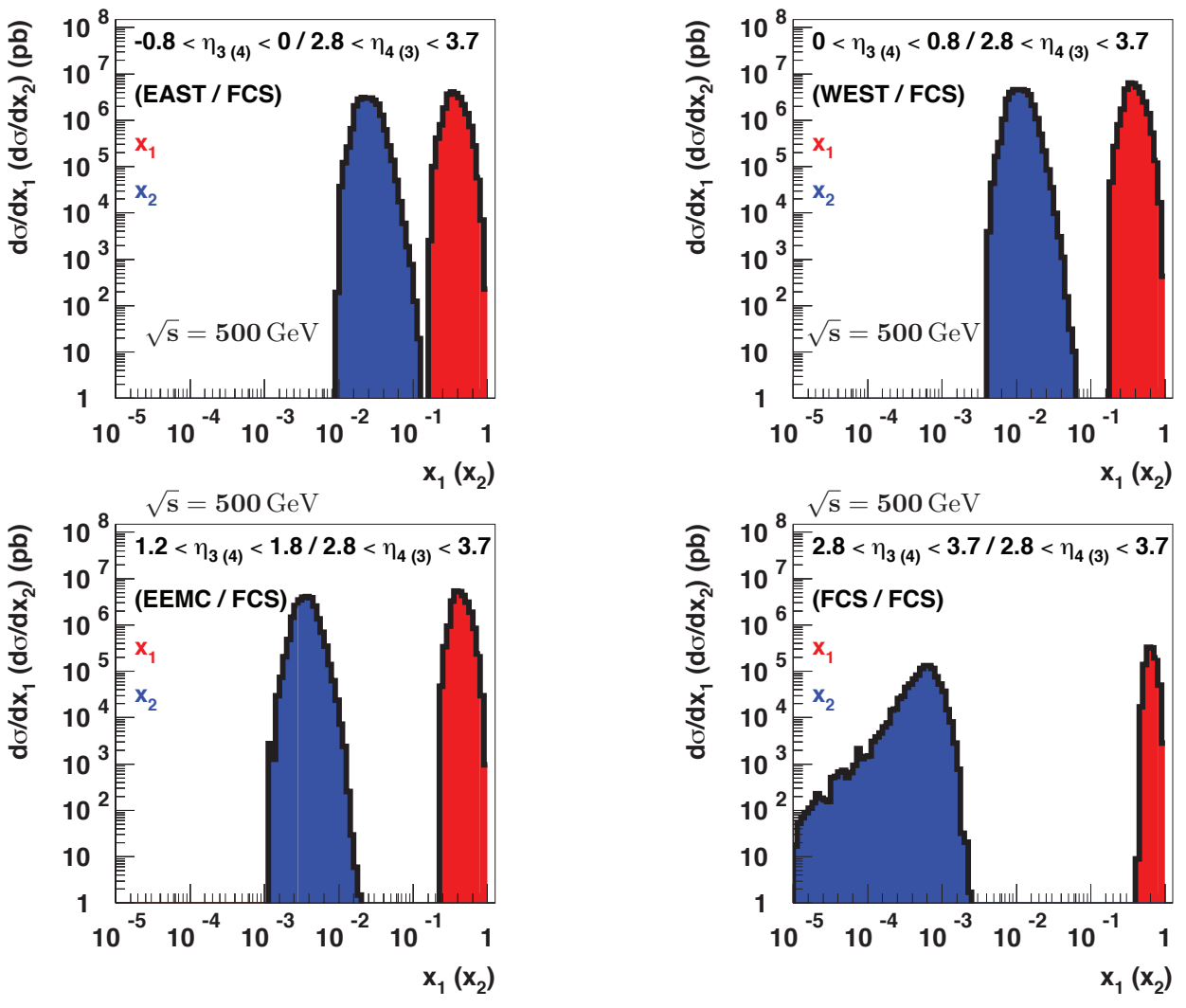

Figure 5-13: $x_1$ / $x_2$ range for the forward STAR acceptance region in $\eta$ of $2.8 < \eta < 3.7$.

# 6 SUMMARY

A myriad of new techniques and technologies made it possible to inaugurate the Relativistic Heavy Ion Collider at Brookhaven National Laboratory as the world's first high-energy polarized proton collider in December 2001. This unique environment provides opportunities to study the polarized quark and gluon spin structure of the proton and QCD dynamics at a high energy scale and is therefore complementary to existing semi-inclusive deep inelastic scattering experiments.

Therefore the polarized proton beam program at RHIC has and will continue to address several overarching questions:

- ***What is the nature of the spin of the proton?***

RHIC has in the last years completed very successful polarized $p$+$p$ runs both at $\sqrt{s}$ = 200 GeV and 500(510) GeV. The measurement of the gluon polarization in a longitudinally polarized proton has been a major emphasis. Data from the RHIC run in 2009 have **for the first time shown a nonzero contribution of the gluons to the proton spin**. The integral of $\Delta g(x,Q^2=10\ \mathrm{GeV}^2)$ in the region $x > 0.05$ is $\mathbf{0.20^{+0.06}_{-0.07}}$ at 90% C.L. Figure 3-6 and Figure 3-7 show clearly that the recent preliminary and upcoming RHIC data are expected to reduce the present uncertainties on the truncated integral even further by about a factor of 2 at $x_{min} = 10^{-3}$ (for more details see Section 3.1).

The production of $W^{\pm}$ bosons in longitudinally polarized proton-proton collisions serves as an elegant tool to access valence and sea quark helicity distributions at a high scale $Q \sim M_W$ and without the additional input of fragmentation functions as in semi-inclusive DIS. While the valence quark helicity densities are already well known at intermediate $x$ from DIS, the sea quark helicity PDFs are only poorly constrained. The latter are of special interest due to the differing predictions in various models of nucleon structure (see Ref. 13). The 2011 and the high statistics 2012 longitudinally polarized $p$+$p$ data sets provided the first results for $W^{\pm}$ with substantial impact on our knowledge of the light sea (anti-) quark polarizations (see Figure 3-9 and Figure 3-10 and Section 3.2). With the complete data from 2011 to 2013 data sets analyzed by both the PHENIX and STAR experiments the expected uncertainties (Figure 3-11) will one allow to measure the integrals of the $\Delta\bar{u}$ and $\Delta\bar{d}$ helicity in the accessed $x$ range above 0.05 (see Figure 3-13). The uncertainty on the flavor asymmetry for the polarized light quark sea $\Delta\bar{u} - \Delta\bar{d}$ will also be further reduced and a measurement on the $2\sigma$ level will be possible (see Figure 3-12). These results demonstrate that the RHIC $W$ program will lead, once all the recorded data are fully analyzed, **to a substantial improvement in the understanding of the light sea quark polarizations in the nucleon.**

- ***How do quarks and gluons hadronize into final-state particles?***
  ***How can we describe the multidimensional landscape of nucleons and nuclei?***

In recent years, transverse spin phenomena have gained substantial attention as they offer the unique opportunity to expand our current one-dimensional picture of the nucleon by imaging the proton in both momentum and impact parametelongitudinalr space. At the same time we can further understand the basics of color interactions in QCD and how they manifest themselves in different processes. Results from PHENIX and STAR have shown that large transverse spin asymmetries for inclusive hadron production that were seen in $p$+$p$ collisions at fixed-target energies and modest $p_T$ extend to the highest RHIC energies and surprisingly large $p_T$. In the recent years the focus has shifted to observables that will help to separate the contributions from the initial and final state effects, and will give insight to the transverse spin structure of hadrons.

Recent results from transversely polarized data taken in 2006, 2011, and 2012, demonstrate for **the first time that transversity is accessible in polarized proton collisions** at RHIC through observables involving the Collins FF times the quark transversity distribution and the IFF times the quark transversity distribution accessed through single spin asymmetries of the azimuthal distributions of hadrons inside a high energy jet and the azimuthal asymmetries of pairs of oppositely charged pions respectively (see Figure 4-2 and Figure 4-3 and section 4.1) at $\sqrt{s}$ = 200 and 500 GeV.

Among the quantities of particular interest to give insight to the transverse spin structure of hadrons is the "Sivers function", because it encapsulates the correlations between a parton's transverse momentum inside the proton and the proton spin. It was found that the Sivers function is not universal in hard-scattering processes, which has as its physical origin what can visually be described as a rescattering of the struck parton in the color field of the remnant of the polarized proton.

The experimental test of this non-universality is one of the big remaining questions in hadronic physics and it is deeply connected to our understanding of QCD factorization. RHIC provides the unique opportunity for the ultimate test of the theoretical concept of TMDs, factorization, evolution and non-universality, by measuring $A_{\mathrm{N}}$ for $W^{\pm}$, $Z^{0}$ boson, *DY* production, and direct photons (for details see Section 4.2).

- ***What is the nature of the initial state in nuclear collisions?***

Using RHIC's world wide unique capability of polarised $p^{\uparrow}$+A collisions gives the unique opportunity to progress our quest to understand QCD processes in Cold Nuclear Matter by studying the dynamics of partons at very small and very large momentum fractions $x$ in nuclei, and at high gluon-density to investigate the existence of nonlinear evolution effects. Scattering a polarized probe on a saturated nuclear wave function provides a unique way of probing the gluon and quark transverse momentum distributions. In particular, the single transverse spin asymmetry $A_N$ may provide access to an elusive nuclear gluon distribution function, which is fundamental to the CGC formalism. In particular the nuclear dependence of $A_N$ may shed important light on the strong interaction dynamics in nuclear collisions (for details see Section 4.3).

To bring our understanding about these overarching questions in QCD at the very high or low region of Bjorken x to a new level new high precision data in an extended kinematic regime are needed, therefore both PHENIX and STAR are planning detector upgrades at forward rapidities, which are anticipated to come online at the start of the 2020s (for details see Section 5.1).

In summary, time and again, spin has been a key element in the exploration of fundamental physics. Spin-dependent observables have often revealed deficits in the assumed theoretical framework and have led to novel developments and concepts. The RHIC spin program plays a key role in this by using spin to study how a complex many-body system such as the proton arises from the dynamics of QCD.

# 7 APPENDIX

## 7.1 KINEMATIC VARIABLES

| Variable | Description |
|---|---|
| $x$ | longitudinal momentum fraction |
| $x_B$ | Bjorken scaling variable |
| $x_T$ | $x_T=2p_T/\sqrt{s}$ |
| $x_F$ | Feynman-$x$ ($x_F \sim x_1$-$x_2$) |
| $Q$ | virtuality of the exchanged photon in DIS |
| $s$ | The squared collision energy $s = 4\ E_{p1}E_{p2}$ |
| $\sqrt{s}$ | Center-of-mass energy |
| $p_T$ | Transverse momentum of final state particles, i.e. jet, hadrons |
| $k_T$ | Transverse momentum of partons |
| $j_T$ | Momentum of a hadron transverse to the jet thrust axis |
| $b_T$ | Transverse position of parton inside the proton |
| $\xi$ | Parton skewness: $x_B/(2- x_B)$ |
| $\eta$ | Pseudorapidity |
| $y$ | rapidity |
| $\cos\theta$ | $\theta$:polar angle of the decay electron in the partonic c.m.s., with $\theta > 0$ in the forward direction of the polarized parton |
| $\phi_s$, $\phi_h$ | Azimuthal angles of the final state hadron and the transverse polarization vector of the nucleon with respect to the proton beam |
| $\phi$ | Azimuthal angles of the final state hadron with respect to the proton beam |

Table 7-1: Definition of all kinematic variables used in the document

## 7.2 RHIC SPIN PUBLICATIONS

1. The RHIC Spin Program: Achievements and Future Opportunities
   E.C. Aschenauer et al., arXiv:1304.0079
   Citations: 28
2. Plans for the RHIC Spin Physics Program
   G. Bunce et al., http://www.bnl.gov/npp/docs/RHICst08_notes/spinplan08_really_final_060908.pdf

***ANDY:***

1. Cross Sections and Transverse Single-Spin Asymmetries in Forward Jet Production from Proton Collisions at √s=500 GeV.
   L. Bland et al, arXiv:1304.1454
   Citations: 19

***BRAHMS:***

1. Cross-sections and single spin asymmetries of identified hadrons in p+p at √s=200 GeV.
   J.H. Lee and F. Videbaek arXiv:0908.4551.
   Citations: 7
2. Single transverse spin asymmetries of identified charged hadrons in polarized p+p collisions at √s=62.4 GeV.
   I. Arsene et al. Phys. Rev. Lett. 101 (2008) 042001.
   **Citations: 100**

## *pp2pp:*

1. Double Spin Asymmetries $A_{NN}$ and $A_{SS}$ at √s=200 GeV in Polarized Proton-Proton Elastic Scattering at RHIC.
   pp2pp Collaboration, Phys. Lett. B647 (2007) 98
   Citations: 8
2. First Measurement of $A_N$ at √s=200 GeV in Polarized Proton-Proton Elastic Scattering at RHIC.
   pp2pp Collaboration, Phys. Lett. B632 (2006) 167
   Citations: 13
3. First Measurement of Proton-Proton Elastic Scattering at RHIC.
   pp2pp Collaboration, Phys. Lett. B579 (2004) 245
   Citations: 25

## *PHENIX:*

1. Inclusive cross sections, charge ratio and double-helicity asymmetries for $\pi^+$ and $\pi^-$ production in p+p collisions at √s=200 GeV.
   PHENIX Collaboration, accepted by Phys. Rev. D; arXiv:1409.1907.
   Citations: --
2. Cross Section and Transverse Single-Spin Asymmetry of η-Mesons in $p^\uparrow$+p Collisions at √s=200 GeV at Forward Rapidity.
   PHENIX Collaboration, Phys.Rev. D90 (2014) 072008
   Citations: 5
3. Inclusive double-helicity asymmetries in neutral pion and eta meson production in  collisions at √s=200 GeV.
   PHENIX Collaboration, Phys.Rev. D90 (2014) 012007.
   Citations: 9
4. Measurement of transverse-single-spin asymmetries for midrapidity and forward-rapidity production of hadrons in polarized p+p collisions at √s=200 and 62.4 GeV.
   PHENIX Collaboration, Phys. Rev. D 90, 012006
   Citations: 15
5. Inclusive cross section and single transverse spin asymmetry for very forward neutron production in polarized p+p collisions at √s=200  GeV.
   PHENIX Collaboration, Phys.Rev. D88 (2013) 3, 032006.
   Citations: 5
6. Double Spin Asymmetry of Electrons from Heavy Flavor Decays in p+p Collisions at √s=200 GeV.
   PHENIX Collaboration, Phys.Rev. D87 (2013) 012011.
   Citations: 6
7. Cross sections and double-helicity asymmetries of midrapidity inclusive charged hadrons in p+p collisions at √s=62.4 GeV.
   PHENIX Collaboration, Phys.Rev. D86 (2012) 092006.
   Citations: 8
8. Cross section and double helicity asymmetry for $\eta$ mesons and their comparison to neutral pion production in p+p collisions at √s=200 GeV.
   PHENIX Collaboration, Phys.Rev. D83 (2011) 032001.
   Citations: 27
9. Measurement of Transverse Single-Spin Asymmetries for J/Ψ Production in Polarized p+p Collisions at √s=200 GeV.
   PHENIX Collaboration, Phys.Rev. D82 (2010) 112008, Erratum-ibid. D86 (2012) 099904.
   Citations: 23
10. Event Structure and Double Helicity Asymmetry in Jet Production from Polarized p+p Collisions at √s=200 GeV.
    PHENIX Collaboration, Phys.Rev. D84 (2011) 012006.
    Citations: 13

11. Cross Section and Parity Violating Spin Asymmetries of $W^{\pm}$ Boson Production in Polarized p+p Collisions at $\sqrt{s}$=500 GeV.
    PHENIX Collaboration, Phys.Rev.Lett. 106 (2011) 062001.
    Citations: 48
12. Double-Helicity Dependence of Jet Properties from Dihadrons in Longitudinally Polarized p+p Collisions at $\sqrt{s}$=200 GeV.
    PHENIX Collaboration, Phys.Rev. D81 (2010) 012002.
    Citations: 3
13. Inclusive cross section and double helicity asymmetry for $\pi^0$ production in p+p collisions at $\sqrt{s}$=62.4 GeV.
    PHENIX Collaboration, Phys.Rev. D79 (2009) 012003.
    **Citations: 74**
14. The Polarized gluon contribution to the proton spin from the double helicity asymmetry in inclusive $\pi^0$ production in polarized p+p collisions at $\sqrt{s}$=200 GeV.
    PHENIX Collaboration, Phys.Rev.Lett. 103 (2009) 012003.
    **Citations: 81**
15. Inclusive cross-section and double helicity asymmetry for $\pi^0$ production in p+p collisions at $\sqrt{s}$=200 GeV: Implications for the polarized gluon distribution in the proton.
    PHENIX Collaboration, Phys.Rev. D76 (2007) 051106.
    **Citations: 190**
16. Improved measurement of double helicity asymmetry in inclusive midrapidity $\pi^0$ production for polarized p+p collisions at $\sqrt{s}$=200 GeV.
    PHENIX Collaboration, Phys.Rev. D73 (2006) 091102.
    Citations: 42
17. Measurement of transverse single-spin asymmetries for mid-rapidity production of neutral pions and charged hadrons in polarized p+p collisions at $\sqrt{s}$=200 GeV.
    PHENIX Collaboration, Phys.Rev.Lett. 95 (2005) 202001.
    **Citations: 147**
18. Double helicity asymmetry in inclusive mid-rapidity $\pi^0$ production for polarized p+p collisions at $\sqrt{s}$=200 GeV.
    PHENIX Collaboration , Phys.Rev.Lett. 93 (2004) 202002.
    **Citations: 77**
19. Mid- rapidity neutral pion production in proton proton collisions at $\sqrt{s}$ = 200-GeV
    PHENIX Collaboration, Phys.Rev.Lett. 91 (2003) 241803.
    **Citations: 325**

***STAR:***

1. Measurement of longitudinal spin asymmetries for weak boson production in polarized proton-proton collisions at RHIC.
   STAR Collaboration, Phys.Rev.Lett. 113 (2014) 072301.
   Citations: 5
2. Precision Measurement of the Longitudinal Double-spin Asymmetry for Inclusive Jet Production in Polarized Proton Collisions at $\sqrt{s}$ =200 GeV,
   STAR Collaboration, arXiv: 1405.5134 [hep-ex], submitted to PRL.
   Citations: 6
3. Neutral pion cross section and spin asymmetries at intermediate pseudorapidity in polarized proton collisions at $\sqrt{s}$=200 GeV,
   STAR Collaboration, Phys. Rev. D **89** (2014) 012001.
   Citations: 5
4. Single Spin Asymmetry $A_N$ in Polarized Proton-Proton Elastic Scattering at $\sqrt{s}$=200 GeV.
   STAR Collaboration, Phys. Lett. B 719 (2013) 62.
   Citations: 7

5. Transverse Single-Spin Asymmetry and Cross-Section for $\pi^0$ and $\eta$ Mesons at Large Feynman-x in Polarized p+p Collisions at √s=200 GeV.
   STAR Collaboration, Phys. Rev. D 86 (2012) 51101.
   Citations: 24
6. Longitudinal and transverse spin asymmetries for inclusive jet production at mid-rapidity in polarized p+p collisions at √s=200 GeV.
   STAR Collaboration, Phys. Rev. D 86 (2012) 32006.
   Citations: 30
7. Measurement of the W → e ν and Z/γ*→$e^+e^-$ Production Cross Sections at Mid-rapidity in Proton-Proton Collisions at √s=500 GeV.
   STAR Collaboration, Phys. Rev. D 85 (2012) 92010.
   Citations: 15
8. Measurement of the parity-violating longitudinal single-spin asymmetry for $W^{\pm}$ boson production in polarized proton-proton collisions at √s=500 GeV,
   STAR Collaboration, Phys. Rev. Lett. 106 (2011) 62002.
   Citations: 39
9. Longitudinal double-spin asymmetry and cross section for inclusive neutral pion production at midrapidity in polarized proton collisions at √=200 GeV.
   STAR Collaboration, Phys.Rev. D80 (2009) 111108.
   Citations: 20
10. Longitudinal Spin Transfer to Lambda and anti-Lambda Hyperons in Polarized Proton-Proton Collisions at √s=200 GeV.
    STAR Collaboration, Phys.Rev. D80 (2009) 111102.
    Citations: 5
11. Forward Neutral Pion Transverse Single Spin Asymmetries in p+p Collisions at √s=200 GeV.
    STAR Collaboration, Phys.Rev.Lett. 101 (2008) 222001.
    **Citations: 123**
12. Longitudinal double-spin asymmetry for inclusive jet production in p+p collisions at √s=200 GeV.
    STAR Collaboration, Phys.Rev.Lett. 100 (2008) 232003.
    **Citations: 104**
13. Measurement of transverse single-spin asymmetries for di-jet production in proton-proton collisions at √s=200 GeV.
    STAR Collaboration, Phys.Rev.Lett. 99 (2007) 142003.
    Citations: 36
14. Longitudinal double-spin asymmetry and cross section for inclusive jet production in polarized proton collisions at √s=200 GeV.
    STAR Collaboration, Phys.Rev.Lett. 97 (2006) 252001.
    **Citations: 178**
15. Cross-sections and transverse single spin asymmetries in forward neutral pion production from proton collisions at √s=200 GeV.
    STAR Collaboration, Phys.Rev.Lett. 92 (2004) 171801.
    **Citations: 277**

# 7.3 BIBLIOGRAPHY

[1] M. Klein and R. Yoshida, Prog. Part. Nucl. Phys. 62 (2008) 343.
P. Newman and M. Wing, Rev. Mod. Phys. 86 (2014) 1037.
[2] B.I. Abelev et al. (STAR Collaboration), Phys. Rev. Lett. 100, 232003 (2008).
[3] D. de Florian, R. Sassot, M. Stratmann, and W. Vogelsang, Phys. Rev. Lett. 101, 072001 (2008).
D. de Florian, R. Sassot, M. Stratmann, and W. Vogelsang, Phys. Rev. D 80, 034030 (2009).
[4] D. de Florian, R. Sassot, M. Stratmann, and W. Vogelsang, Phys. Rev. Lett. 113 (2014) 012001.
[5] G. Bunce, N. Saito, J. Soffer, and W. Vogelsang, *Ann. Rev. Nucl. Part. Sci.* **50** (2000) 525.
[6] PHENIX Collaboration, Phys.Rev. D90 (2014) 012007.
[7] STAR Collaboration, arXiv: 1405.5134 [hep-ex], submitted to PRL
[8] E. Leader, A.V. Sidorov, and D.B. Stamenov, Phys. Rev. D 82, 114018 (2010).
[9] The NNPDF Collaboration: E.R. Nocera, R.D. Ball, St. Forte, G. Ridolfi and J. Rojo, Nucl.Phys. B887 (2014) 276, arXiv:1406.5539
[10] Ch. Dilks, Spin 2014 Conference, https://indico.cern.ch/event/284740/session/27/contribution/126/material/slides/0.pdf
[11] M. Glück et al., Phys. Rev. D 63, 094005 (2001).
[12] PHENIX-Collaboration, http://www.phenix.bnl.gov/WWW/plots/show_plot.php?editkey=p1169
[13] W.C. Chang and J.C. Peng, Prog.Part.Nucl.Phys. 79 (2014) 95, 1406.1260.
[14] PHENIX Collaboration, Phys.Rev.Lett. 106 (2011) 062001.
STAR Collaboration, Phys. Rev. Lett. **106**, 062002 (2011).
[15] STAR Collaboration, Phys.Rev.Lett. 113 (2014) 072301; arXiv:1404.6880
[16] E.R. Nocera, Proceedings DIS-2014, https://inspirehep.net/record/1327109/files/PoS(DIS2014)204.pdf
[17] S. Fazio, talk at DIS-2014, http://indico.cern.ch/event/258017/session/11/contribution/219/material/slides/0.pptx
[18] H.-L. Lai et al., Phys.Rev. D82 (2010) 074024
[19] J. Campbell, K. Ellis, C. Williams, MCFM - Monte Carlo for FeMtobarn processes, http://mcfm.fnal.gov.
[20] R.S. Towell et al., Phys. Rev. D64, 052002, 2001
[21] A. V. Efremov and O. V. Teryaev, Sov. J. Nucl. Phys. 36, 140 (1982) [Yad. Fiz. 36, 242 (1982)];
Phys. Lett. B 150, 383 (1985).
J.-W. Qiu and G. F. Sterman, Phys. Rev. Lett. 67, 2264 (1991); Nucl. Phys. B 378, 52 (1992);
Phys. Rev. D 59, 014004 (1999)
[22] J. C. Collins and A. Metz, Phys.Rev.Lett. 93 (2004) 252001.
[23] T.C. Rogers and P.J. Mulders, Phys.Rev. D81 (2010) 094006
[24] J. Ralston and D.Soper, Nucl. Phys. **B 152**, 109 (1979).
[25] R. Jaffe and X. Ji, Nucl. Phys. **B 375**, 527 (1992).
[26] P. Mulders and R. Tangerman, Nucl. Phys. **B 461**, 197 (1996).
[27] J. C. Collins, Nucl. Phys. **B 396**, 161 (1993).
[28] J. C. Collins, S. F. Heppelmann, and G. A. Ladinsky, Nucl. Phys. **B 420**, 565 (1994).
[29] A. Airapetian *et al.* (HERMES Collaboration), Phys. Rev. Lett. **94**, 012002 (2005); **103**, 152002 (2009);
Phys. Lett. **B 693**, 11 (2010).
[30] V. Y. Alexakhin *et al.* (COMPASS Collaboration), Phys. Rev. Lett. **94**, 202002 (2005);
E. S. Ageev et al. (COMPASS Collaboration), Nucl. Phys. **B 765**, 31 (2007);
M. G. Alekseev et al. (COMPASS Collaboration), Phys. Lett. **B 673**, 127 (2009);
M. G. Alekseev et al. (COMPASS Collaboration), Phys. Lett. **B 692**, 240 (2010);
C. Adolphetal. (COMPASS Collaboration), Phys. Lett. **B 717**, 376 (2012);
N. Makke (COMPASS Collaboration), arXiv:1403.4218.
[31] A. Airapetian *et al.* (HERMES Collaboration), JHEP **0806**, 017 (2008).
[32] C. Adolph *et al.* (COMPASS Collaboration), Phys. Lett. **B 713**, 10 (2012).
[33] R. Seidl *et al.* (Belle Collaboration), Phys. Rev. Lett **96**, 232002 (2006); Phys. Rev. **D 86**, 039905(E) (2012).
[34] A. Vossen *et al.* (Belle Collaboration), Phys. Rev. Lett. **107**, 072004 (2011).
[35] M. Anselmino *et al.,* Phys. Rev. **D 75**, 054032 (2007);
Nucl. Phys. B, Proc. Suppl. **191**, 98 (2009);
Phys. Rev. **D 87**, 094019 (2013).
[36] A. Bacchetta, A. Courtoy, and M. Radici, Phys. Rev. Lett. **107**, 012001 (2011).
[37] F.Yuan, Phys. Rev. Lett. 100, 032003 (2008); Phys.Rev.D77 074019 (2008)
[38] U. D'Alesio, F. Murgia, and C. Pisano, Phys. Rev. D 83, 034021 (2011)
[39] A. Bacchetta and M. Radici, Phys. Rev. D 70, 094032.
[40] Kevin Adkins, Proceedings Spin 2014
[41] A. Vossen, *Il Nuovo Cimento* **C 35**, 59 (2012).

[42] Y. Pan, Proceedings Spin 2014
[43] Z.-B. Kang, Phys. Rev. **D 83**, 036006 (2011).
[44] R. L. Jaffe and X. Ji, Phys. Rev. Lett. **67**, 552 (1991); Nucl. Phys. **B 375**, 527 (1992).
[45] L. Adamczyk et al., Phys. Rev. **D 89** (2014) 012001; arXiv:1309.1800
[46] B. I. Abelev *et al.*, Phys. Rev. Lett. **101**, 222001 (2008), arXiv: 0801.2990
[47] L. Adamczyk et al., Phys. Rev. **D 86** (2012) 051101; arXiv:1205.6826
[48] K. Kanazawa, Y. Koike, A. Metz, D. Pitonyak, arXiv:1404.1033.
[49] S. Heppelmann, CIPANP-2012,
https://indico.triumf.ca/getFile.py/access?contribId=349&sessionId=44&resId=0&materialId=slides&confId=1383
[50] B. I. Abelev *et al.* (STAR Collaboration), Phys. Rev. Lett. **101**, 222001 (2008).
[51] M. M. Mondal, talk at DIS-2014:
http://indico.cern.ch/event/258017/session/6/contribution/216/material/slides/1.pptx
[52] $A_N$DY collaboration, arXiv:1304.1454
[53] L. Gamberg, Z.-B. Kang, and A. Prokudin, Phys.Rev.Lett. 110 (2013) 23, 232301
[54] Z.-B Kang, J.-W. Qiu, W. Vogelsang, and F. Yuan, Phys.Rev. D83 (2011) 094001
[55] STAR Beam Use Request for Run-2015 and Run-2016: https://drupal.star.bnl.gov/STAR/starnotes/public/sn0606
[56] STAR Collaboration, "A polarized p+p and p+A program for the next years"
https://drupal.star.bnl.gov/STAR/starnotes/public/sn0605
[57] S. S. Adler et al. (PHENIX Collaboration), Phys. Rev. Lett. 95, 202001 (2005)
[58] M. Anselmino, U. D'Alesio, S. Melis, and F. Murgia, Phys. Rev. D 74, 094011 (2006).
[59] P. Sun and F. Yuan, Phys. Rev. D. 88 (2013) 114012
[60] P. Sun, J. Isaacson, C.-P.Yuan and F. Yuan aeXiv:1406.3073
[61] M.G. Echevarria, A. Idilbi, Z.-B. Kang and I. Vitev, Phys. Rev. **D 89** (2014) 074013, arXiv:1401.5078
[62] J. Collins, arXiv:1409.5408
[63] Z.-B. Kang and J.-W. Qiu, Phys. Rev. Lett. **103** (2009) 172001, arXiv:0903.3629
[64] A. Metz and J. Zhou, Phys. Lett. B **700** (2011) 11, arXiv:1006.3097
[65] E.A. Hawker et al., Phys. Rev. Lett. **80**, 3715 (1998)
[66] PHENIX Beam Use Request for Run-2015 and Run-2016:
https://indico.bnl.gov/materialDisplay.py?materialId=0&confId=764
[67] L. Gamberg, Z.-B. Kang, and A. Prokudin, Phys. Rev. Lett. 110, 232301 (2013).
[68] H. Weigert, Prog. Part. Nucl. Phys. **55** (2005) 461.
E. Iancu and R. Venugopalan, hep-ph/0303204.
F. Gelis, E. Iancu, J. Jalilian-Marian, and R. Venugopalan, Ann. Rev. Nucl. Part. Sci. 60 (2010) 463.
Y. V. Kovchegov and E. Levin, Quantum Chromodynamics at High Energy. Cambridge University Press, 2012.
J. Jalilian-Marian and Y. V. Kovchegov, Prog. Part. Nucl. Phys. **56** (2006) 104.
[69] Y.V. Kovchegov and M.D. Sievert, Phys. Rev. **D 86**, 034028 (2012),
Erratum-ibid. **D 86**, 079906 (2012)
arXiv:1201.5890
[70] A. Metz and J. Zhou, Phys. Rev. **D 84** (2011) 051503.
F. Dominguez, J.-W. Qiu, B.-W. Xiao, and F. Yuan, Phys.Rev. **D 85** (2012) 045003; J. Jalilian-Marian, A. Kovner, L. D. McLerran, and H. Weigert, Phys. Rev. **D 55** (1997) 5414.
[71] Y. V. Kovchegov and A. H. Mueller, Nucl. Phys. **B529** (1998) 451.
F. Dominguez, C. Marquet, B.-W. Xiao, and F. Yuan, Phys. Rev. **D 83** (2011) 105005.
[72] D. Boer, A. Dumitru, and A. Hayashigaki, Phys. Rev. **D 74** (2006) 074018.
D. Boer and A. Dumitru, Phys. Lett. **B556** (2003) 33.
D. Boer, A. Utermann, and E. Wessels, Phys. Lett. **B671** (2009) 91.
[73] Z.-B. Kang and F. Yuan, Phys. Rev. **D 84** (2011) 034019.
[74] Y. V. Kovchegov and M. D. Sievert, Phys. Rev. **D 86** (2012) 034028.
[75] J.-W. Qiu, talk at the workshop on "Forward Physics at RHIC", RIKEN BNL Research Center, BNL, 2012.
[76] D. W. Sivers, Phys. Rev. **D 41** (1990) 83.
[77] J. C. Collins, Phys. Lett. **B536** (2002) 43.
[78] M. Walker *et al.*, arXiv:1107.0917.
[79] L. Gamberg and Z.-B. Kang, arXiv:1208.1962.
[80] D. Mueller, Fortschr. Phys. **42** (1994) 171;
X.-D. Ji, Phys. Rev. Lett. 78, 610 (1997); J. Phys. **G24** (1998) 1181;
A.V. Radyushkin, Phys.Lett. B380 (1996) 417;
M. Burkardt, Phys.Rev. D62 (2000) 071503; Erratum-ibid. D66 (2002) 119903
[81] S. Klein and J. Nystrand, hep-ph/0310223.
[82] T. Toll and T. Ullrich, Phys.Rev. **C 87** (2013) 024913

[83] The PHENIX Collaboration, "Future Opportunites for p+p and p+A at RHIC",
http://www.phenix.bnl.gov/phenix/WWW/publish/dave/sPHENIX/pp_pA_whitepaper.pdf
[84] C. Aidala et al. sPHENIX: An Upgrade Concept from the PHENIX Collaboration.2012. arXiv:1207.6378.
[85] RHIC Collider Projections (FY 2014 – FY 2022), http://www.rhichome.bnl.gov/RHIC/Runs/RhicProjections.pdf
[86] C. Aidala et al. sPHENIX: An Upgrade Concept from the PHENIX Collaboration.2012. arXiv:1207.6378.
[87] A. Adare et al. Concept for an Electron Ion Collider (EIC) detector built around the BaBar solenoid. 2014. arXiv:1402.1209.
[88] COMPASS-II Proposal, 2010.
URL: http://wwwcompass.cern.ch/compass/proposal/compass-II_proposal/compass-II_proposal.pdf.
[89] D. Sivers, Phys. Rev. **D 41**, 83 (1990); 43, 261 (1991).
[90] A. V. Efremov and O. V. Teryaev, Sov. J. Nucl. Phys. **36**, 140 (1982); Phys. Lett. **150B**, 383 (1985); J.-W. Qiu and G. Sterman, Phys. Rev. Lett. **67**, 2264 (1991); Phys. Rev. **D 59**, 014004 (1998).
[91] D. Boer and P. J. Mulders, Phys. Rev. **D 57**, 5780 (1998).
[92] T. Rogers, Phys. Rev. **D 88**, 1, 014002 (2013).
[93] E.C. Aschenauer et al., eRHIC Design Study: An Electron-Ion-Collider at BNL, arXiv:1409:1633 (2014)
A. Accardi et al., Electron Ion Collider: The Next QCD Frontier - Understanding the glue that binds us all, arXiv:1212.1701 (2014).

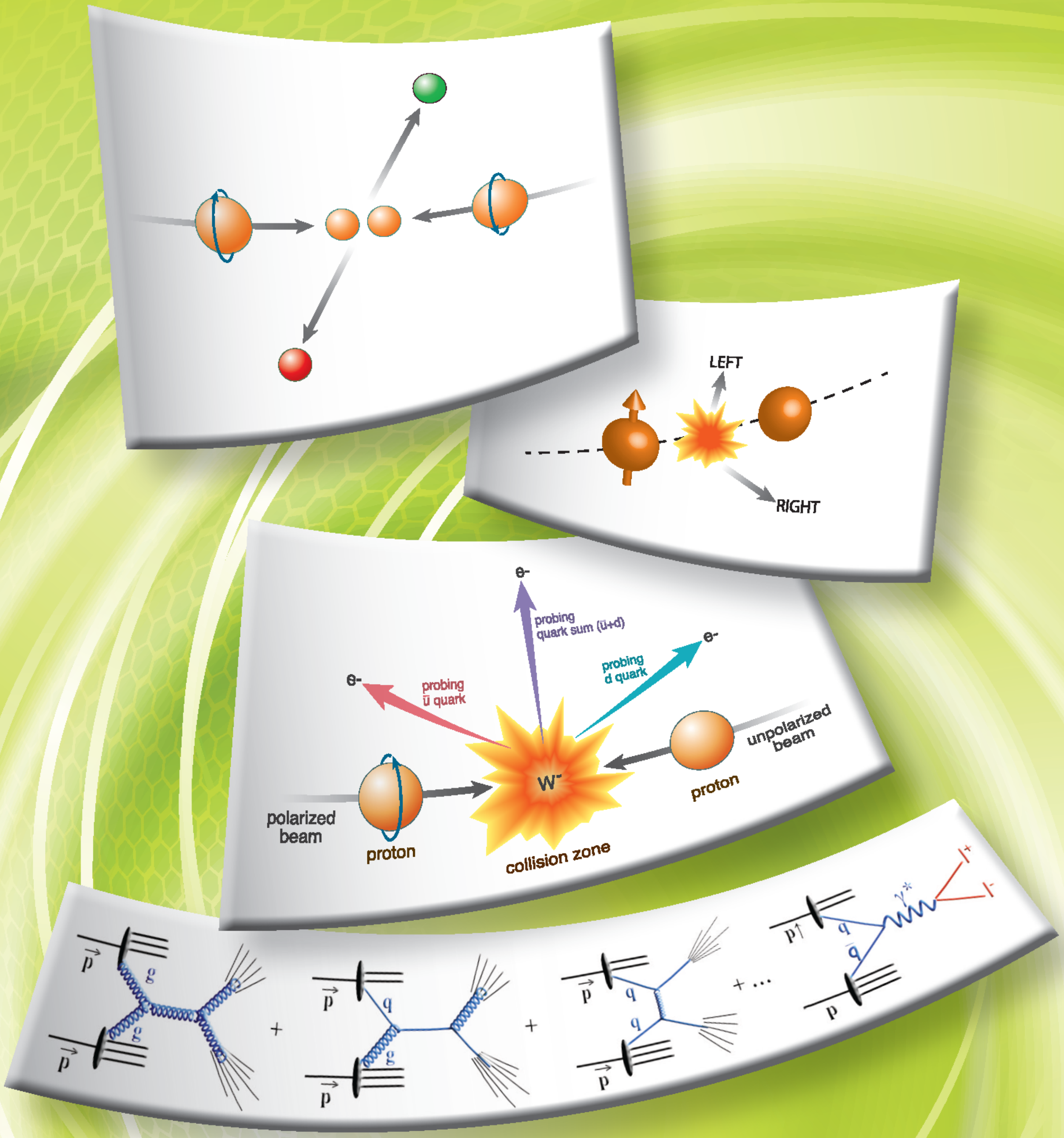

BROOKHAVEN
NATIONAL LABORATORY

U.S. DEPARTMENT OF ENERGY | Office of Science